\documentclass{emulateapj}

\shorttitle{MW Formation Through Halo Substructure. I.}
\shortauthors{Schlaufman et al.}
\submitted{Received 2009 June 29; accepted 2009 August 10}

\begin{document}

\title{Insight Into the Formation of the Milky Way Through Cold Halo
Substructure. I. The ECHOS of Milky Way Formation}

\author{Kevin C. Schlaufman\altaffilmark{1,5},
Constance M. Rockosi\altaffilmark{1,6,7}, Timothy C. Beers\altaffilmark{2},
Dmitry Bizyaev\altaffilmark{3}, Howard Brewington\altaffilmark{3},
Young Sun Lee\altaffilmark{2}, Viktor Malanushenko\altaffilmark{3},
Elena Malanushenko\altaffilmark{3}, Dan Oravetz\altaffilmark{3},
Kaike Pan\altaffilmark{3}, Audrey Simmons\altaffilmark{3}, Stephanie
Snedden\altaffilmark{3}, and Brian Yanny\altaffilmark{4}}

\altaffiltext{1}{Astronomy and Astrophysics Department, University of
California, Santa Cruz, CA 95064; kcs@ucolick.org and crockosi@ucolick.org}
\altaffiltext{2}{Dept. of Physics and Astronomy, CSCE: Center for the Study of
Cosmic Evolution, and JINA: Joint Institute for Nuclear Astrophysics, Michigan
State University, E. Lansing, MI 48824; beers@pa.msu.edu and lee@pa.msu.edu}
\altaffiltext{3}{Apache Point Observatory, Sunspot, NM 88349; dmbiz, hjbrew,
viktorm, elenam, doravetz, kpan, asimmons, and snedden@apo.nmsu.edu}
\altaffiltext{4}{Fermi National Accelerator Laboratory, P.O. Box 500, Batavia,
IL 60510, USA; yanny@fnal.gov}
\altaffiltext{5}{NSF Graduate Research Fellow}
\altaffiltext{6}{University of California Observatories}
\altaffiltext{7}{Packard Fellow}

\begin{abstract}
We identify ten -- seven for the first time -- elements of cold halo
substructure (ECHOS) in the volume within 17.5 kpc of the Sun in the
inner halo of the Milky Way.  Our result is based on the observed spatial
and radial velocity distribution of metal-poor main sequence turnoff
(MPMSTO) stars in 137 Sloan Extension for Galactic Understanding and
Exploration (SEGUE) lines of sight.  We point out that the observed radial
velocity distribution is consistent with a smooth stellar component of
the Milky Way's inner halo overall, but disagrees significantly at the
radial velocities that correspond to our detections.  We show that all
of our detections are statistically significant and that we expect no
false positives.  These ECHOS represent the observable stellar debris of
ancient merger events in the stellar accretion history of the Milky Way,
and we use our detections and completeness estimates to infer a formal
upper limit of $0.34^{+0.02}_{-0.02}$ on the fraction of the MPMSTO
population in the inner halo that belong to ECHOS.  Our detections
and completeness calculations also suggest that there is a significant
population of low fractional overdensity ECHOS in the inner halo, and we
predict that 1/3 of the inner halo (by volume) harbors ECHOS with
MPMSTO star number densities $n \approx 15$ kpc$^{-3}$.  In addition,
we estimate that there are of order 10$^3$ ECHOS in the entire inner halo.
ECHOS are likely older than known surface brightness substructure,
so our detections provide us with a direct measure of the accretion
history of the Milky Way in a region and time interval that has yet to
be fully explored.  In concert with previous studies, our result suggests
that the level of merger activity has been roughly constant over the past
few Gyr and that there has been no accretion of single stellar systems
more massive than a few percent of a Milky Way mass in that interval.
\end{abstract}


\section{Introduction}

The stellar halo of our Galaxy is an excellent place to study the
residuals of its formation, because the timescale over which those
residuals disappear is long relative to the other parts of the Galaxy.
Moreover, the stellar populations in the halo of our Galaxy are
predominantly old and metal-poor.  This fact implies that most of
the stars in the halo date to the earliest stages of the Milky Way's
formation.  Indeed, \citet{egg62} used the dynamical and chemical
signature of high proper-motion stars to deduce that metal-poor
stars in the solar neighborhood are preferentially on radial orbits.
They interpreted this as the result of star formation during a rapid
collapse of the nascent Milky Way after its own self-gravity took
over from universal expansion.  Later studies of globular clusters like
those of \citet{sea78} revealed that the distribution of globular cluster
metallicities beyond eight kpc from the center of the Galaxy is broad and
independent of galactocentric radius.  They also found that differences in
color-magnitude morphology were uncorrelated with metallicity, contrary
to expectations of a smooth collapse model.  \citet{sea78} interpreted
these observations as evidence of multiple episodes of star formation
in the halo over an extended period.  Detailed classical studies of our
Galaxy as a whole \citep[e.g.][]{bah80,gil83}, and of the stellar halo
in particular, showed that the halo was not well described by the same
$\rho^{1/4}$ power-law that describes the bulge \citep[e.g.][]{mor93}.
Instead, the halo follows a power-law $\rho^{\alpha}$ with $\alpha
\approx -3.5$ \citep[e.g.][]{harr76,zin85} and is roughly spherically
symmetric at large galactocentric radii but more oblate closer in
\citep[e.g.][]{pre91,chi00}.  These classical observations implied that
stars in halo were generally, but not always \citep[cf.][]{rat85,maj96},
smoothly distributed in both coordinate-space and velocity-space.

The hierarchical model of structure formation \citep[e.g.][]{pre74,whi78}
in a $\Lambda$CDM universe can make predictions that match
observations of large-scale structure and galaxy clustering,
as well as many characteristics of individual galaxies
\citep[e.g.][]{bul05,rob05,spr05,bow06,cro06,fon06}.  In general,
the agreement between theory and observation is best at the largest
scales where the physics is dominated by cosmology and dark matter.
However, that agreement becomes increasingly model dependent on the
smaller galactic and sub-galactic scales where baryons are important.
Theorists have taken several approximate approaches to better understand
the formation of Milky Way analog halos: cosmological dark matter only
$n$-body simulations with live potentials \citep{die07,die08,spr08},
cosmological dark matter only $n$-body simulations coupled to
semi-analytic models \citep{del08}, cosmological dark matter only $n$-body
simulations without live potentials but with added resolution coupled to
semi-analytic models \citep{hard01,bul05,rob05,fon06}, as well as full
cosmological hydrodynamic simulations \citep{abad03a,abad03b,gov07}.
While each theoretical approach has its relative advantages and
disadvantages, for the time being there is no way to self-consistently
track all of the necessary baryon physics important in galaxy formation
over a large dynamic range in spatial scale.  Nevertheless, a broad
consensus has emerged: all of these approaches suggest that the inner
halo formed early from the accretion of relatively massive protogalaxies
into the nascent Milky Way \citep[e.g.][]{bul05,abad06} and that
violent relaxation played a major role in producing the classically
observed kinematically smooth inner halo \citep[e.g.][]{die05}.
There is also growing evidence that stars formed in the disk of the
nascent Milky Way contribute to the stellar populations in the inner
halo \citep[e.g.][]{zol09}.  In addition, all of these calculations
predict the presence of substructure in the inner halo as a result of
accretion events more recent than the last episode of violent relaxation.
All of the referenced simulations are of Milky Way analogs, not the
Milky Way itself, so the observations meant to verify these predictions
must be statistical.  We are further limited by current technology
to observational comparisons of the Milky Way's halo, its satellites,
and Local Group companions with these theoretical models to test their
predictions for galaxy formation on small scales.

Fortunately, the halo of our own Galaxy provides us with an excellent
observational example for tests of this hierarchical formation scenario
\citep[for a recent review see][]{hel08}.  The halo's resolved stellar
populations are bright enough to study both photometrically and
spectroscopically over large fields-of-view with modern telescopes
and instrumentation.  The Sloan Digital Sky Survey \citep[SDSS
-][]{fuk96,gun98,yor00,hog01,smi02,pie03,ive04,gun06,tuc06} is
one such study that has made significant contributions to our
understanding of the Galaxy.  The earliest results to come out
of the SDSS \citep{ive00,yan00,che01} confirmed many classical
results with high precision, and at same time provided more evidence
that the halo was not entirely homogeneous.  Those hints have been
followed-up with many more detailed studies.  As a result of the SDSS
and other modern large-scale surveys there is now strong evidence
of substructure in the halo of the Milky Way from star count maps
\citep{tot98,tot00,ive00,yan00,ode01,viv01,gil02,new02,roc02,maj03,yan03,
roc04,duf06,bel06,gri06a,gri06b,viv06,bel07,bell08,jur08,gri09,wat09},
kinematic information
\citep{chi98,hel99b,chi00,kep07,ive08,kle08,sea08,kle09,smi09,sta09},
and chemical abundances \citep{ive08}.  The data from these large-scale
surveys have pushed the field beyond individual detections to systematic
statistical searches in which both detections and non-detections are
meaningful and can strongly inform theoretical models.  \citet{bell08}
used photometric overdensities in projected SDSS star counts to
statistically quantify the degree of substructure in the outer halo.
They showed that while the classical symmetric smooth model is a poor
match to the observations, the amount of substructure in the outer halo is
consistent with that expected from $\Lambda$CDM simulations.  Likewise,
\citet{caro07} revealed the two-component (inner/outer) nature of the
halo, a property that is naturally explained in the hierarchical model.
Accordingly, now that the existence of substructure in the halo of
our own Galaxy has been well-established, our challenge is to use the
observations of substructure to better understand the current state of
the Milky Way \citep[e.g.][]{hel04,joh05,mez05,fel06,hel06,wil09}, to inform
models of its formation, and to study the stellar populations and thereby
the star formation processes at work in accreted protogalaxies and the
disk of a nascent Milky Way.

It was recognized at least as early as \citet{egg62} that searching for
substructure in the halo of our own Galaxy using kinematic information
would be very informative, because the long dynamical times in the halo
ensure that substructure remains kinematically distinct for Gyr.  This is
in spite of potential degeneracies between progenitor mass, velocity
dispersion, and accretion time \citep[e.g.][]{joh98,hel99a,joh08}.
Ideally, that kinematic information would include all six phase-space
coordinates.  Indeed, surveys of the solar neighborhood using the
precise proper motions available in existing data sets have constrained
both the absolute number of dynamically distinct features in the local
halo as well as the fraction of the local halo population that belong
to those features \citep[e.g.][]{hel99a,hel99b,gou03}.  More recently,
\citet{mor09} analyzed available local volume data and suggested that
the data implied that violent relaxation was not efficient in the early
Milky Way.

Unfortunately, existing proper motion catalogs extending beyond 10 kpc
\citep[e.g. USNO-B -][]{mon03} can only provide 3 mas yr$^{-1}$ precision
\citep{mun04}.  That precision corresponds to a tangential velocity error
of about 140 km s$^{-1}$ at 10 kpc, a value at least as large as the halo
velocity dispersion at that distance.  Note that this error estimate
also neglects the fact that imprecise distance estimates can degrade
accuracy further.  At the same time, today's ground-based radial velocity
surveys like the SDSS, the Sloan Extension for Galactic Understanding
and Exploration \citep[SEGUE -][]{yan09,all08,lee08a,lee08b}, and
the RAdial Velocity Experiment \citep[RAVE -][]{ste06,zwi08} provide
precise radial velocities independent of any uncertainty in distance.
In the case of SEGUE, the radial velocity precision is at least an
order-of-magnitude better at 10 kpc than the tangential velocity
precision provided by USNO-B.  In the future, these radial velocity
surveys will be complimented by 10$^9$ space-based proper motions and
10$^8$ radial velocities from the Gaia satellite \citep{per01,lin08}.
Nevertheless, a search for substructure in radial velocity is timely
and beyond 10 kpc radial velocities will remain the most precisely
estimated phase-space component for the foreseeable future.  Moreover,
the information content of radial velocity data relative to tangential
velocity data is substantial \citep[e.g.][]{bov09}.

Determining the origin of substructures detected using only radial
velocity data is theoretically intractable because the properties
of the substructure left behind by an accretion event observable in
radial velocity are degenerate in progenitor mass, velocity dispersion,
orbital properties, and time since accretion.  Recently accreted,
massive, and compact substructure progenitors are likely to produce
broad radial velocity features while long ago accreted, low mass, and
diffuse substructure progenitors are likely to produce narrow radial
velocity features.  Nevertheless, the substructure left behind by a
long ago accreted massive progenitor can effectively masquerade as the
substructure that results from the more recent disruption of a less
massive progenitor, or as the substructure left behind by the even more
recent accretion of an even less massive but more diffuse progenitor.
Precisely measured proper motions, parallaxes, or metallicities and
$\alpha$-enhancements derived from spectroscopy can potentially break
these degeneracies and play an important role in connecting observations
of cold substructure to the properties of their progenitors.

In this paper, we describe a systematic, statistical search for elements
of cold halo substructure (ECHOS) in the inner halo using the SDSS-II Data
Release 7 radial velocity data.  Radial velocities are the most precisely
measured property of SEGUE stars in the inner halo at distances greater
than about 10 kpc.  Restricting our search to radial velocities allows us
to avoid the difficult-to-determine, heteroskedastic, and model-dependent
systematic errors potentially associated with both proper motions
and spectrophotometric parallaxes at that distance.  From this point
forward, ECHOS are defined as that substructure that manifests itself
as an overdensity in radial velocity-space.  Note that ECHOS are not
necessarily equivalent to the tidal streams that have been discovered as
photometric overdensities; to emphasize that point, we do not use the word
stream to describe any of our detections.  ``Cold" specifically implies
that the inherent radial velocity dispersion of the substructure we seek
is less than (or of order) our SEGUE radial velocity error estimates at
a given S/N.  In other words, cold substructure is that substructure for
which the inherent velocity dispersion is unresolved (or barely resolved)
in radial velocity in the SEGUE data.  Our search for cold substructure
has at least two advantages relative to a search for warmer substructure.
First, the radial velocity scale of the search is set naturally by
the observational errors and not artificially at some arbitrary value.
Second, as we will subsequently show, our sensitivity to substructure is
not dependent on the velocity dispersion of that substructure.  Our search
will not necessarily be sensitive to substructure that is now so well
mixed that its constituent stars are no longer close to each other in
velocity or position along their orbit; that is, we are not sensitive to
the substructure that is now so dispersed that it can only be recovered by
examining integrals of motion \citep[e.g.][]{hel99a,hel99b,kle08,kle09}.
Our search will also not necessarily be sensitive to early accretion
events that experienced the effects of violent relaxation.  Nevertheless,
our search for ECHOS still at least partially exploits the desirable
property of collisionless systems that substructure remains coherent in
velocity-space for much longer than it remains coherent in position-space.
Therefore, we are likely to find in our search volume substructures that
are on average older than those substructures discovered as photometric
overdensities in the same volume.  Our search technique can naturally
be used to estimate our completeness, and that property enables us to
extrapolate our result to determine the fraction of MPMSTO stars in the
inner halo that belong to ECHOS.  That estimate can reveal the level
of merger activity over the past few Gyr.  Subsequent spectroscopic
follow-up of the stars in individual ECHOS could probe through chemical
abundances the physics of the high redshift universe and the star
formation environments in the disk of the nascent Milky Way and in the
massive protogalaxies accreted by the Milky Way over the past few Gyr.
Finally, theoretical hierarchical models of Milky Way analog formation in
a $\Lambda$CDM universe make predictions about the existence, properties,
and degree of halo substructure that are testable by our search, and our
results have the potential to uniquely inform models of galaxy formation.

This paper is organized as follows: in \S2 we describe the SEGUE survey
and the data we use in this analysis.  In \S3 we describe how we identify
ECHOS in the observations, how we estimate our false positive rate and
our completeness, and we present our detections.  In \S4 we generalize
our results to the entire inner halo.  In \S5 we discuss the implications
of our findings for the formation of the Milky Way.  We summarize our
conclusions in \S6.

\section{Data}

The SEGUE survey obtained 240,000 moderate-resolution ($R \approx
1800$) fiber-fed spectra of Milky Way stars in the magnitude interval
$14.0<g<20.3$. The spectroscopic targets were selected using SDSS
photometry and additional SEGUE $ugriz$ imaging data at low Galactic
latitude and in the South Galactic Cap.  The SEGUE spectra were taken
on 212 pointings spread sparsely over the 11,663 square degrees in the
combined imaging surveys.  Each pointing covers a field of seven square
degrees, three degrees in diameter, and has 1180 science targets taken on
two 640-fiber spectroscopic ``plates''.  The targets are split between
a bright and faint plate in each pair at $r=17.8$ (for $g-r<0.55$).
The instrumentation, data processing pipelines, survey strategy, along
with radial velocity and atmospheric parameter accuracies are described
in \citet{yan09}, \citet{lee08a,lee08b}, \citet{all08}, and the SDSS-II
DR7 paper \citep{abaz09}.

The radial velocity accuracy of the SEGUE data is discussed in detail
in \citet{yan09}.  The systematic uncertainty is estimated using repeat
observations.  There are 20 randomly selected stars with magnitudes $r
\approx 17.8$ repeated on the bright and faint plates in each pair.
The two plates in each pointing are observed independently, often
on different nights, lunations, or even years; so these duplicates
are a fair test of the velocity zero-point shifts.  From these and
an additional 12 plates with mostly stellar targets repeated in the
course of the SDSS and SEGUE, we find that the mean plate-to-plate
variation is zero with standard deviation 1.8 km s$^{-1}$. We use the
duplicates and a realistic noise model applied to SEGUE spectra observed
at high resolution \citep{all08} to characterize the uncertainty in the
radial velocities as a function of signal to noise.  We find that the
uncertainty for stars at the mean $g-r$ color of our sample and [Fe/H]
$\approx$ -1.5 (typical of the inner halo) is 5.3 km s$^{-1}$ at $g=18$
and rises to 20 km s$^{-1}$ at $g=20.3$.

In this analysis, we use the subset of SEGUE targets selected as
metal-poor main sequence turnoff (MPMSTO) stars.  These stars are good
tracers of the inner halo because of their relatively large luminosity and
high number density as compared with more evolved stars.  At $g=20$ our
MPMSTO stars have distances greater than 12 kpc, which for most Galactic
latitudes is well out of the disk and in a region dominated by the halo.

We draw our MPMSTO sample from two of the SEGUE target selection
categories.  The first category is the ``F turnoff" stars as described
in \citet{yan09}.  This is a UV-excess selection in the $ugr$
color-color diagram, and is designed to preferentially pick out blue,
less line-blanketed halo stars rather than more metal-rich thick disk
stars. At moderate and high Galactic latitude the thick disk stars are
the majority population at the turnoff for magnitudes brighter than about
$r=18.5$.  The selection criteria are $g<20.3, 0.4<u-g<1.4, 0.2<g-r<0.7,
-0.7< P1(s)<-0.25$ where $P1(s)$ is defined to be a color parallel to
the blue branch of the $ugr$ stellar locus, $0.91(u-g)+0.415(g-r)-1.28$
\citep[see][]{hel03}.  Up to 200 fibers per pointing were allocated to
objects in this selection category.  If there were more 200 candidates,
as was usually the case, the spectroscopic targets were randomly selected
from the candidate list with some preference for brighter and bluer stars.

The second SEGUE target category we drew from for this analysis was
the red half of the ``BHB" selection.  Because blue horizontal branch
(BHB) stars are rare and valuable tracers of the distant halo, the SEGUE
selection used a generous red limit of $g-r<0.2$.  This has significant
overlap with the main sequence turnoff, and the reddest fraction of BHB
candidates selected this way contains very few true BHBs.  We add the
red BHB targets with $0.1<g-r<0.2$ to the F turnoff targets to create
our MPMSTO sample.  The selection criteria for both the BHB and F
turnoff categories evolved slightly over the course of the survey, but
not enough to significantly alter the global properties of our sample.
In any case, the SEGUE targets were selected homogeneously along each
individual line of sight.  Since we conduct our search and evaluate our
detection efficiency independently for each line of sight, any changes
between lines of sight should not affect our analysis.

We use an M~13 globular cluster fiducial from \citet{an08}, augmented at
the faint end by the M~13 fiducial in \citet{cle08}, to obtain approximate
distances to our MPMSTO stars.  We do this to determine the range in
distance and Galactic coordinates over which the SEGUE MPMSTO stars sample
the halo.  We use a cluster fiducial rather than a theoretical isochrone
because the former is likely to be a better match to the data at the
turnoff; we choose M~13 because it is near the mean metallicity of the
halo \cite[e.g.][]{rya91a,rya91b,caro07}.  We emphasize that we do not
use these distances in our search for substructure.  We find that 95\%
of the MPMSTO stars in the SEGUE sample are within 17.5 kpc of the Sun.
We therefore impose a maximum distance limit of 17.5 kpc on the sample.
To minimize contamination from the bulge and thick disk, we further
restrict the sample to be in the inner halo.  We define the inner halo
as stars that are: more than ten kpc from the center of the Galaxy, at
vertical distance $|z|$ more than four kpc from the Galactic plane, and
at distances from the Sun less than 17.5 kpc.  From this point onward,
when we refer to a MPMSTO star, we mean one that is in the inner halo as
defined above.  This is consistent with the work of \citet{caro07,caro09},
who showed that at a distance of 17.5 kpc from the Sun the halo is
dominated by their inner halo component.  Of the 43,000 stars in our
combined MPMSTO sample, 10,739 are in the inner halo as defined above.
We further impose line of sight specific distance limits that contain 95\%
of the photometrically selected MPMSTO candidates.  We use these 95\%
limits to eliminate outliers that could significantly skew the distance
thresholds.  The median number of MPMSTO star spectra per line of sight
in our final sample is 77.

The cuts on Galactocentric radius and distance from the plane effectively
remove the brightest objects, with 95\% of the final sample at $r > 18.26$.
For the final inner halo sample, the effect of the color selections
described above was such that 95\% of the stars have $-0.59<P1(s)<-0.19,
0.63< u-g<1.1$, and $0.12<g-r<0.34$. For the subset of our inner halo
sample brighter than about $g = 19$ we can use the metallicities from
the SEGUE stellar parameters pipeline \citep{lee08a} to estimate
the metallicity bias introduced by the UV-excess and blue $g-r$
selection cuts.  We find the [Fe/H] distribution is approximately
Gaussian, with mean -1.62 and standard deviation 0.5.  These are
consistent with other estimates of the halo metallicity distribution
\citep{rya91a,rya91b,carn96,all06} from which we conclude that any
metallicity bias in our sample is not serious.  At magnitudes $r>19$,
the errors in the $u$ magnitude increase rapidly.  The average SDSS $u$
psf magnitude error for stars at $g=19$ at the mean color of our sample
is 0.06.  The increasing $u$ error causes the UV-excess selection to
become inefficient at faint magnitudes, reducing any bias even further.

\section{Identification of Cold Substructure}

We search along each SEGUE line of sight for statistically significant
differences between the observed MPMSTO star radial velocity distribution
and the radial velocity distribution that would result from SEGUE
observations of a smooth inner halo.  In our Monte Carlo approach to
the identification of ECHOS in the inner halo, we first generate a mock
catalog of synthetic stars distributed according to published empirical
models for the position and velocity distributions of inner halo stars.
We then sample the mock catalog line of sight by line of sight in the
same way that SEGUE sampled the Milky Way's inner halo to obtain the
radial velocity distribution that would result from observing a smooth
inner halo.  We employ this derived radial velocity distribution as our
null hypothesis.  We use two independent algorithms to systematically
search for radial velocities at which there are statistically significant
excesses of MPMSTO stars relative to the smooth model.  The fact that our
search is both systematic and statistical means that we can accurately
quantify our false positive rate and completeness.  From that information
we can derive an upper limit on the fraction of MPMSTO stars in the inner
halo that belong to ECHOS, as well as an estimate for the absolute number
of ECHOS like those we find in the inner halo.

\subsection{Simulation of the Radial Velocity Distribution of the Inner Halo's
            Smooth Component as Viewed From the Sun}

To construct our null hypothesis we simulate the radial velocity
distribution of the smooth inner halo as viewed by an observer in the
plane of our Galaxy eight kpc from its center and moving with the Sun.
We include a detailed description of the procedure we use in Appendix
A and briefly describe our method here.  We start by creating a very
large mock catalog of synthetic stars, each with its own position
and velocity coordinate drawn randomly from empirically determined
distributions for the galactocentric position-space and velocity-space
structure of the smooth inner halo.  Specifically, we use a spherically
symmetric distribution in galactocentric position-space $\rho \propto
r^{\alpha}$  with $\alpha = -3.5$ \citep[e.g.][]{mor00,yan00,bell08}
and a galactocentric velocity ellipsoid selected to match previous
observations of the inner halo in the range of galactocentric radii we
search \citep[e.g.][]{som97,sir04a,sir04b,xue08}

\begin{eqnarray}\label{eq1}
\mathbf{\Sigma}_{r,\theta,\phi} & = & \left( \begin{array}{ccc}
120^2 & 0 & 0 \\
0 & 100^2 & 0 \\
0 & 0 & 100^2 \end{array} \right)
\end{eqnarray}

\begin{figure}
\plotone{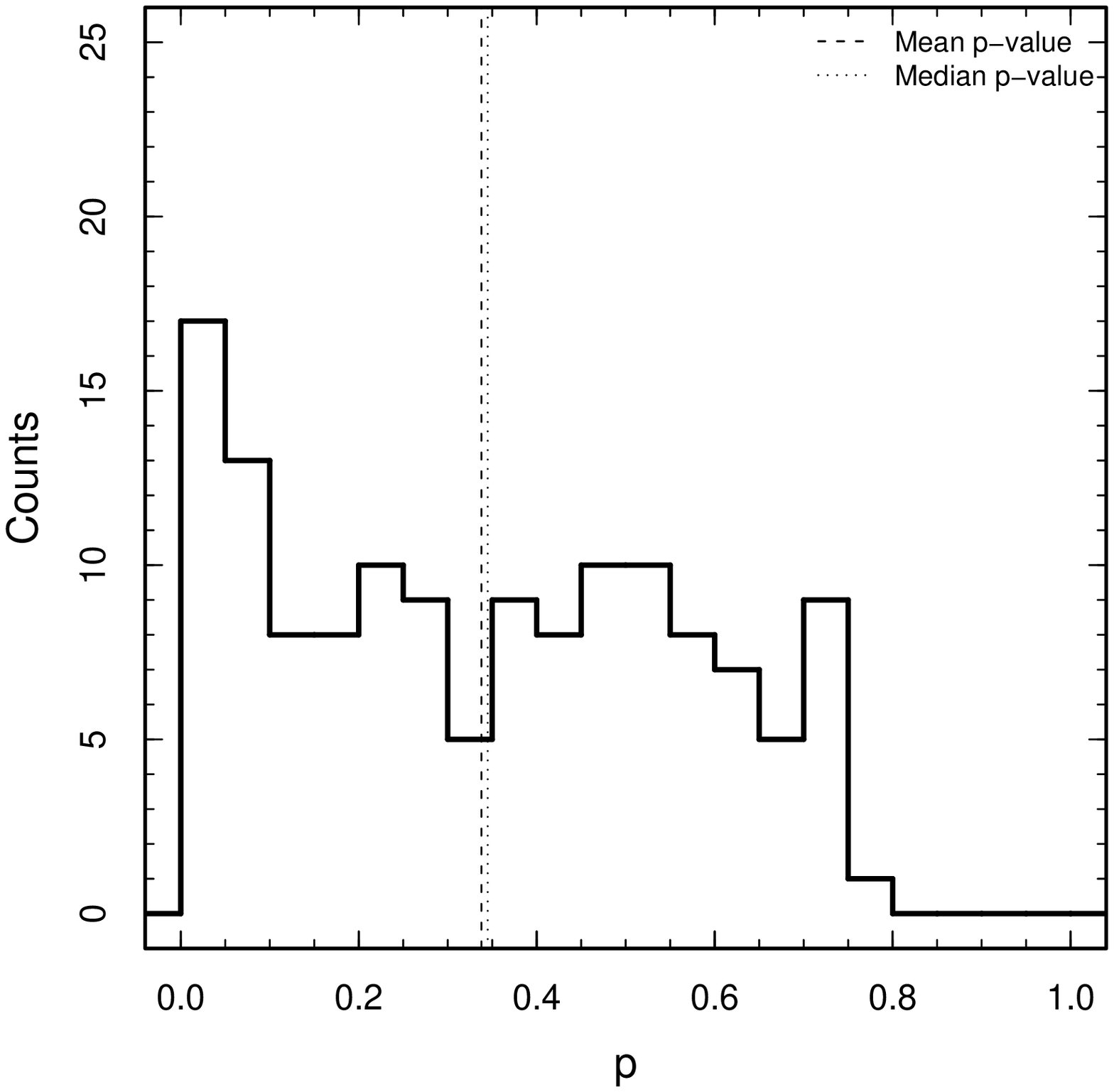}
\caption{The $p$-value distribution from line of sight by line of
sight Kolmogorov-Smirnov (KS) tests comparing the observed SEGUE
MPMSTO star radial velocity distributions with the radial velocity
distributions derived from our mock catalog for that line of sight.
Recall that $p$-values from the KS test give the probability that the
two data sets under comparison are drawn from the same distribution.
Therefore, $p$-values $\gtrsim 0.05$ are usually a sign that both samples
under comparison are plausibly drawn from the same parent distribution.
As such, we see no reason to reject the radial velocity distribution
produced by observing our mock catalog for the majority of the lines
of sight we study.  The pile-up at small $p$-values is caused by the
presence of substructure along those lines of sight.  One caveat is that
we are not necessarily sensitive to an inner halo entirely populated by
an ensemble of very diffuse substructures, as the velocity distribution
of the MPMSTO sample in that scenario could very well resemble the
distribution that results from a kinematically smooth model.\label{fig1}}
\end{figure}

\noindent
The velocity ellipsoid results from diagonalizing the symmetric stress
tensor in the Jeans equations \citep[e.g.][]{bin87}; the numerical
values in Equation~(\ref{eq1}) are in units of km s$^{-1}$.  Our final
result is insensitive to the power-law exponent and the inner halo
velocity ellipsoid.  We tried $-4 < \alpha < -3$ and 100 km s$^{-1}$ $ <
\sigma_r < $ 120 km s$^{-1}$ and arrived at the same set of detections;
the exact parameters of the power-law and the velocity ellipsoid are
not as important as the functional forms themselves.  To ensure that
the smooth model is a reasonable null hypothesis for a substructure
search, for each line of sight we use a two-sample Kolmogorov-Smirnov
(KS) test on the observed MPMSTO radial velocity distribution and on the
radial velocity distribution derived from observing our smooth model.
Large $p$-values for a great majority of our lines of sight would
demonstrate that the smooth model is not too gross an approximation of
the radial velocity distribution of the inner halo.  Figure~\ref{fig1}
indicates that most lines of sight have KS $p$-values $\gtrsim$ 0.05
characteristic of a common parent distribution for both observations.
Those lines of sight with small $p$-values are those lines of sight
along which we subsequently identify prominent ECHOS.  In other words,
for the majority of our lines of sight we find that the smooth model is
not obviously incorrect; for those lines of sight where it is obviously
incorrect as indicated by the KS tests, we subsequently find ECHOS with
more sensitive algorithms discussed in the next subsection.  In any
case, we do not use the results of these KS tests in our search for
substructure and we find that our observed radial velocity distributions
give us no reason to reject our fiducial null hypothesis.  However,
we cannot reject the possibility that the inner halo is entirely made
up of elements of substructure below our sensitivity thresholds that at
our velocity resolution masquerade as a kinematically smooth population.

We find that we need $n_s \sim 10^7$ synthetic stars in the mock catalog
to ensure that we have at least an order-of-magnitude more synthetic stars
than SEGUE MPMSTO stars along a given line of sight.  The extra synthetic
stars are necessary to robustly sample the radial velocity distribution
along each line of sight.  We then project the full three-dimensional
velocity of each synthetic star onto the line of sight between the star
and an observer in the plane of the Galaxy eight kpc from its center and
moving with the space velocity of the Sun \citep{deh98}.  As a result,
we can characterize the radial velocity distribution of the smooth inner
halo as viewed from Sun.  We sample the radial velocity distribution of
the mock catalog in exactly the same way that SEGUE sampled the radial
velocity distribution of MPMSTO stars in the inner halo of the Galaxy.
Everything is the same: we sample the same number of radial velocities
from the mock catalog that SEGUE obtained along each line of sight and we
use the line of sight specific heliocentric distance thresholds discussed
in \S2 to ensure that the radial velocity distribution we obtain from the
mock catalog is an accurate representation of the smooth inner halo along
that line of sight.  It is important to keep in mind that fact that the
radial velocity distribution observed by SEGUE in kinematically smooth
halo -- our null hypothesis -- is not a strong function of the distance
distribution of synthetic stars.  As a result, our detections themselves
are not sensitive to any small biases present in the selection of our
halo tracer population.  Thus, any possible small bias in the SEGUE data
to nearer MPMSTO stars will not affect our result.  In short, the radial
velocity distribution that results from sampling our mock catalog matches
as closely as possible the distribution that SEGUE would have observed in
a kinematically smooth inner halo.  Statistically significant departures
from this null hypothesis are the substructure we seek.

\subsection{Detection Algorithms}

For every SEGUE line of sight, we compare the simulated radial velocity
distribution of the smooth inner halo described in \S3.1 with the observed
MPMSTO star radial velocity distribution derived for that line of sight.
We analyze each 3$^{\circ}$ diameter SEGUE line of sight independently.
To see why, recall that the Sagittarius tidal stream is about 5$^{\circ}$
wide and is by far the most substantial known element of high latitude
substructure.  The Orphan stream is about 2$^{\circ}$ wide and is more
representative of the currently known substructure population as a whole.
Thus the 3$^{\circ}$ diameter SEGUE plates are likely well matched to
the angular size of inner halo substructure.  In addition, the Sculptor
dwarf is the largest dwarf spheroidal galaxy in projected area on the sky
and it subtends only about 0.4 square degrees.  Individual SEGUE lines
of sight are separated by 10-20 degrees from their nearest neighbor, and
the angular correlation function between SEGUE lines of sight peaks at
about 80$^{\circ}$.  Therefore, the characteristic angular scale of the
objects we seek is much less than the typical pairwise line of sight
angular separation.  For that reason, we assert that the short axis
of any potential element of substructure is unlikely to intersect two
different 3$^{\circ}$ diameter SEGUE lines of sight.  Along the long axis
of potential streams, the characteristic 80$^{\circ}$ angular distance
between lines of sight ensures that the velocity gradient along the
orbit of the stream will produce distinct velocity peaks.  Without a
model for the accretion history of the Milky Way that allows us to a
priori associate multiple lines of sight with distinct velocity peaks,
analyzing multiple lines of sight together does not provide any extra
sensitivity to individual elements of substructure.  Taken as whole,
these numbers suggest that each SEGUE line of sight probes a distinct
piece of the inner halo and that data obtained along one line of sight
is unlikely to add significantly to the detectability of an element
of substructure in a neighboring line of sight.  At the same time,
our completeness calculations in \S3.5 indicate that along almost all
individual lines of sight we have enough MPMSTO radial velocities to
both characterize the background smooth halo and to resolve ECHOS.

ECHOS in the observed radial velocity distribution manifest themselves
as relative overdensities at their mean radial velocities.  Our task is
therefore to differentiate the radial velocity overdensities that result
from the presence of ECHOS along a line of sight from those that are
produced by chance through random sampling of the smooth inner halo.
Note that we do not use our admittedly imprecise spectrophotometric
parallaxes as we search for substructure.  We only use them to determine
the lower and upper bounds of the observed column along each line of
sight.  We use that information to sample our mock catalog to determine
the radial velocity distribution of a smooth inner halo along that line
of sight.  That is, we only used the spectrophotometric parallaxes in the
construction of our null hypothesis.  We describe the two algorithms we
employed to solve this problem in the following two subsections.

\subsubsection{The Bin Algorithm}

Our first approach is to compare the radial velocity histogram derived
from the observed MPMSTO sample with an equivalent observation of
our mock catalog.  First, we compute a histogram that describes the
distribution of MPMSTO star radial velocities along a given line of
sight, and we use bootstrap resampling to quantify the uncertainty in the
number of counts in each histogram bin.  We use bins of 20 km s$^{-1}$
width both because that is approximately our median radial velocity
precision at the faint end of our sample and because that width minimizes
Poisson noise while still giving us sensitivity to cold substructure.
We repeatedly resample our mock catalog along that line of sight to
determine the median histogram and the associated distributions for the
number of counts in each bin that would result under the null hypothesis.
We can then compare the two histograms and flag velocity bins that have
a statistically significant overabundance of SEGUE MPMSTO stars relative
to the mock catalog.  Note that once we identify candidate ECHOS we
could estimate the number of MPMSTO stars that belong to that element
of substructure and then redo our calculation, taking into account the
reduced number of MPMSTO stars that we suspect belong to the smooth
background along that line of sight.  We choose not do this because it
could produce overly optimistic significance levels for our detections.
We describe this algorithm in detail in Appendix B.

\subsubsection{The Peak Algorithm}

Our second approach is to compare the steepness of the cumulative
distribution function (CDF) as a function of radial velocity as derived
from the observed MPMSTO sample with an equivalent observation of our
mock catalog.  First, we compute the CDF that describes the distribution of
MPMSTO radial velocities along a given line of sight.  Next, we repeatedly
resample our mock catalog along that line of sight to determine the
average CDF that would result under the null hypothesis.  Recall that
the CDF $F(x)$ of a discrete data set $x_1, x_2, \ldots, x_n$
drawn from the discrete random variable $X$ is a monotonically increasing
function that is discontinuous at each $x_i$ and mathematically defined as

\begin{eqnarray}
F(x) & = & P(X \leq x) = \sum_{x_i \leq x} P(X = x_i)
\end{eqnarray}

\noindent
Therefore, an overdensity of stars at the mean radial velocity of
an ECHOS would appear as a much steeper set of steps in the observed CDF
than was expected at that velocity.  We can calculate line of sight and
radial velocity specific significance thresholds for cold substructure by
quantifying how steep we expect steps to be at that radial velocity under
the null hypothesis.  Then we can flag any set of steps with steepness
statistic $\Theta(v_r)$ above its local threshold as a potential ECHOS.
As with the bin algorithm, we do not iteratively adjust the number of
stars in the smooth background component of the inner halo for those
lines of sight where we detect ECHOS. We describe this algorithm in
detail in Appendix C.

\subsection{Expected False Positive Rate}

The insight into the formation of the Milky Way that we hope to gain from
our search for ECHOS depends crucially on both the number of candidate
ECHOS we find and on their properties.  In order to be certain that our
candidate ECHOS are real features in the inner halo (and not just chance
projections in radial velocity-space), we estimate our false positive
rate with a Monte Carlo simulation.  We generate two independent mock
catalogs denoted $M_1$ and $M_2$ each with more than 10$^7$ synthetic
stars and radial velocities distributed as described in Appendix A.
For each SEGUE line of sight we randomly select from $M_2$ a sample $S_r$
of $n$ synthetic stars equal to the number of SEGUE MPMSTO star spectra
available along that line of sight from the $m \gg n$ synthetic stars
available.  We treat this sample $S_r$ as the data and put it aside.
We can then analyze the data in $S_r$ using the algorithms described in
\S3.2.1 and \S3.2.2 with mock catalog $M_1$ as our null hypothesis.

We examine all 137 lines of sight ten times and count the number of
detections -- all of which are chance projections in radial velocity-space
that falsely appear to be ECHOS.  We find that the expected number
of false positives produced by the basic bin algorithm described in
\S3.2.1 is less than one over our entire search and therefore requires no
further additions.  We find that the expected number of false positives
produced by the basic peak algorithm described in \S3.2.2 is greater than
one over our entire search.  To understand why, recall that we analyze
137 lines of sight at each of 1,000 points in radial velocity-space at
which there could be a significant detection.  If all of the tests are
independent, the formal probability of at least one false positive is
$P(X) = 1-(1-1/10^4)^{137*1000} \sim 1$.  Fortunately, the tests are not
independent (because of the smoothing we applied) and the peak algorithm
can be modified such that the false positive rate is nearly zero.  We find
through our false positive analysis that to eliminate false positives
the basic peak algorithm needs to be extended in the following ways:

\begin{enumerate}
   \item
   The use of the smooth model as our null hypothesis is most likely
   to break down far from the median radial velocity along a line of
   sight (where the sampling is very sparse).  We require that the
   radial velocity associated with a candidate ECHOS falls within an
   interval centered on the median radial velocity of the smooth inner
   halo component containing 95\% of the synthetic stars from the mock
   catalog along that line of sight.
   \item
   The finite size of our mock catalogs and the number of Monte Carlo
   iterations we can computationally afford is an issue.  In the limit of
   an infinitely large mock catalog and an infinite number of Monte Carlo
   iterations, the significance contours we derive would be perfectly
   smooth with no small-scale fluctuations.  However, since both our
   mock catalog and Monte Carlo simulation are of finite size we need to
   ensure that the small-scale fluctuations present in our significance
   thresholds do not lead to false positives.  Therefore we impose one
   more requirement: we fit a Gaussian to the lower bound of our 1 in
   10$^{4}$ significance region and increases its amplitude to ensure
   that it is an upper-bound for all of the small-scale structure in the
   significance contours.  We require that $\Theta(v_r)$ must be above
   this upper-bound for a detection.
\end{enumerate}

\noindent
Our analysis indicates that the expected number of false positives
with these properties is less than one over our entire search.  We call
detections that meet these criteria class I peak detections and the basic
peak algorithm including these criteria the class I peak algorithm.
From this point forward, we define high-confidence detections as all
candidate ECHOS identified by either the bin algorithm or the class I
peak algorithm.  Our analysis indicates that we expect less than one false
positive in our list of high confidence detections.  Therefore, together
the bin detections in Table~\ref{tbl-1} and class I peak detections in
Table~\ref{tbl-2} provide the definitive list of promising candidate
ECHOS for further study.

We note that the stringent requirements necessary to ensure an expected
number of false positives less than one over our entire search also
eliminates many candidates that are likely real ECHOS.  While those
strict requirements are necessary to isolate individual ECHOS at
the highest confidence needed for follow-up investigation, they can
result in a biased view of the entire population of ECHOS.  Therefore,
we also report in Table~\ref{tbl-3} the subset of peak detections that
are formally significant at more than the 1 in 10$^4$ level but not so
significant to ensure that the entire list is free from false positives;
we call these class II peak detections.  Our analysis indicates that
false positives are almost always associated with candidates that have
very few stars within an interval centered on the peak of $\Theta(v_r)$
with width equal to our median velocity resolution -- about 11.5 km
s$^{-1}$.  For that reason, we can reduce the level of contamination in
Table~\ref{tbl-3} to an acceptable level by varying the required number
of MPMSTO stars within that radial velocity interval.  We find that if we
require more than ten MPMSTO stars within that velocity interval, then the
expected number of false positives in the 21 rows of Table~\ref{tbl-3} is
less than three.  We call the basic peak algorithm with this additional
criteria the class II peak algorithm.  We believe that the population
of candidate ECHOS in Table~\ref{tbl-3} is perhaps more representative
of the ECHOS population as a whole, but that individual entries in that
list are potentially unreliable.

\subsection{An Example}

\begin{figure}
\epsscale{.40}
\plotone{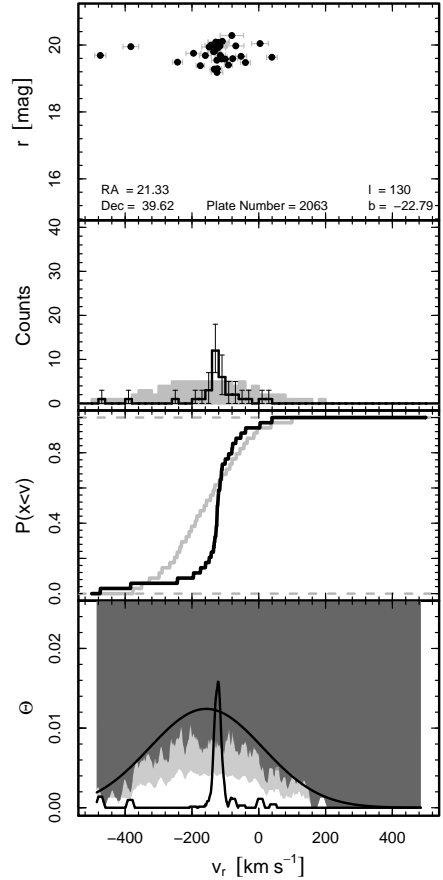}
\caption{The data and our analyses for the line of sight along which we
found the element of cold substructure B-1 from Table~\ref{tbl-1}, PCI-3
from Table~\ref{tbl-2}, and PCII-3 from Table~\ref{tbl-3}.  In the top
panel we plot radial velocity versus $r$-magnitude.  In the same panel
we also plot our estimated radial velocity errors as the gray horizontal
bars (the photometric errors are much smaller than the plotted points).
In the second panel we plot in black a histogram derived from the observed
MPMSTO star radial velocity distribution with bin-width 20 km s$^{-1}$
and 95\% confidence intervals from bootstrap resampling.  In the same
panel we also plot in gray an area that corresponds to a 95\% confidence
region for the histogram that would result from a SEGUE observation of
our mock catalog.  As a result, a significant bin is one for which the
error bar on the black histogram does not intersect the gray region.
In the third panel we plot in black the CDF of the observed MPMSTO star
radial velocity distribution and we plot in gray the average CDF of the
simulated radial velocity distribution obtained from our mock catalog.
In the fourth panel we plot in black our $\Theta(v_r)$ statistic while we
plot the 1 in 10$^2$ significance region in light gray, the 1 in 10$^3$
significance region in medium gray, and the 1 in  10$^4$ region in the
dark gray.  The white region is significant at less than the 1 in 10$^2$
level; note that the medium and dark gray regions are nearly coincident,
emphasizing the extreme non-Gaussianity of our $\Theta(v_r)$ statistic.
The black Gaussian curve is an upper-bound that we use to ensure the
small-scale fluctuations in our significance contours do not lead to
false positives.  Note the significant feature in all panels at $v_r
\approx -121$ km s$^{-1}$.\label{fig2}}
\end{figure}

\begin{figure}
\epsscale{0.9}
\plottwo{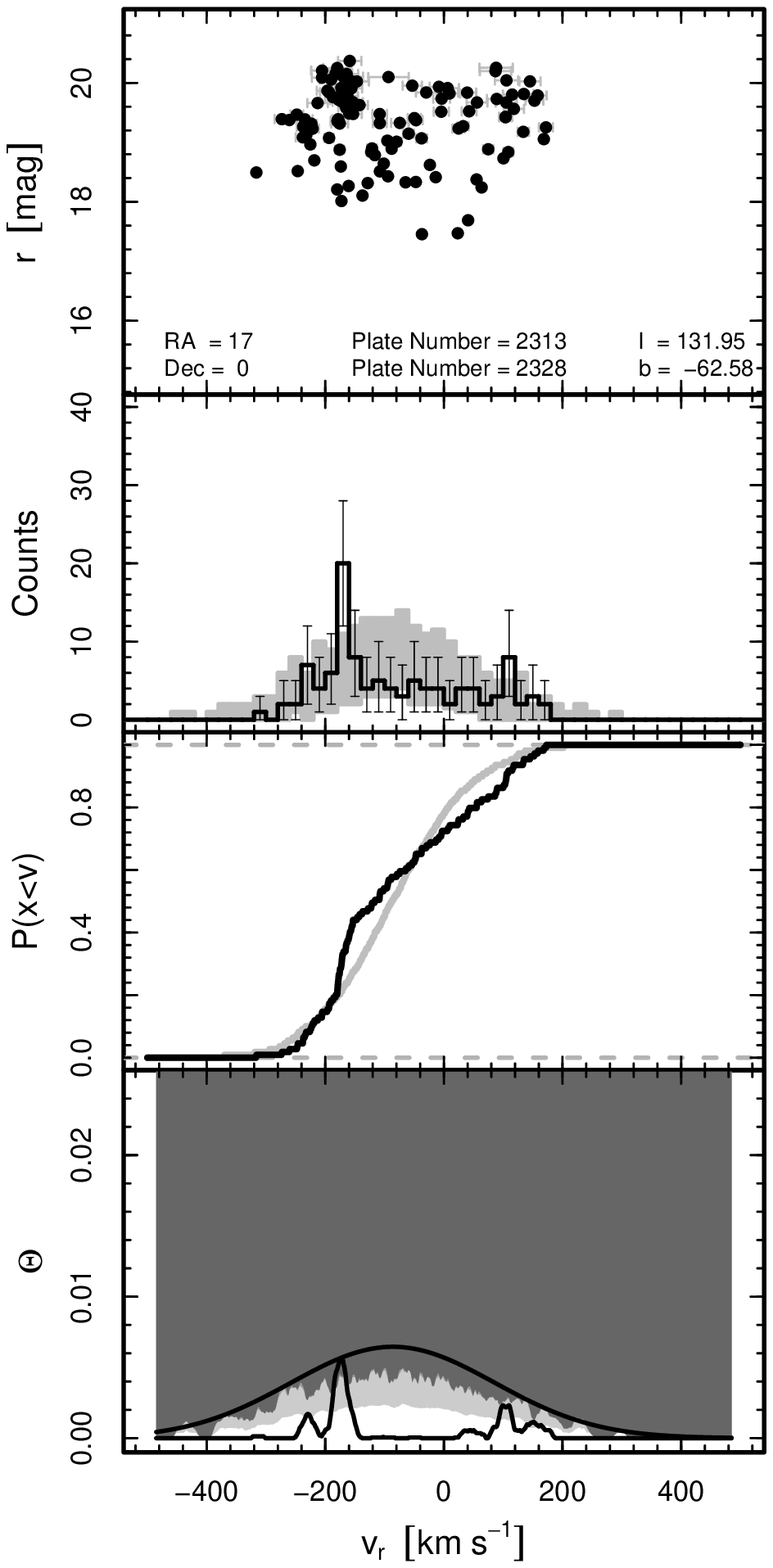}{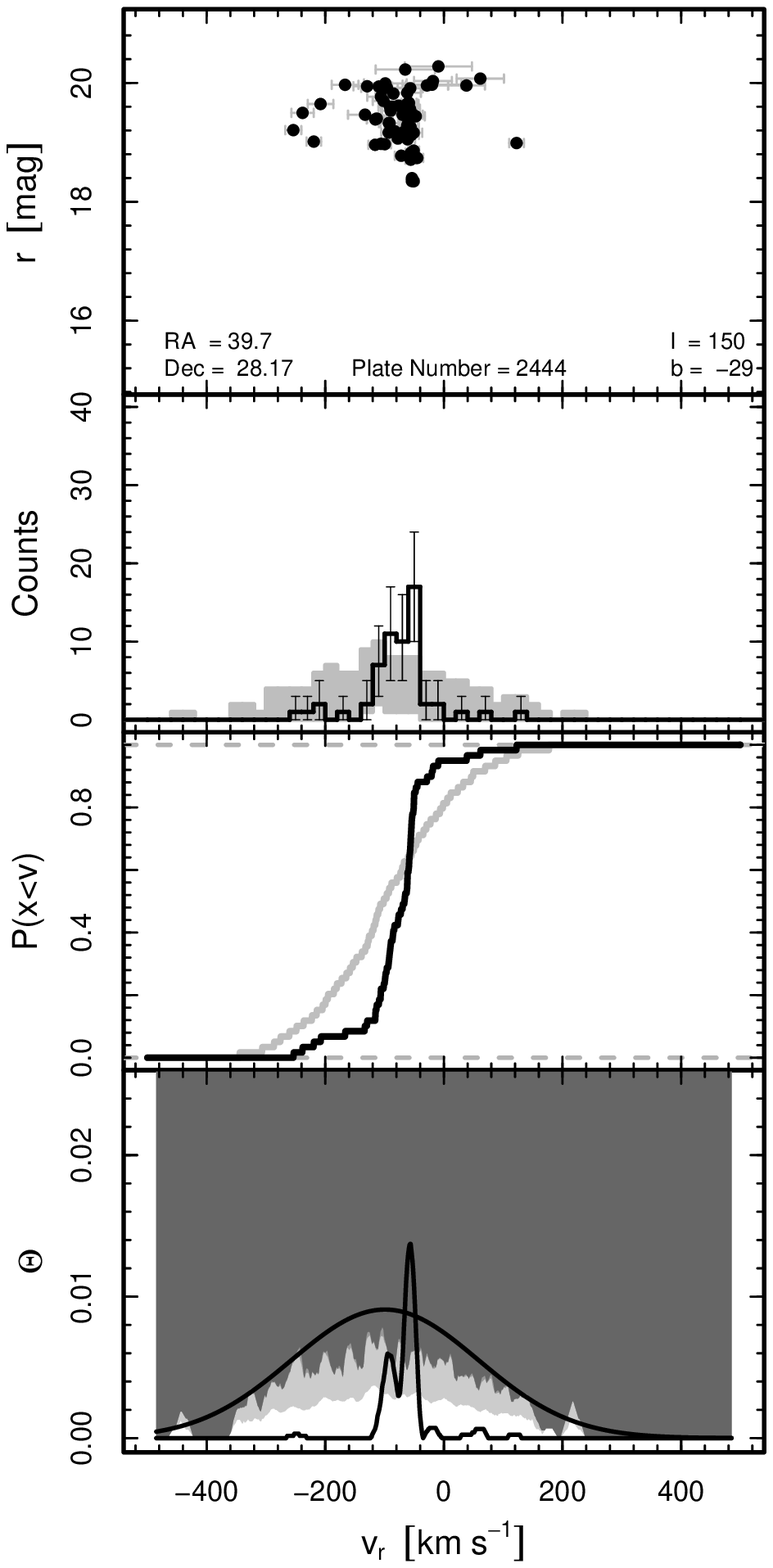}
\caption{In the top panel we plot radial velocity versus $r$-magnitude.
In the same panel we also plot our estimated radial velocity errors as
the gray horizontal bars (the photometric errors are much smaller than the
plotted points).  In the second panel we plot in black a histogram derived
from the observed MPMSTO star radial velocity distribution with bin-width
20 km s$^{-1}$ and 95\% confidence intervals from bootstrap resampling.
In the same panel we also plot in gray an area that corresponds to a
95\% confidence region for the histogram that would result from a SEGUE
observation of our mock catalog.  As a result, a significant bin is
one for which the error bar on the black histogram does not intersect
the gray region.  In the third panel we plot in black the CDF of the
observed MPMSTO star radial velocity distribution and we plot in gray the
average CDF of the simulated radial velocity distribution obtained from
our mock catalog.  In the fourth panel we plot in black our $\Theta(v_r)$
statistic while we plot the 1 in 10$^2$ significance region in light gray,
the 1 in 10$^3$ significance region in medium gray, and the 1 in  10$^4$
region in the dark gray.  The white region is significant at less than
the 1 in 10$^2$ level; note that the medium and dark gray regions are
nearly coincident, emphasizing the extreme non-Gaussianity of our
$\Theta(v_r)$ statistic.  The black Gaussian curve is an upper-bound that
we use to ensure that the small-scale fluctuations in our significance
contours do not lead to false positives.  \emph{Left}: Data and
analyses for the line of sight along which we found the element of cold
substructure B-2 from Table~\ref{tbl-1} and PCII-5 from Table~\ref{tbl-3}.
\emph{Right}: Data and analyses for the line of sight along which we
found the element of cold substructure B-3 from Table~\ref{tbl-1}, PCI-4
from Table~\ref{tbl-2}, and PCII-8 from Table~\ref{tbl-3}.  This line
of sight is also expected to intersect the Monoceros stream.\label{fig3}}
\end{figure}

\begin{figure}
\epsscale{1.0}
\plottwo{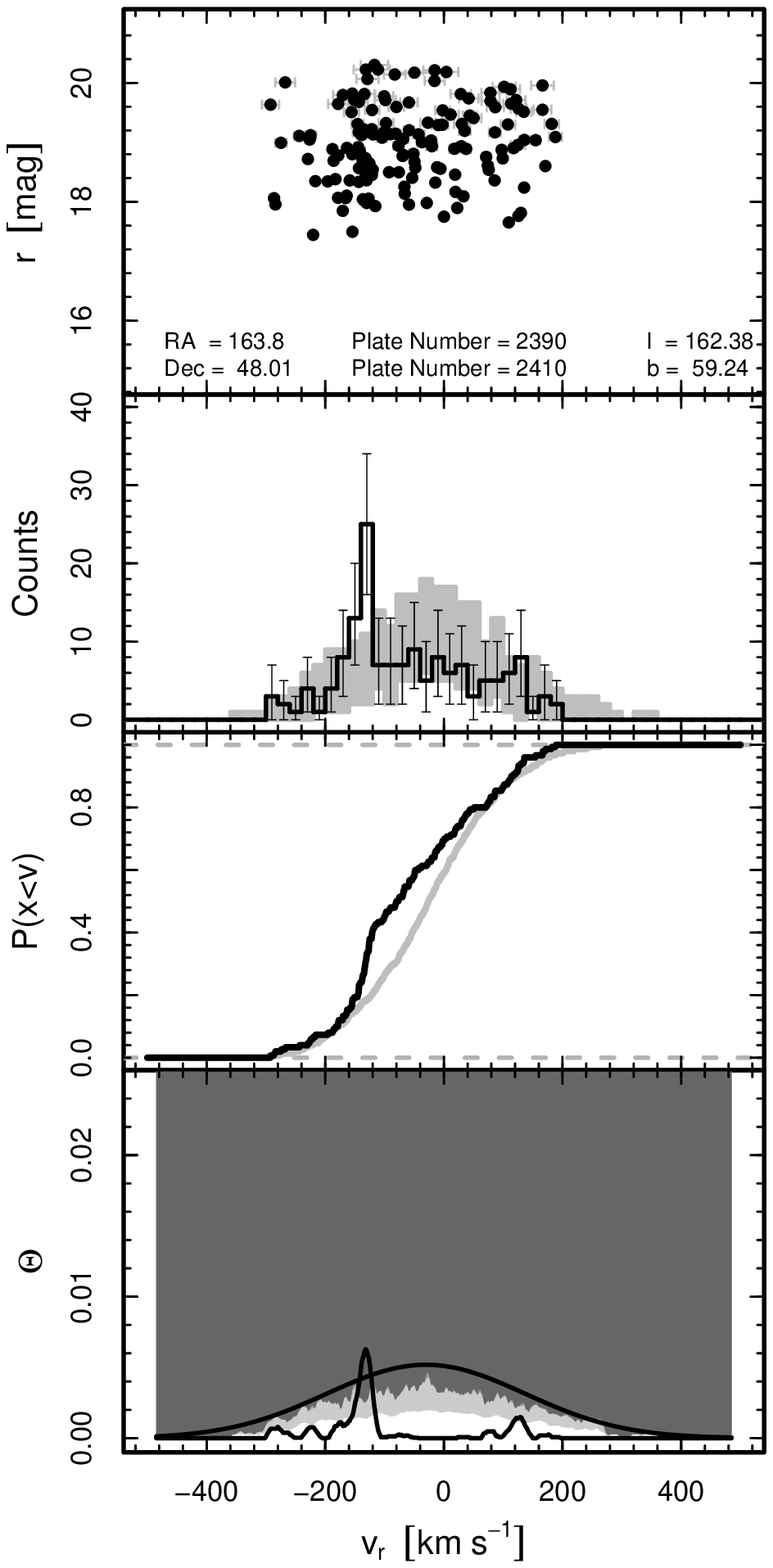}{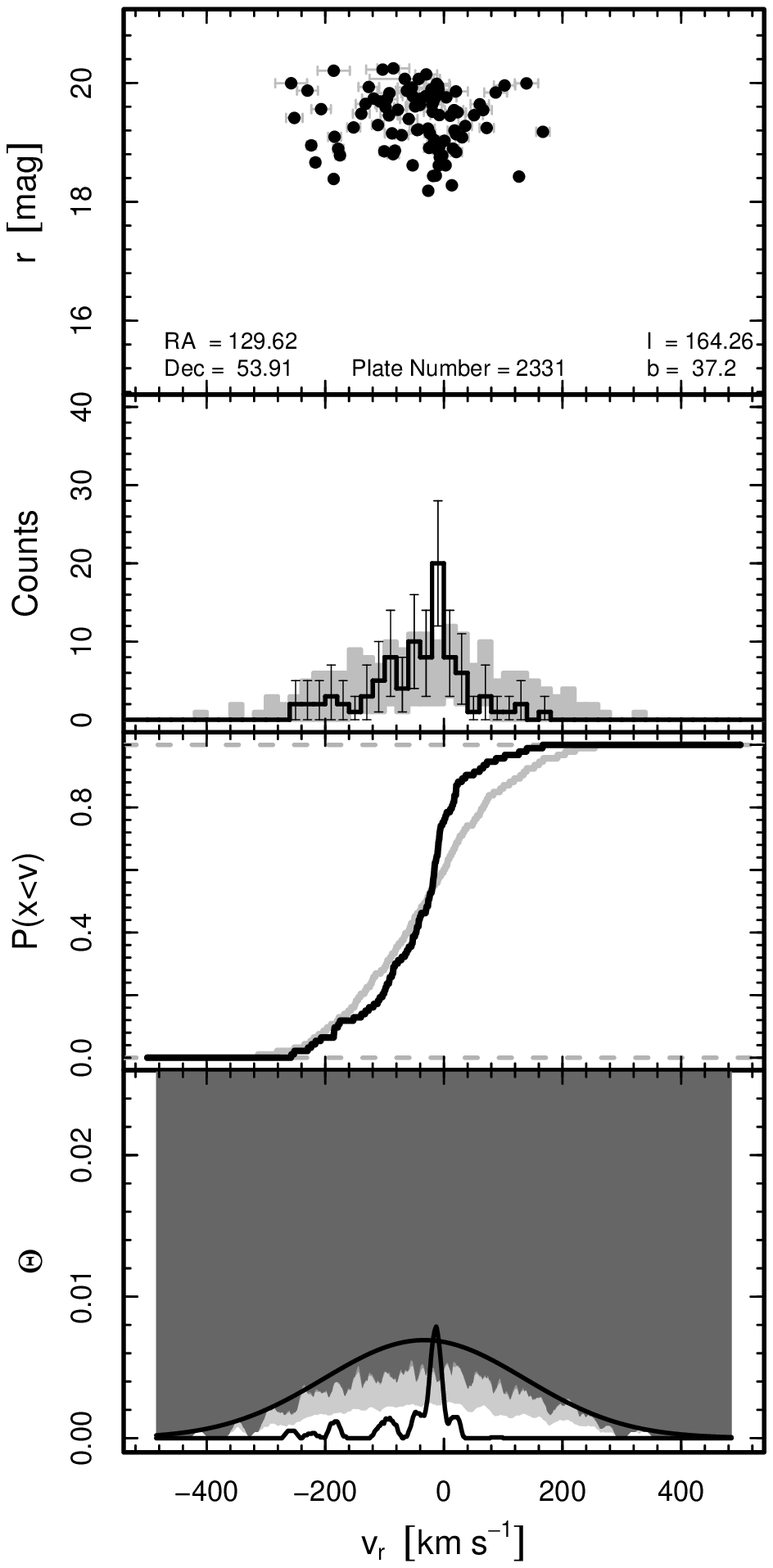}
\caption{\emph{Left}: Data and analyses for the line of sight along which
we found the element of cold substructure B-4 from Table~\ref{tbl-1},
PCI-5 from Table~\ref{tbl-2}, and PCII-10 from Table~\ref{tbl-3}.  This
line of sight is also expected to intersect the \citet{gri06b} stream.
\emph{Right}: Data and analyses for the line of sight along which we
found the element of cold substructure B-5 from Table~\ref{tbl-1},
PCI-6 from Table~\ref{tbl-2}, and PCII-11 from Table~\ref{tbl-3}.
See the caption to Figure~\ref{fig3} for a detailed description of this
type of figure.\label{fig4}}
\end{figure}

\begin{figure}
\plottwo{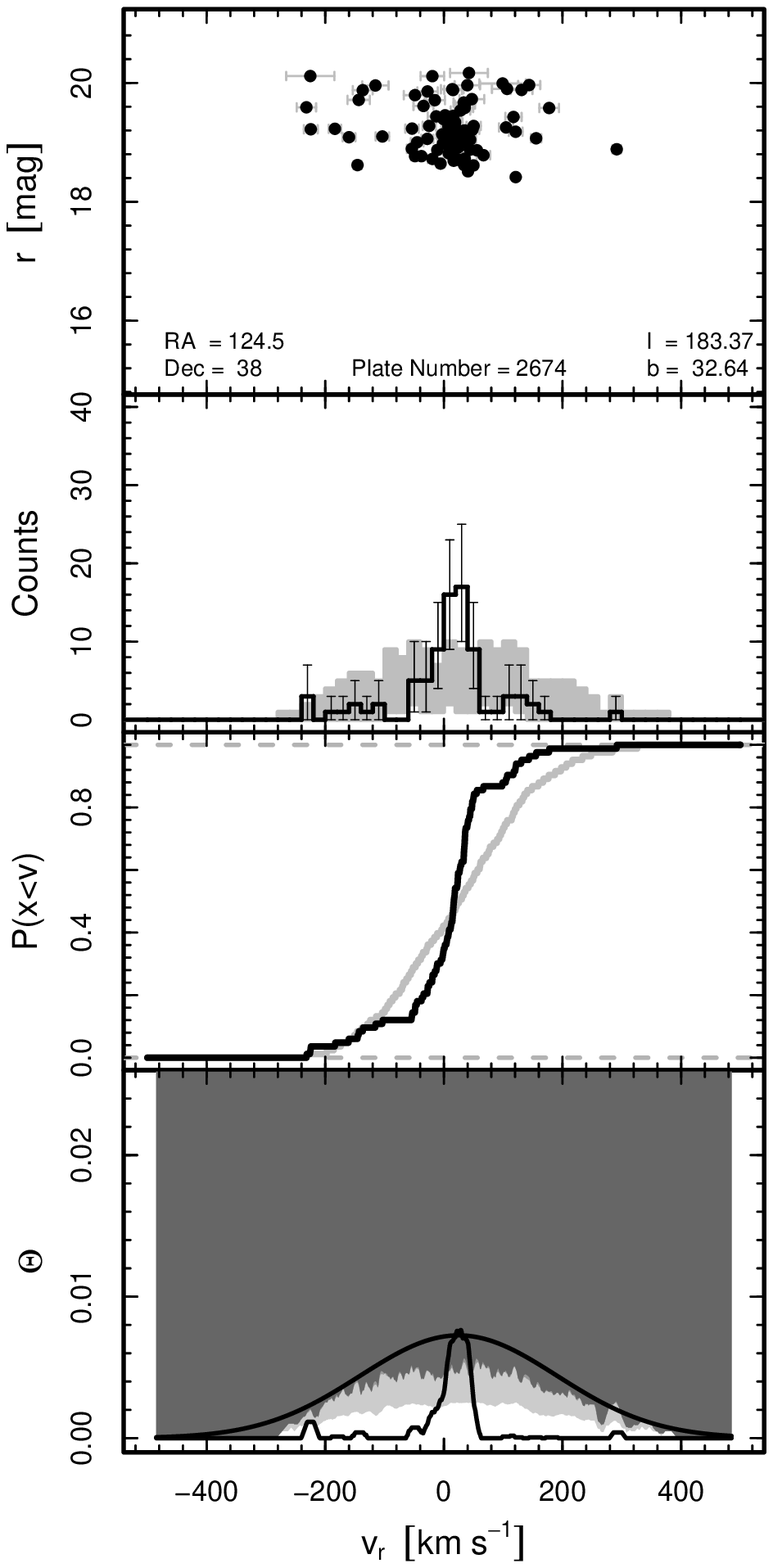}{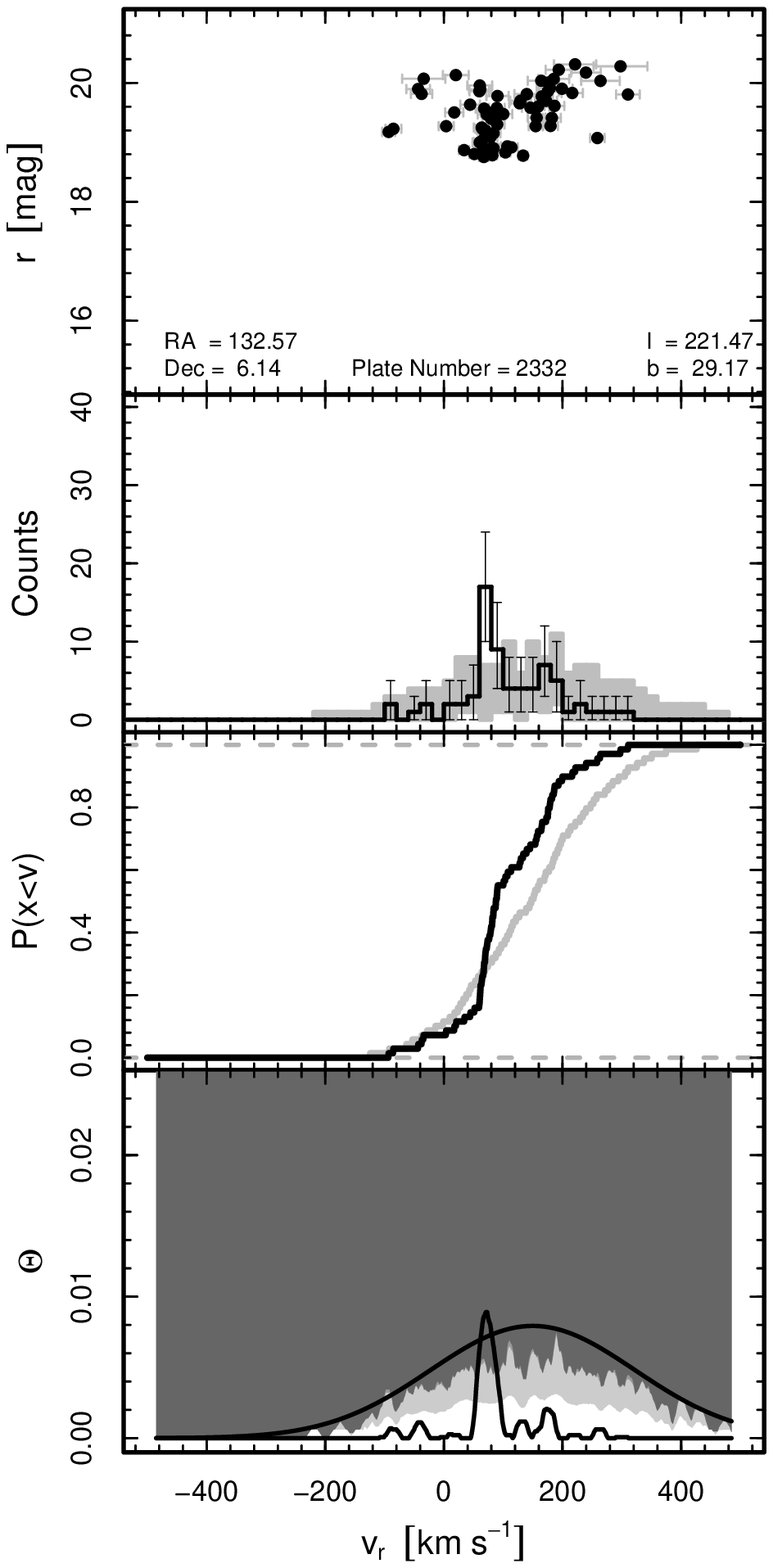}
\caption{\emph{Left}: Data and analyses for the line of sight along which
we found the element of cold substructure B-6 from Table~\ref{tbl-1},
PCI-7 from Table~\ref{tbl-2}, and PCII-12 from Table~\ref{tbl-3}.
\emph{Right}: Data and analyses for the line of sight along which we found
the element of cold substructure B-7 from Table~\ref{tbl-1}, PCI-8 from
Table~\ref{tbl-2}, and PCII-20 from Table~\ref{tbl-3}.  See the caption
to Figure~\ref{fig3} for a detailed description of this type of figure.
\label{fig5}}
\end{figure}

\begin{figure}
\plottwo{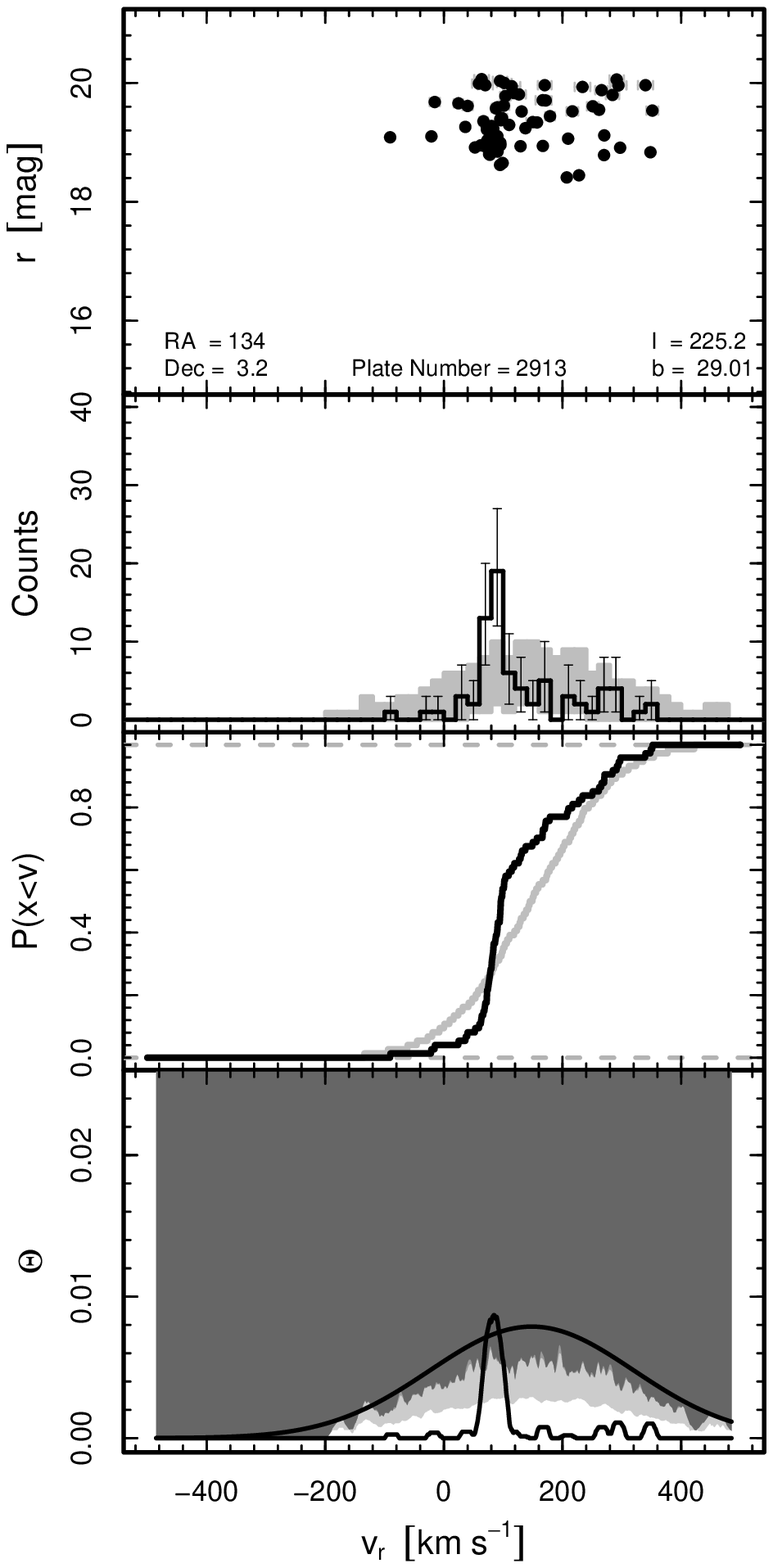}{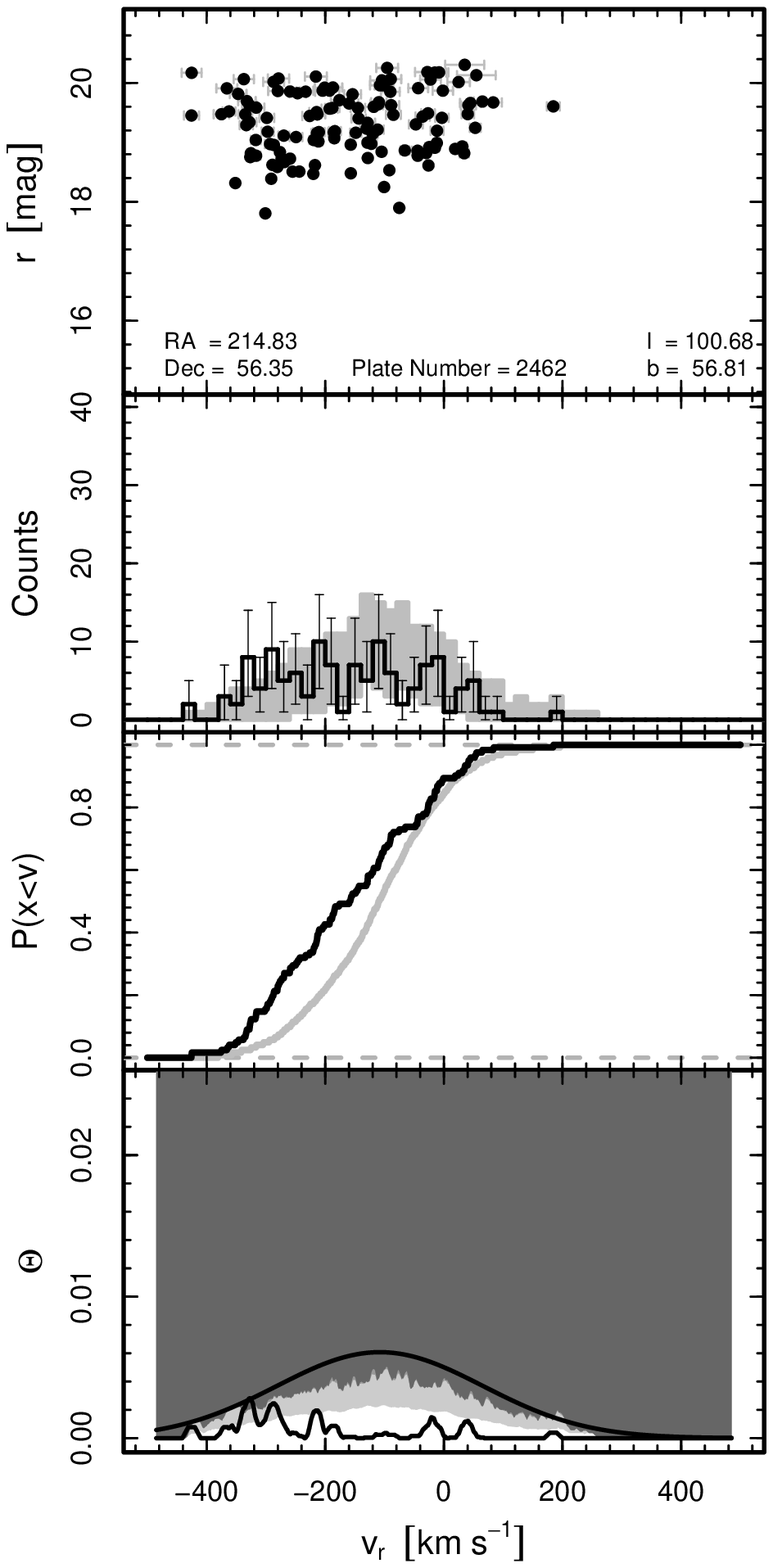}
\caption{\emph{Left}: Data and analyses for the line of sight along which
we found the element of cold substructure B-8 from Table~\ref{tbl-1},
PCI-9 from Table~\ref{tbl-2}, and PCII-21 from Table~\ref{tbl-3}.
This line of sight is also expected to intersect the Monoceros stream.
\emph{Right}: Data and analyses for the line of sight along which we
found the element of cold substructure PCI-1 from Table~\ref{tbl-2}.
See the caption to Figure~\ref{fig3} for a detailed description of this
type of figure.\label{fig6}}
\end{figure}

\begin{figure}
\plottwo{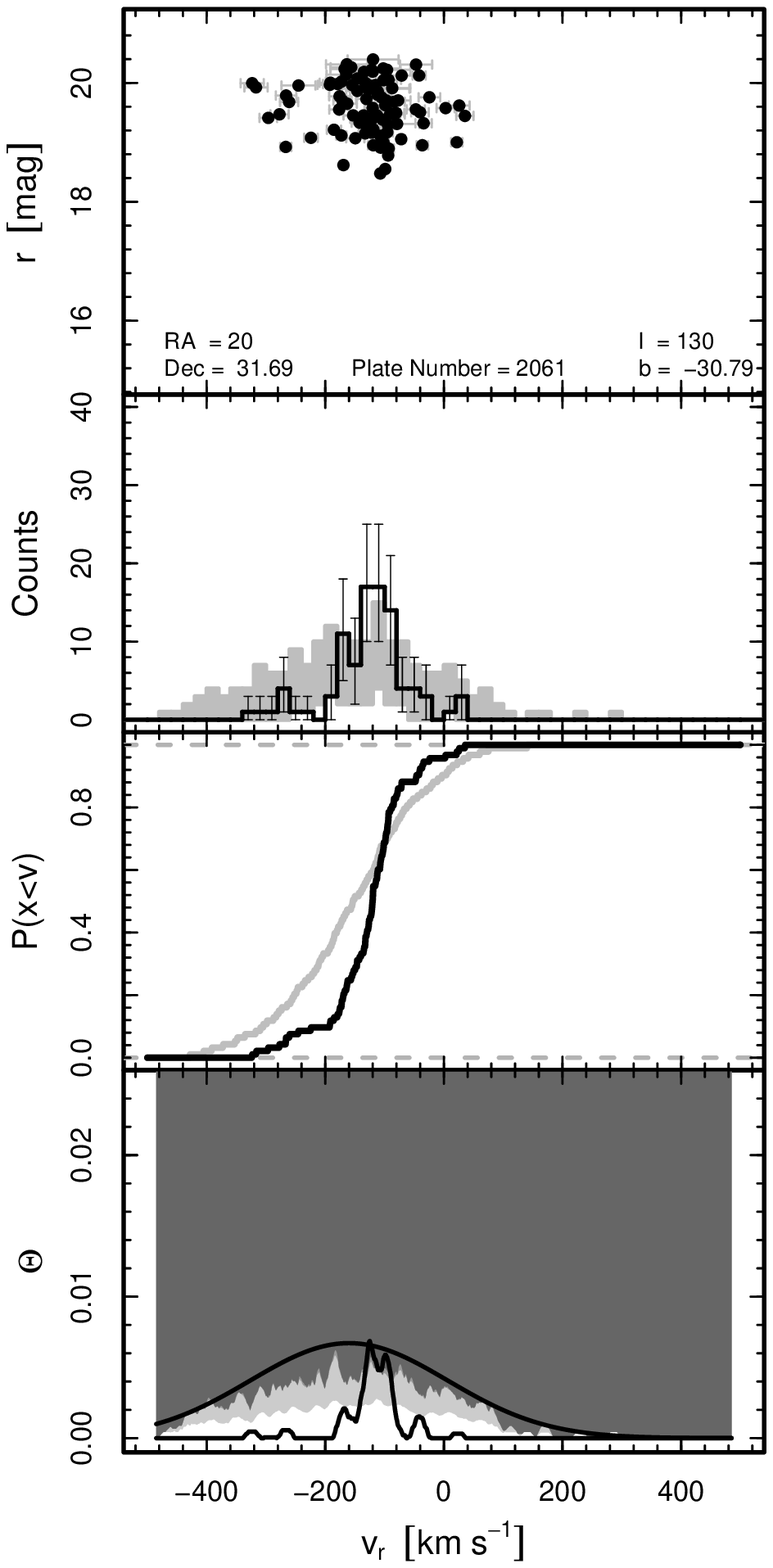}{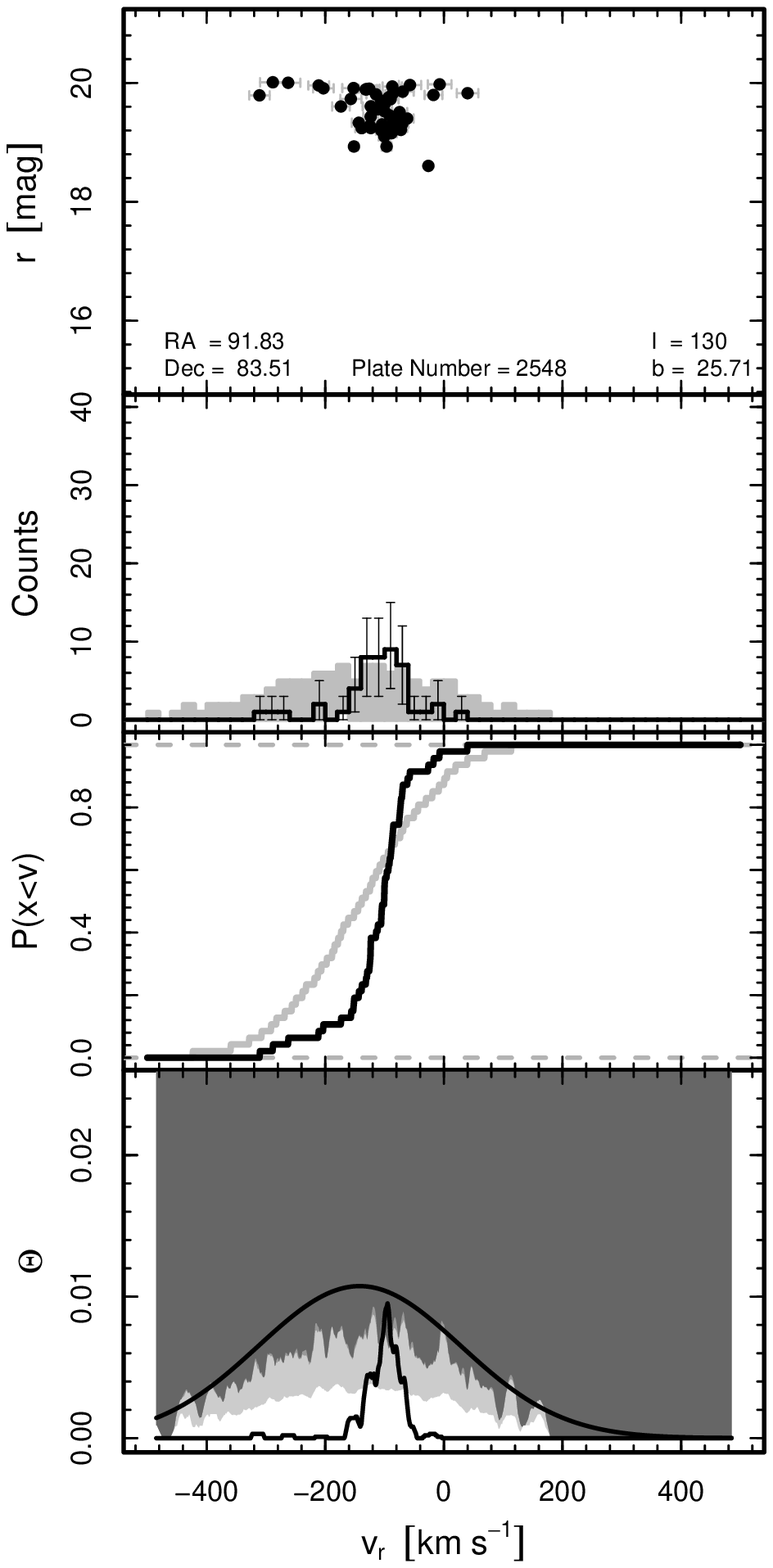}
\caption{\emph{Left}: Data and analyses for the line of sight along which
we found the element of cold substructure PCI-2 from Table~\ref{tbl-2}
and PCII-1 and PCII-2 from Table~\ref{tbl-3}.  \emph{Right}: Data and
analyses for the line of sight along which we found the element of
cold substructure PCII-4 from Table~\ref{tbl-3}.  See the caption to
Figure~\ref{fig3} for a detailed description of this type of figure.
\label{fig7}}
\end{figure}

\begin{figure}
\plottwo{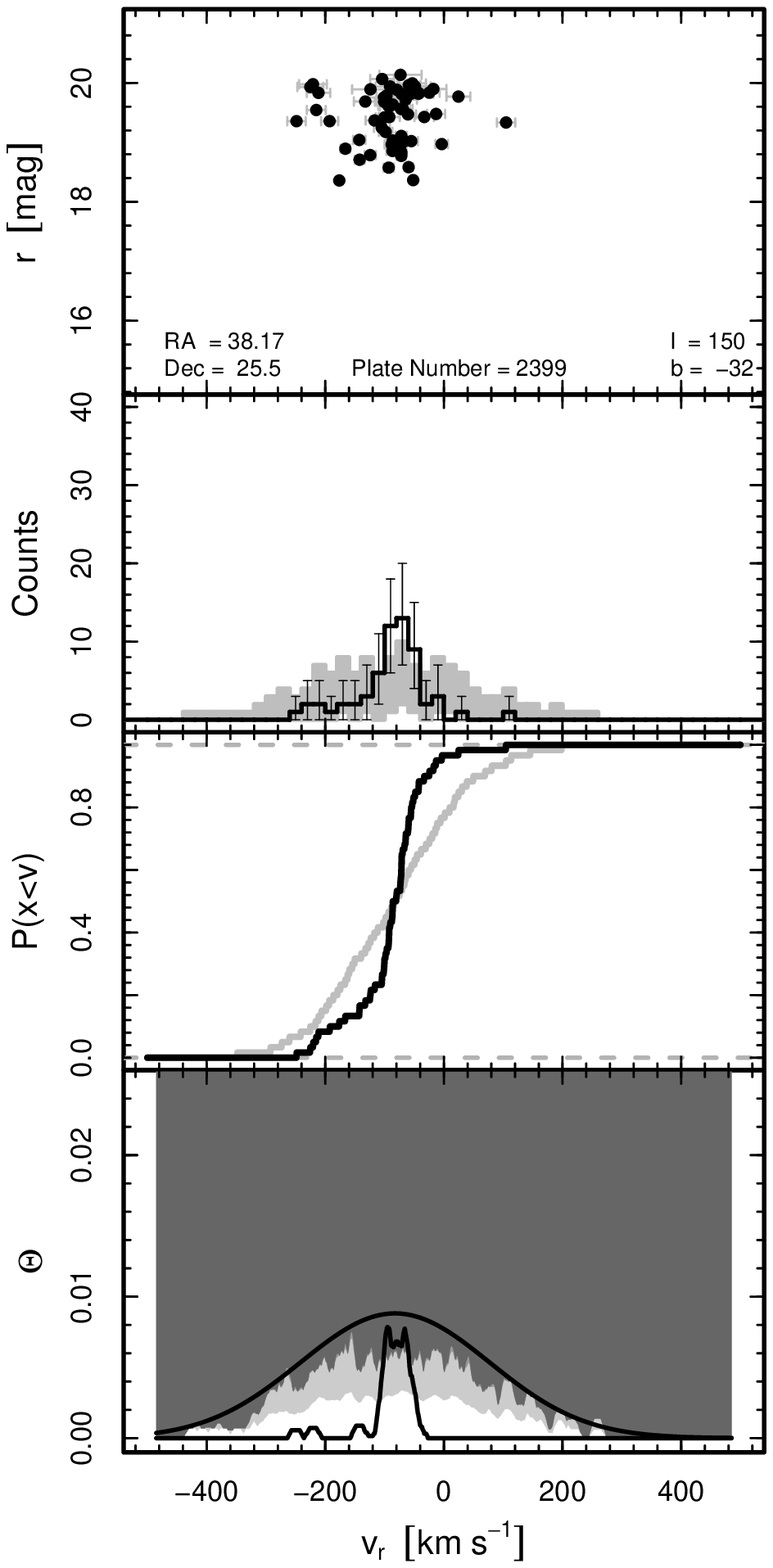}{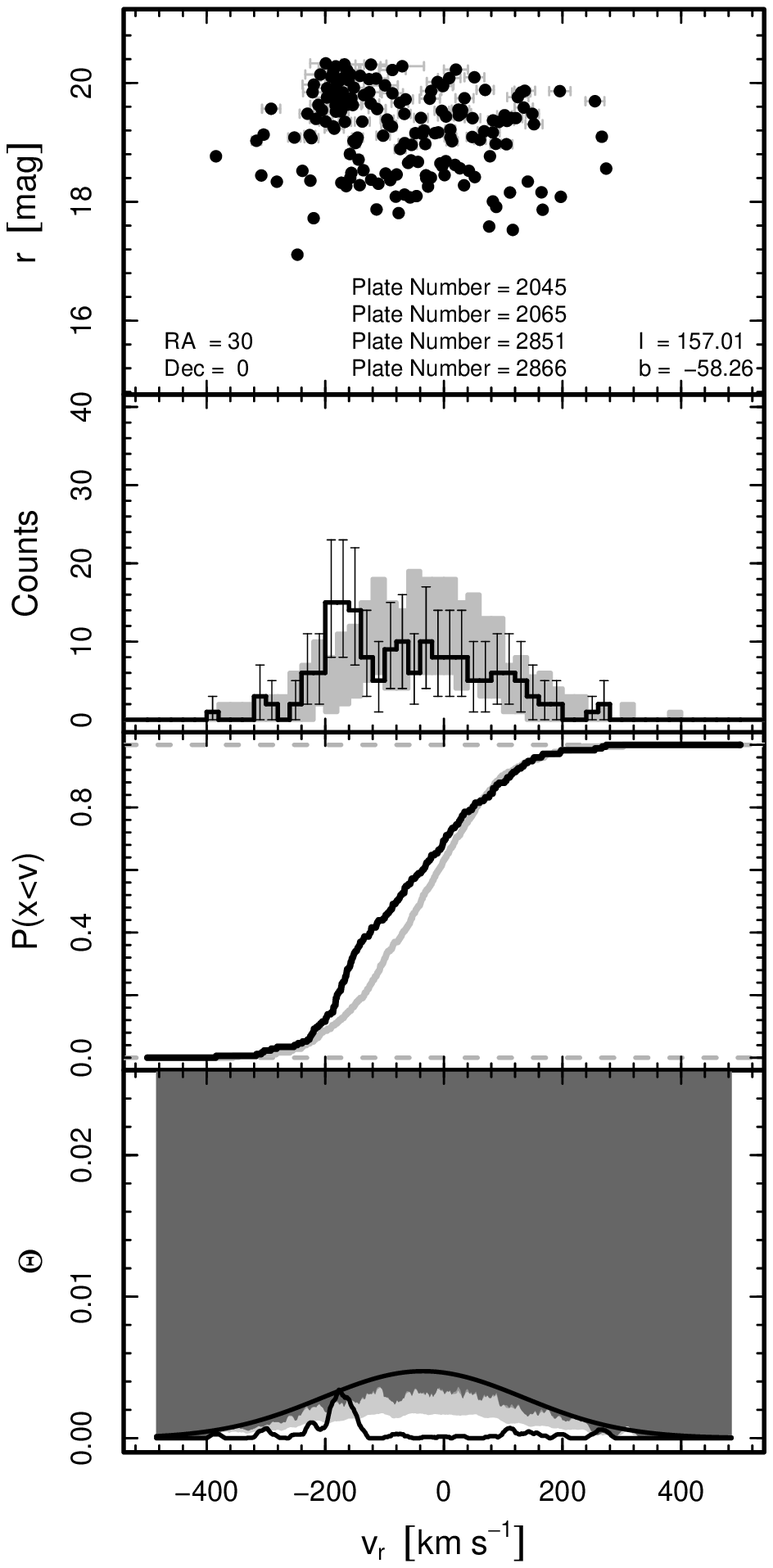}
\caption{\emph{Left}: Data and analyses for the line of sight along
which we found the element of cold substructure PCII-6 and PCII-7
from Table~\ref{tbl-3}.  \emph{Right}: Data and analyses for the line
of sight along which we found the element of cold substructure PCII-9
from Table~\ref{tbl-3}.  See the caption to Figure~\ref{fig3} for a
detailed description of this type of figure.\label{fig8}}
\end{figure}

\begin{figure}
\plottwo{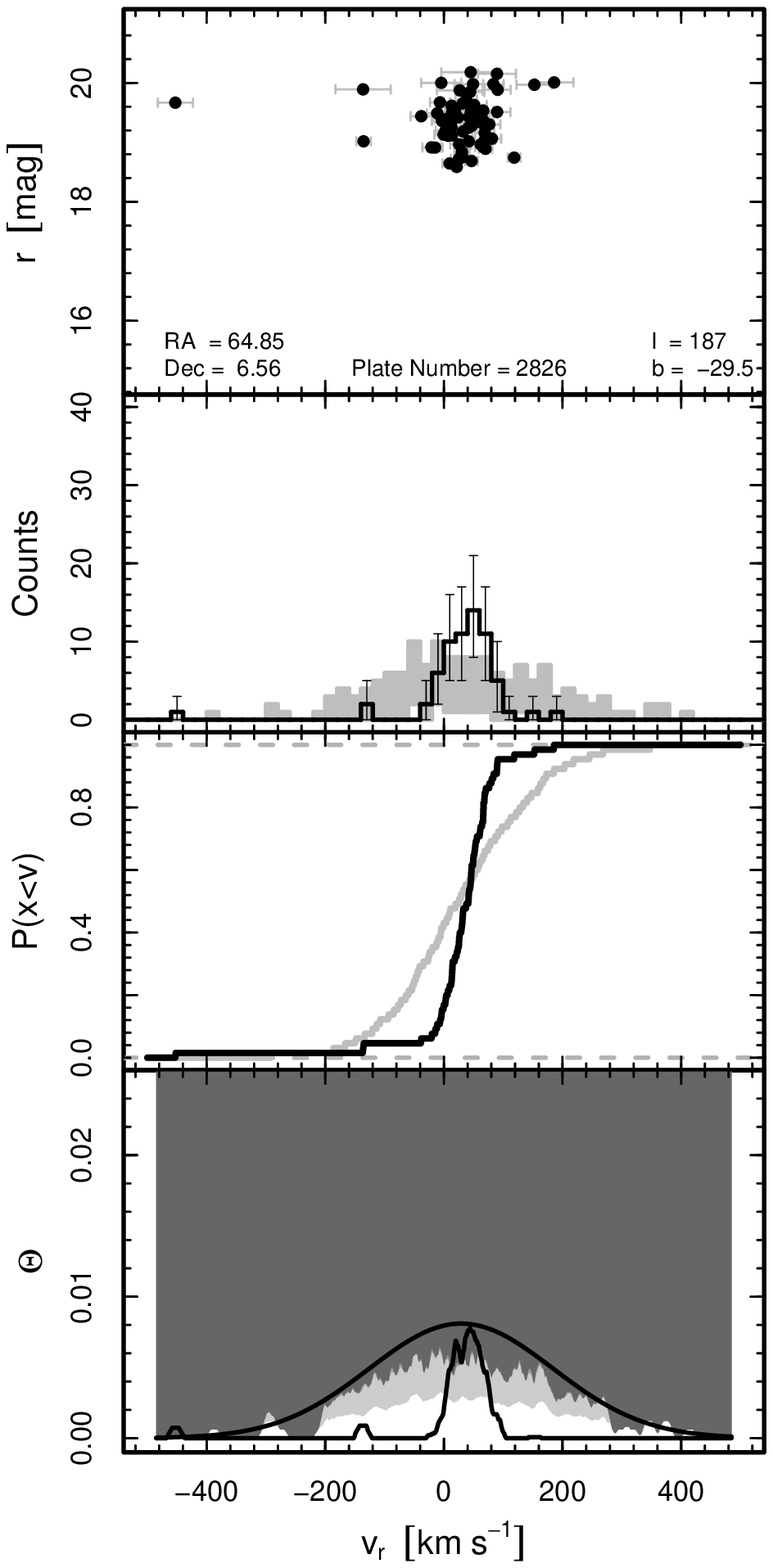}{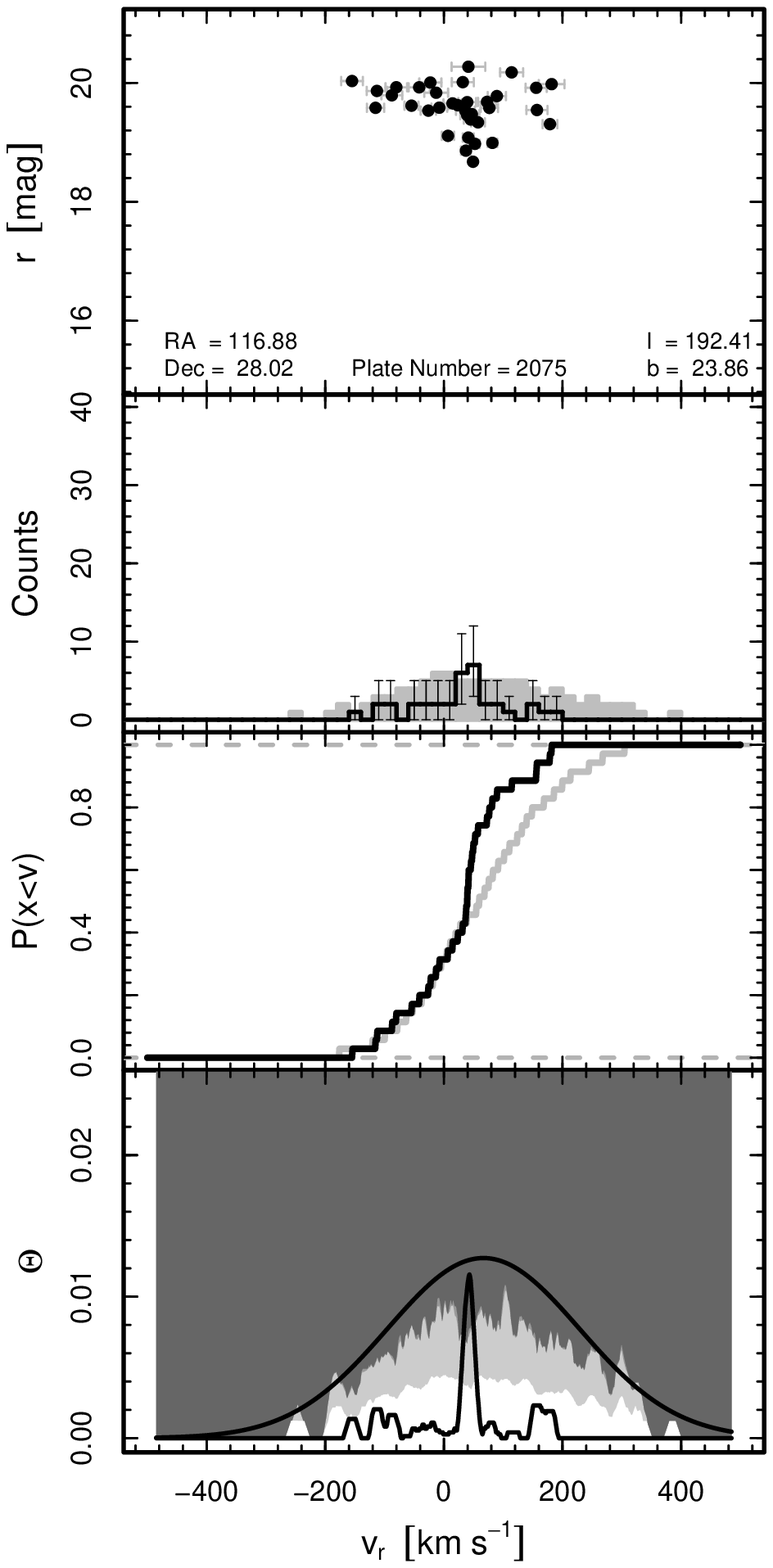}
\caption{\emph{Left}: Data and analyses for the line of sight along
which we found the element of cold substructure PCII-13 and PCII-14
from Table~\ref{tbl-3}.  \emph{Right}: Data and analyses for the line
of sight along which we found the element of cold substructure PCII-15
from Table~\ref{tbl-3}.  See the caption to Figure~\ref{fig3} for a
detailed description of this type of figure.\label{fig9}}
\end{figure}

\begin{figure}
\plottwo{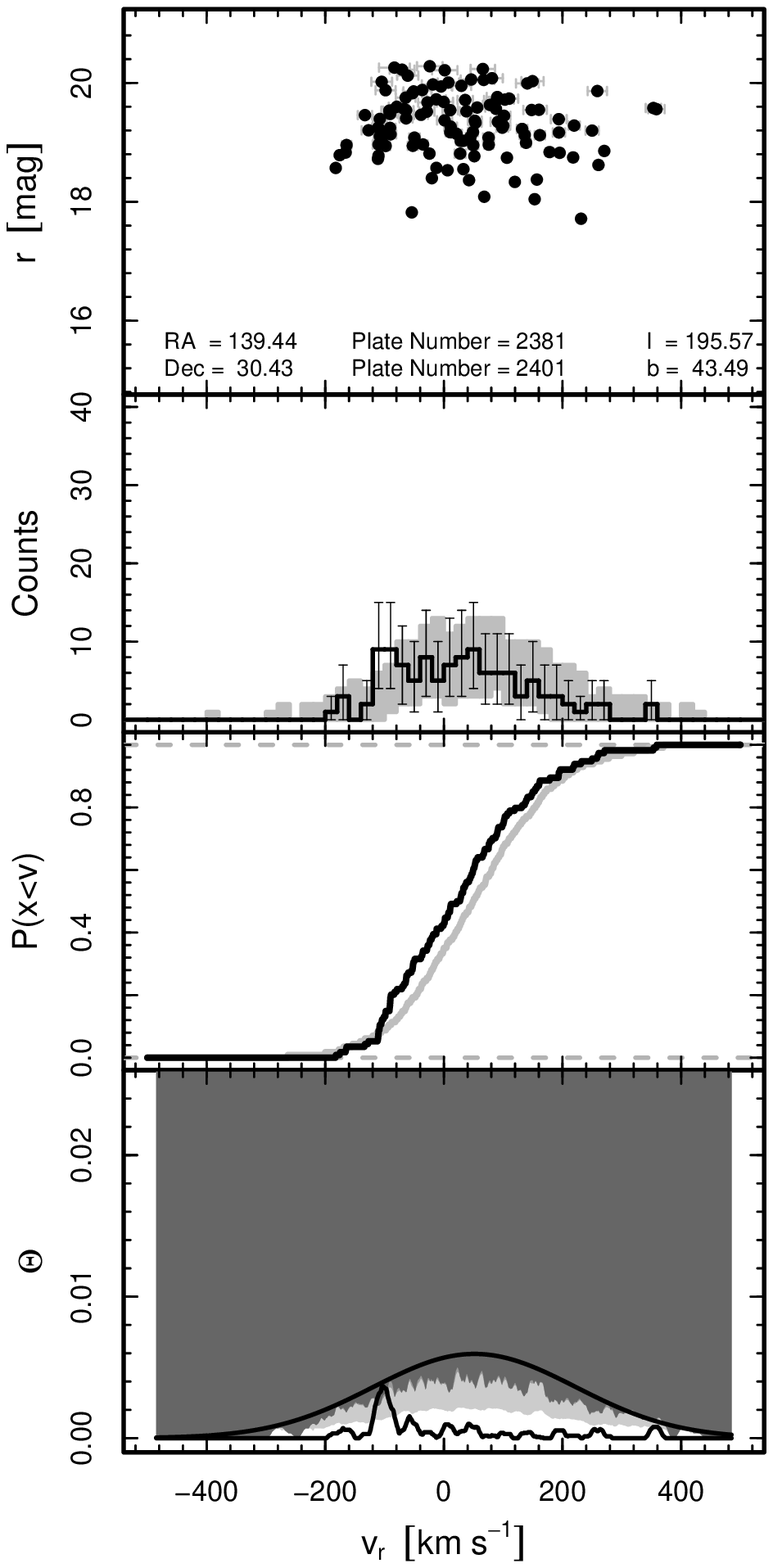}{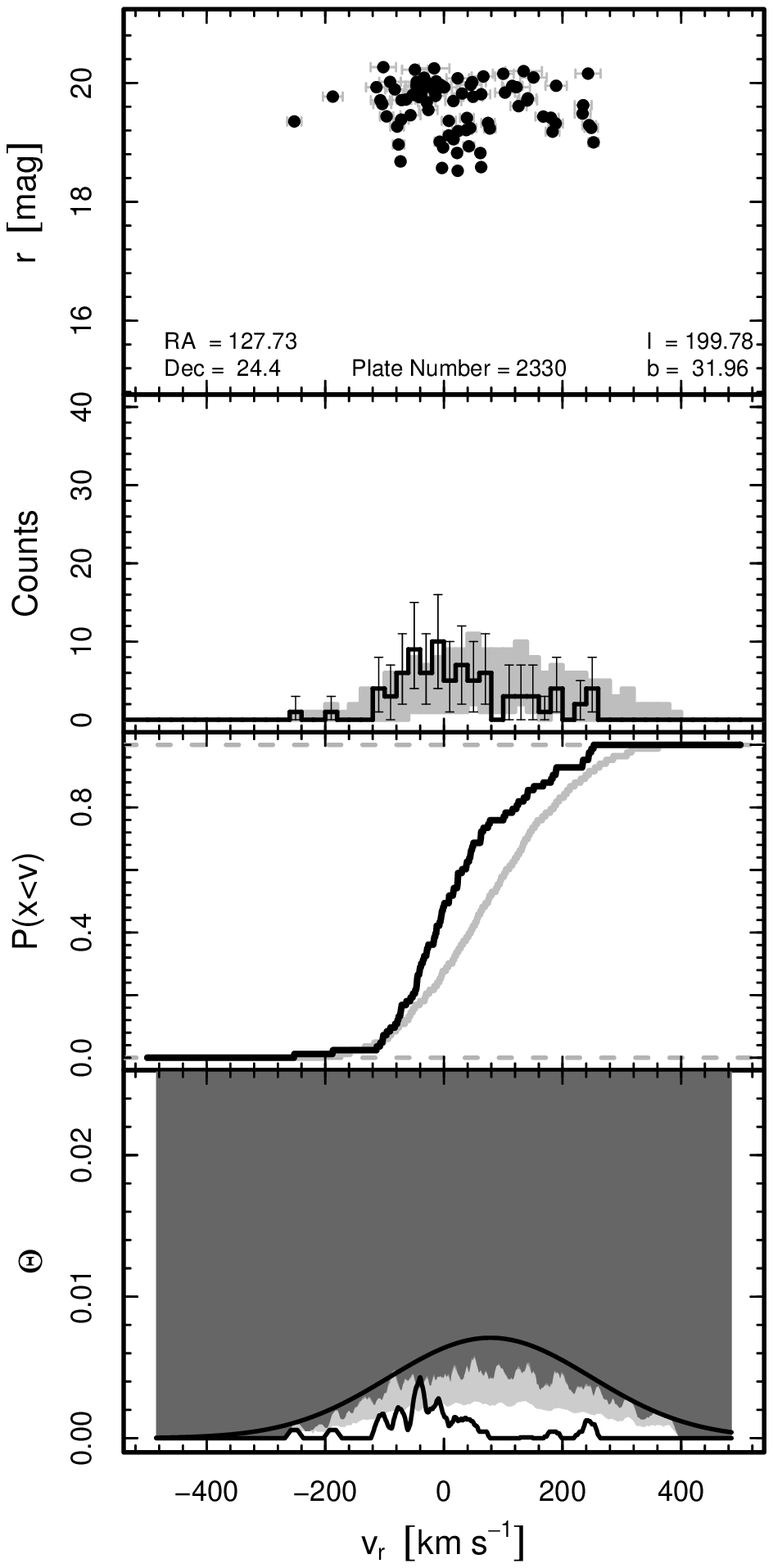}
\caption{\emph{Left}: Data and analyses for the line of sight along which
we found the element of cold substructure PCII-16 from Table~\ref{tbl-3}.
\emph{Right}: Data and analyses for the line of sight along which we
found the element of cold substructure PCII-17 from Table~\ref{tbl-3}.
See the caption to Figure~\ref{fig3} for a detailed description of this
type of figure.\label{fig10}}
\end{figure}

\begin{figure}
\plottwo{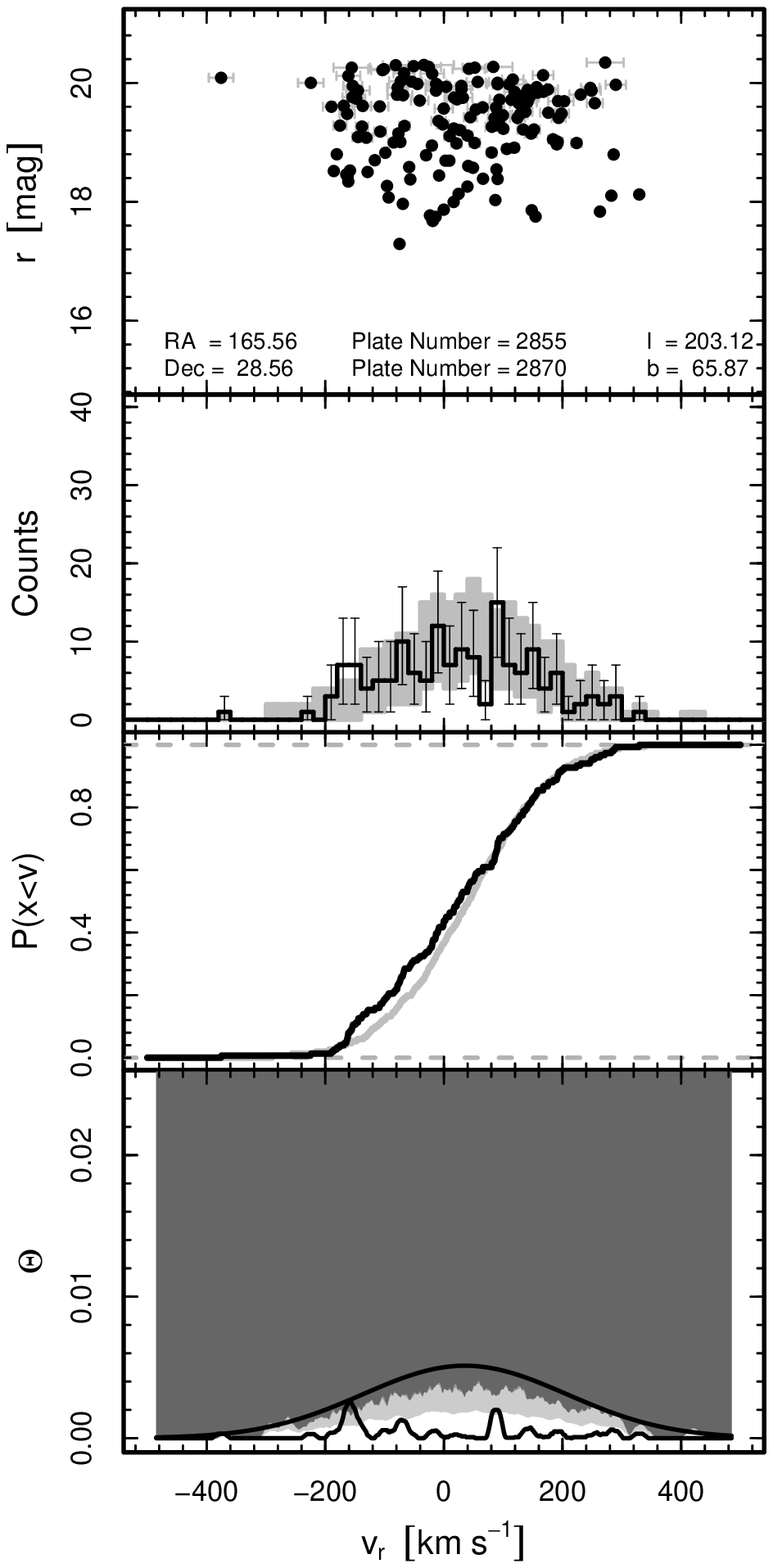}{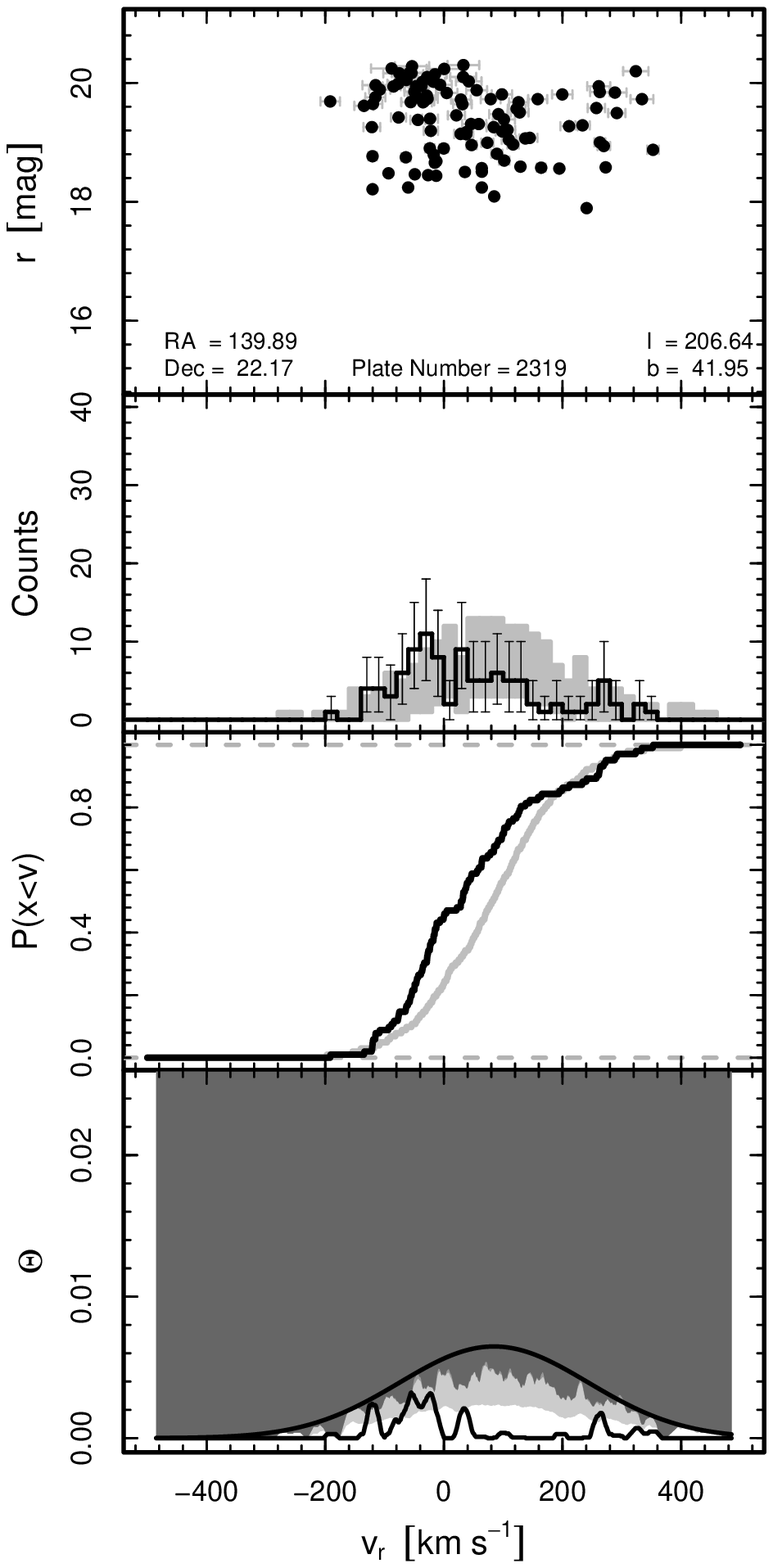}
\caption{\emph{Left}: Data and analyses for the line of sight along which
we found the element of cold substructure PCII-18 from Table~\ref{tbl-3}.
\emph{Right}: Data and analyses for the line of sight along which we
found the element of cold substructure PCII-19 from Table~\ref{tbl-3}.
See the caption to Figure~\ref{fig3} for a detailed description of this
type of figure.\label{fig11}}
\end{figure}

To illustrate these algorithms, consider Figure~\ref{fig2}.  It shows the
radial velocity data and our analyses for the line of sight along which
we found ECHOS B-1 from Table~\ref{tbl-1}, PCI-3 from Table~\ref{tbl-2},
and PCII-3 from Table~\ref{tbl-3}.  The top panel shows radial velocity
versus $r$-magnitude.  We plot radial velocity errors as gray horizontal
bars -- the median error is approximately $11.5$ km s$^{-1}$ -- while the
photometric errors are much smaller than the plotted points.  The second
panel shows in black a histogram derived from the observed SEGUE MPMSTO
star radial velocity distribution with bin-width 20 km s$^{-1}$ and 95\%
confidence intervals from bootstrap resampling.  We also plot in gray
an area that corresponds to a 95\% confidence region for the histogram
that would result from a SEGUE observation of our mock catalog.  As a
result, a significant bin is one for which the error bar on the black
histogram does not intersect the gray region.  The bin centered at
$v_r = -130$ km s$^{-1}$ hosts a significant excesses of MPMSTO stars
relative to the smooth model.  The third panel shows in black the CDF
of the observed MPMSTO star radial velocity distribution and in gray
the average CDF of the simulated radial velocity distribution obtained
from our mock catalog.  Note the large discrepancy between the slope
of the CDF $F'(v_r)$ of the MPMSTO star data and the ``average" CDF of
smooth component $\overline{F_S'}(v_r)$; this indicates that there is
a very significant overdensity in the MPMSTO data at $v_r \approx -120$
km s$^{-1}$.  The fourth panel shows in black our $\Theta(v_r)$ statistic.
We plot the formal 1 in 10$^2$ significance region in light gray, the
formal 1 in 10$^3$ significance region in medium gray, and the formal
1 in 10$^4$ region in the dark gray.  The white region is significant
at less than the 1 in 10$^2$ level; note that the medium and dark gray
regions are nearly coincident, emphasizing the extreme non-Gaussianity of
our $\Theta(v_r)$ statistic.  These significance thresholds are naively
formally equivalent to 2.33-$\sigma$, 3.09-$\sigma$, and 3.72-$\sigma$,
though the distribution from which we derive the significance thresholds
is only defined for positive real numbers and highly non-Gaussian, so
this comparison is not robust.  The heavy black curve is the Gaussian
upper-bound that we use to ensure that the small-scale fluctuations
that result from the finite size of our mock catalog and the finite
number of Monte Carlo iterations we can computationally afford do not
produce false positives.  Our false positive analysis indicates that
the probability that a smooth halo can produce peaks like that peak
observed at $v_r = -121$ km s$^{-1}$ is vanishingly small.  We present
similar plots for all of our detections in order of galactic longitude
in Figures~\ref{fig2} through \ref{fig11}.  The interested reader should
use Tables~\ref{tbl-1}, \ref{tbl-2}, and \ref{tbl-3} (also ordered by
galactic longitude) as a guide to the plots in Figures~\ref{fig2} through
\ref{fig11}.  Note that because we sample from the mock catalog the same
number of MPMSTO stars that were observed by SEGUE along a given line of
sight, any overdensity ensures a subsequent underdensity somewhere else.
Those underdensities are not meaningful.  Also be aware that the apparent
$r$-magnitude inhomogeneities present along some lines of sight do not
necessarily imply position-space substructure.  The reason is that the
transformation of apparent $r$-magnitude into absolute $r$-magnitude is
difficult for individual turnoff stars because isochrones are nearly
vertical in a color-magnitude diagram at the turnoff.  As a result,
small errors in observed color combined with uncertain metallicity can
lead to large systematic uncertainties in photometric parallaxes for
individual MPMSTO stars.

\subsection{A Completeness Estimate}

We would like to quantify the fraction of cold substructure everywhere
in the inner halo of the Milky Way Galaxy, not just along those lines
of sight for which we have candidate ECHOS.  To that end, we need to
know the properties of our detections as well as the properties of the
substructure that our algorithms are not capable of separating form
the smooth inner halo.  Therefore, we calculate our completeness with a
Monte Carlo simulation similar to that described in \S3.3.  This time,
however, we replace a certain fraction of the synthetic stars in $S_r$
with a synthetic ECHOS with known number of stars, physical extent,
radial velocity, and velocity dispersion.  We determine, on average,
how often our algorithms find the inserted ECHOS as a function of the
fraction of the total number of stars along the line of sight that are
a part of the ECHOS.

\subsubsection{Strategy}

For each SEGUE line of sight we perform 1,000 simulations in each of ten
steps in the fraction of the stars along that line of sight in an ECHOS,
from 10\% to 100\%.  We use the fraction of stars in substructure instead
of an absolute number because our detection probabilities also depend
on the number of spectra obtained along a line of sight.  For each
simulation, we select the synthetic ECHOS's mean radial velocity at
random and with uniform probability in the range $v_{\mu} \pm 200$ km
s$^{-1}$ where $v_{\mu}$ is the observed mean radial velocity for the
smooth inner halo at that galactic longitude and latitude.  We give the
inserted ECHOS a three-dimensional Cartesian velocity dispersion chosen
to match (at least to order-of-magnitude) the velocity dispersion of
possible progenitors and  described by a diagonal matrix in which we
select the diagonal entry at random and with uniform probability in
the interval $[0,5/\sqrt{3}]$ km s$^{-1}$.  We then add noise to the
individual radial velocities of the stars that make up the inserted
substructure at a level characteristic of our median estimated radial
velocity errors at the S/N of the individual stars in our detections,
about 11.5 km s$^{-1}$.  We then use our algorithms to determine whether
they detect the synthetic ECHOS.

Considering the significance contours in the bottom panel of
Figure~\ref{fig2}, our sensitivity to a given fractional overdensity
depends on its mean velocity.  That is, an ECHOS with mean velocity far
from the mean velocity of the smooth inner halo is easier to find than
a similar ECHOS with mean velocity close to the mean velocity of the
smooth component.  We attempt to marginalize over this effect by using
a uniform distribution in the synthetic ECHOS's mean radial velocity,
as described in the preceding paragraph.  The uniform distribution is
our attempt to minimize the amount of ``prior information" we include in
the completeness calculation.  This should be a reasonable assumption if
the mean velocities of the real population of ECHOS are equally likely
to lie anywhere in the radial velocity range of the smooth inner halo.
To test this assumption, we compare the difference between the observed
mean velocities of our candidate ECHOS with the median radial velocity of
the smooth component along that line of sight with velocities selected
at random and with uniform probability in the range $v_{\mu} \pm 200$
km s$^{-1}$.  We perform a KS test on the two samples, and repeat the
process 1,000 times drawing a new random set each time.  We find that
the median $p$-value from the KS tests is 0.90; therefore our assumption
that ECHOS are equally likely to be at any radial velocity in an interval
about the mean radial velocity of the smooth model is a good one.  This is
not surprising, as the radial velocities of both ECHOS and of the smooth
background is primarily determined by the relationship between the Sun's
velocity vector in its orbit about the Galaxy and the vector describing
the direction along a given line of sight.  The velocity dispersion we
assign to each inserted ECHOS is an order-of-magnitude estimate of the
velocity dispersion of a real element of cold substructure.  Regardless,
the dominant source of dispersion is random measurement error in the
radial velocities -- which will be larger than the velocity dispersion
of any ECHOS -- and we have a solid understanding of that distribution.
In any case, we tested our algorithms' abilities to detect ECHOS with
velocity dispersions approaching 50 km s$^{-1}$ -- half that of the halo
itself -- and found no significant change in performance.

\begin{figure}
\epsscale{1.0}
\plottwo{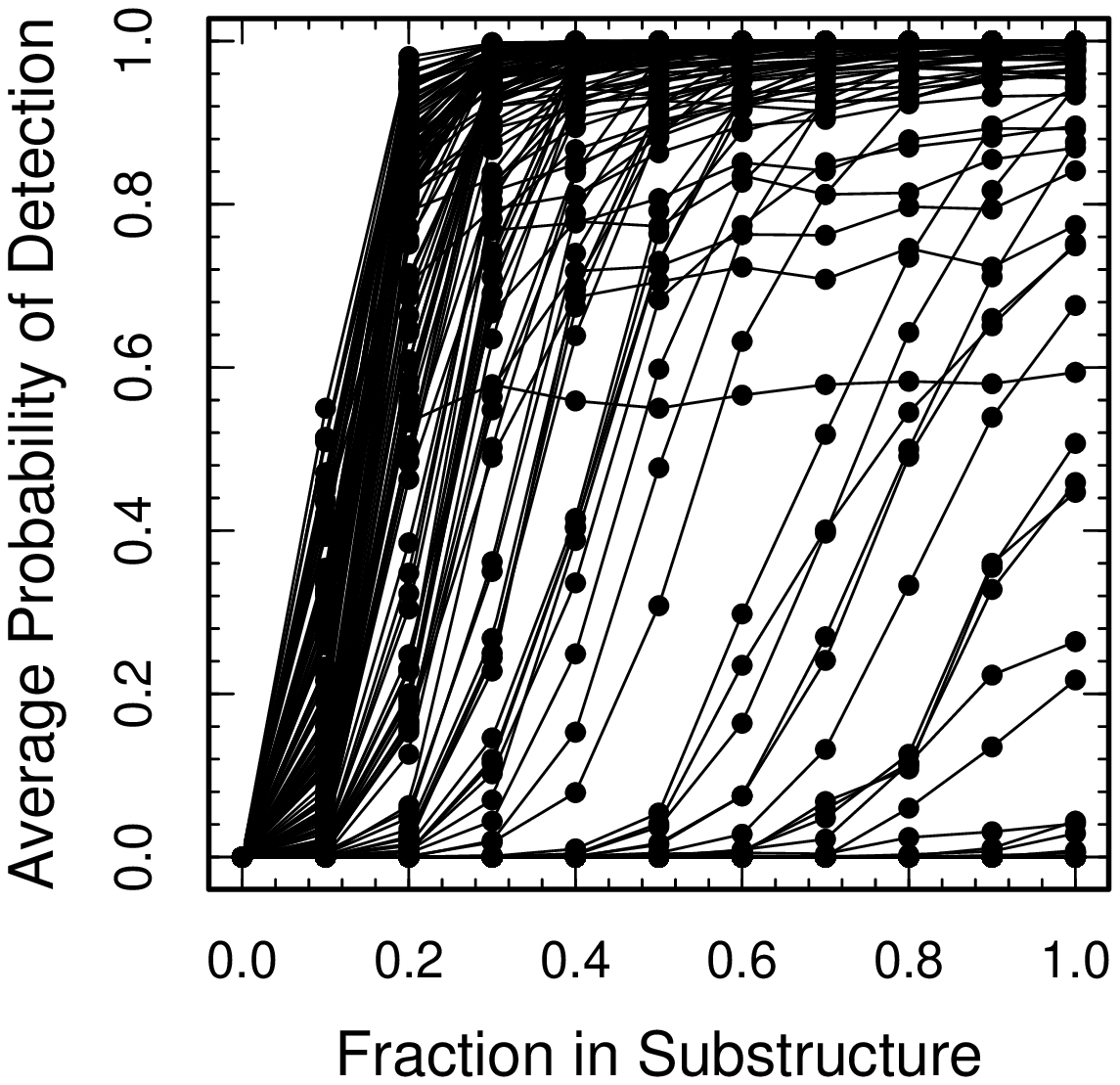}{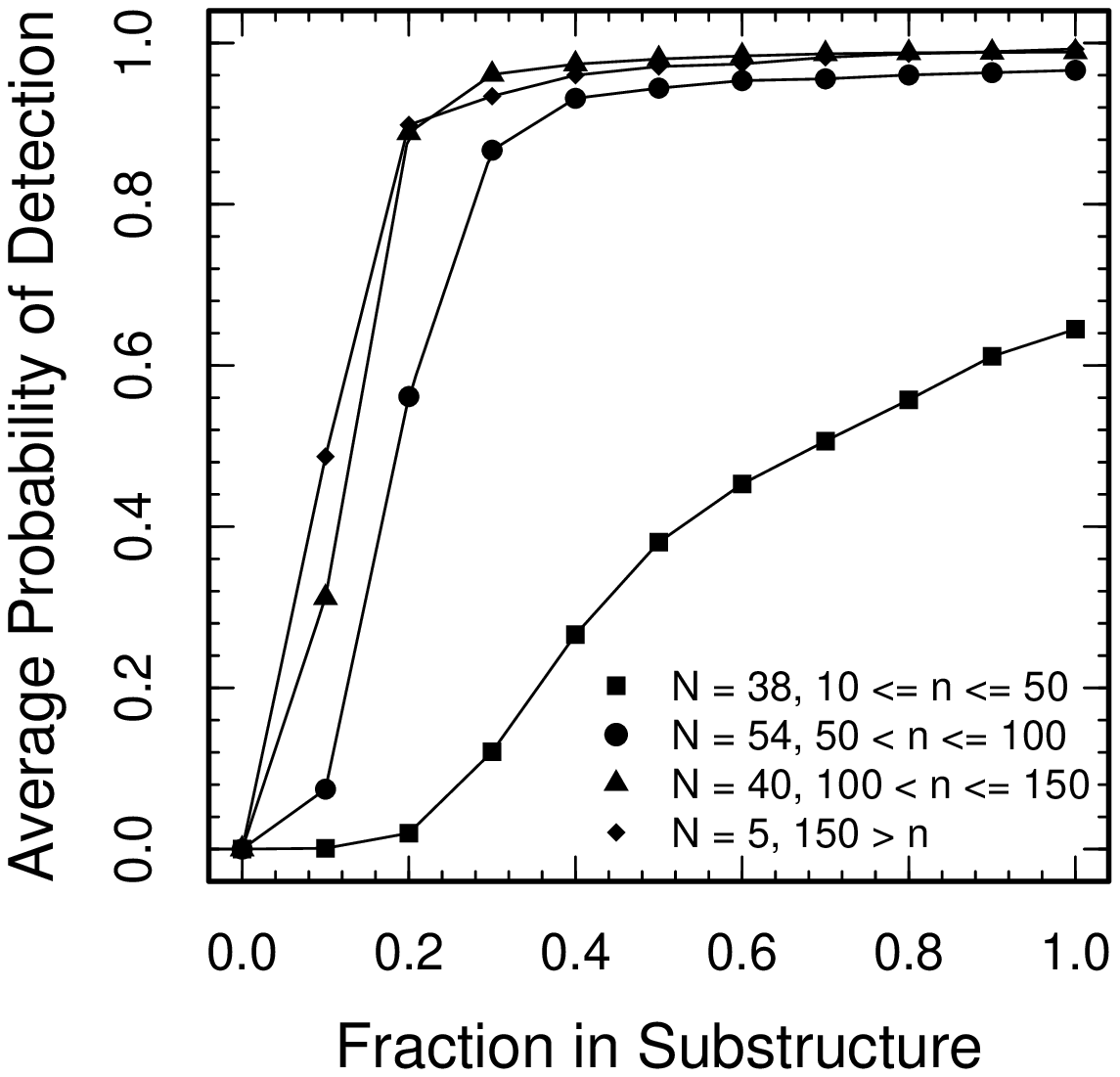}
\caption{The results of our completeness calculation for the class II
peak algorithm.  \emph{Left}: The completeness result for every line
of sight.  \emph{Right}: The average completeness result in bins by the
number of spectra obtained by SEGUE along lines of sight; $N$ is the
number of lines of sight with $n$ spectra.\label{fig12}}
\end{figure}

\begin{figure}
\plottwo{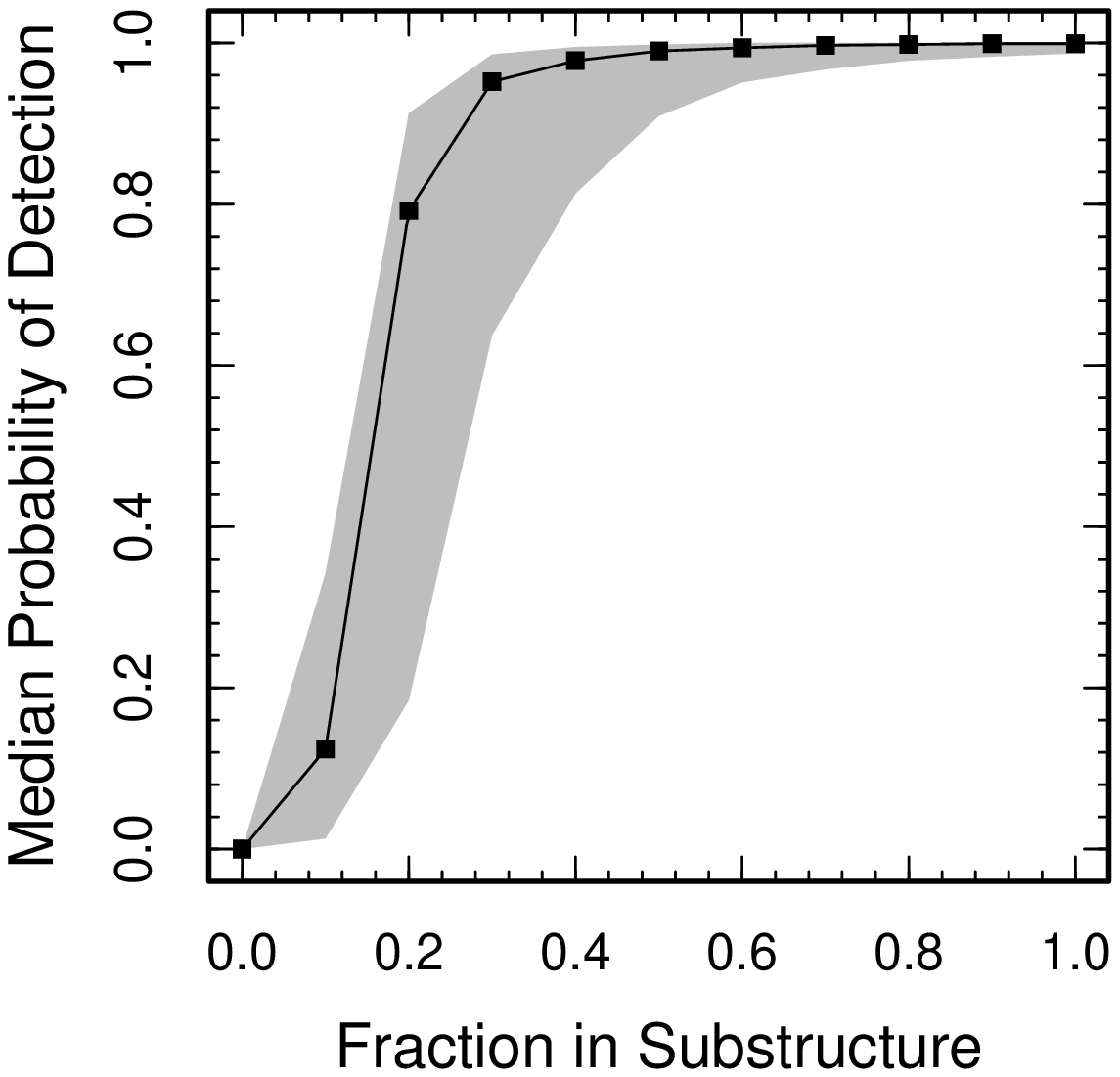}{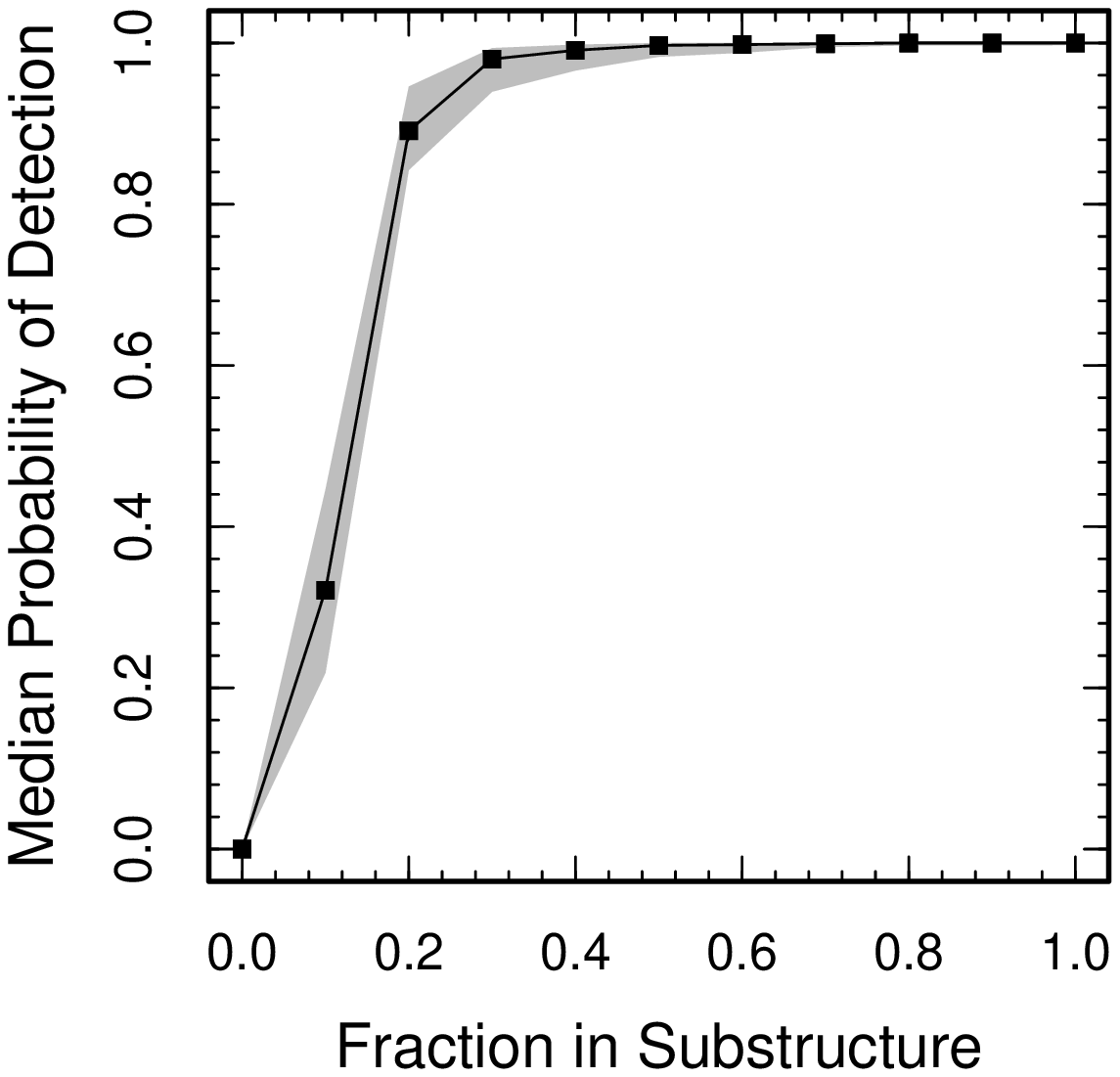}
\caption{The median detection probability for the class II peak algorithm
as a function of the fraction in substructure; the gray area is the
1-$\sigma$ region. \emph{Left}:  For all 115 lines of sight with more
than 30 spectra.  \emph{Right}:  For all 47 lines of sight with more
than 100 spectra.\label{fig13}}
\end{figure}

For each SEGUE line of sight we find that, averaged over the given
distributions for radial velocity and velocity dispersion, the
relationship between probability of detection and fraction of stars
along that line of sight that are part of an ECHOS is nonlinear.  As a
result, we interpolate the results for each line of sight and invert the
function to determine the fraction of the total number of stars observed
along that line of sight that must be part of an ECHOS to ensure that
our algorithms can detect it 95\% of the time.  We illustrate these
calculations for the class II peak algorithm in Figures~\ref{fig12}
and \ref{fig13}; the results for the bin algorithm and the class I
peak algorithm are similar.  Note that for all of the algorithms the
probability of detection in the $N = 99$ lines of sight with more than 50
stars quickly approaches unity as the fraction in substructure increases.
On the other hand, for the $N = 38$ lines of sight with fewer than 50
stars the probability of detection increases much more slowly.  Lines of
sight with less than about 30 stars correspond to those lines in the left
hand panel of Figure~\ref{fig12} that have small average probability of
detection, even with a large fraction in substructure.  More precisely,
for all lines of sight with fewer than 30 stars the average probability
of detection never reaches unity even if every star observed along that
line of sight is in the same ECHOS.  Also note that the performance
of the class II peak algorithm suffers along lines of sight with few
MPMSTO star spectra because of the requirement that each class II peak
detection must have ten MPMSTO radial velocities within one velocity
resolution element of the peak in our $\Theta(v_r)$ statistic.

Table~\ref{tbl-4} lists our upper limits for each line of sight.
These limits are the fraction of spectroscopically observed MPMSTO stars
that could belong to single ECHOS yet still go undetected, on average,
5\% of the time.  The data in Table~\ref{tbl-4} indicates that our
algorithms are unable to detect ECHOS at least 95\% of the time along
lines of sight with fewer than about 30 spectra, even if every star along
that line of sight belongs to an ECHOS.  As a result, we remove these
sparsely sampled lines of sight from our analyses from this point on.

The fact that our completeness calculation only accounts for single ECHOS
again emphasizes the point that our search for single overdensities
in radial velocity-space is not necessarily sensitive to an ensemble
of substructures below our sensitivity thresholds.  We are also
potentially insensitive to diffuse, fully phase-mixed, or violently-relaxed
substructure and not necessarily sensitive to diffuse multiply wrapped
substructure.  We have shown, however, that our completeness is not a
function of velocity dispersion and that we can detect substructure with
velocity dispersion up to 50 km s$^{-1}$ as well as we can detect cold
substructure.  In the end, an ECHOS is detectable if it contributes the
threshold fractional overdensity determined by SEGUE sampling along that
line of sight.  Physically, that means that we can detect ECHOS that have
not spread very far over their orbits.  The fact that detectability is
not sensitive to velocity dispersion implies that we can find ECHOS with
a wide variety of possible orbital configurations, so long as there is a
high enough number density along a single line of sight.  As we will show
in the discussion, in the volume we search we are likely to find older
substructures than those discovered in photometric searches, but younger
substructures than those discovered by 6D searches in the same volume.
We recognize that the time it takes for the debris of a given accretion
event to spread beyond our detectability threshold depends on progenitor
mass, velocity dispersion, and orbit.  Still, for any given set of
progenitor properties our search can delineate its debris for a longer
time after progenitor disruption than photometric searches.  In any case,
we have very precisely calculated our sensitivity thresholds and developed
a search strategy that can be meaningfully applied to theoretical models
to assess their level of agreement with our observations.

\subsection{The Relative Merits of Each Algorithm}

The bin algorithm and the class I peak algorithm are both very unlikely to
produce false positive detections at the cost of a failure to detect less
obvious genuine ECHOS.  On the other hand, the class II peak algorithm
is better at detecting diffuse or more phase-mixed ECHOS at the cost of
an occasional false positive.  In general the class II peak algorithm
is most capable of detecting low-density ECHOS, followed by the class I
peak algorithm and the bin algorithm; Table~\ref{tbl-4} quantifies this
ranking.  Note though that the class II peak algorithm is less capable
than the other algorithms along lines of sight with relatively few stars,
because of the requirement that at least ten MPMSTO stars must be within
an interval of width 11.5 km s$^{-1}$ centered on the velocity of the
detected peak.  That is, the class II peak algorithm cannot detect small
fractional densities along lines of sight with $\sim$10 spectra because a
small fraction of $\sim$10 spectra will always be less than the required
ten MPMSTO.  Recall that the limit of ten MPMSTO stars within 11.5 km
s$^{-1}$ of the peak in $\Theta(v_r)$ was set to minimize false positives.
The peak algorithms can also self-consistently determine the velocity
dispersion of an ECHOS (a characteristic lacking in the bin algorithm)
and they do not require the discretization into bins that causes the
bin algorithm to sometimes split the signal of a genuine ECHOS into two
neighboring bins.

\begin{figure}
\plotone{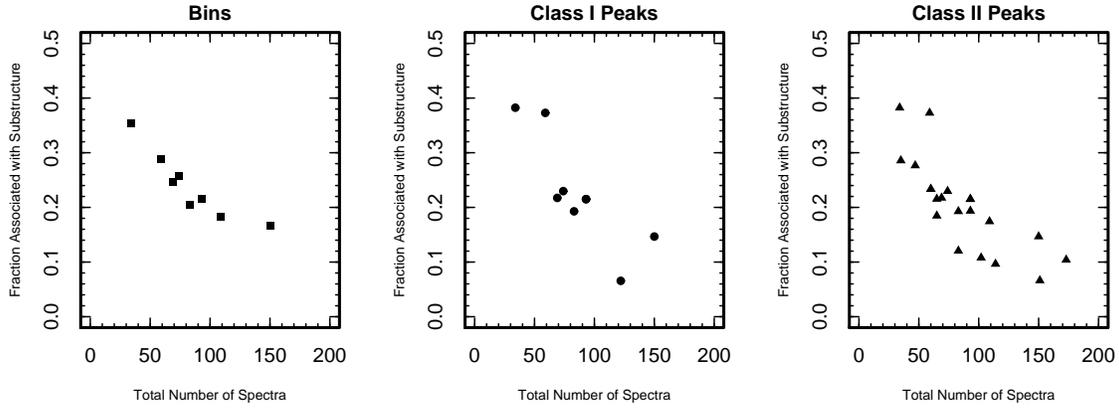} 
\caption{Plots showing the fraction in substructure versus the total
number of spectra obtained along the line of sight where the substructure
was detected for all detections and all methods.  Note that the lower
detected fractional overdensities are associated with a large number of
spectra in agreement with our completeness calculation.\label{fig14}}
\end{figure}

\begin{figure}
\plotone{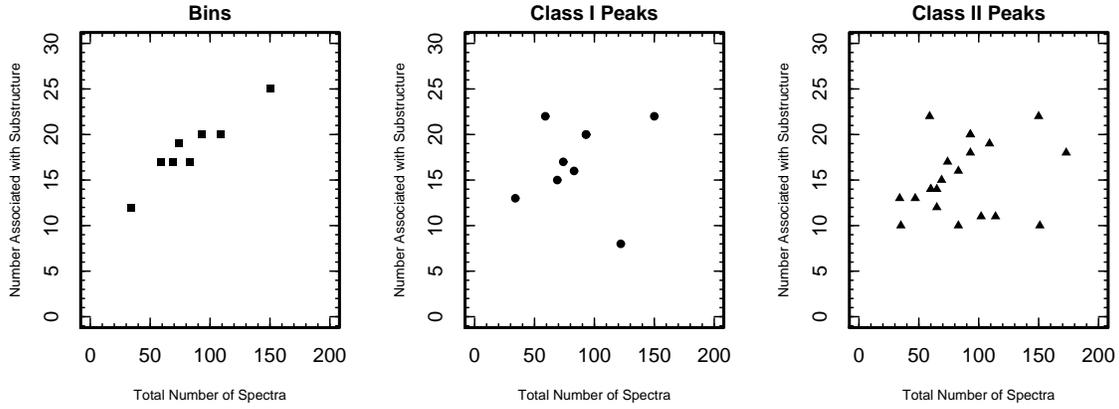}
\caption{Plots showing the number of stars associated with an element
of substructure versus the total number of spectra obtained along the
line of sight where the substructure was detected for all detections and
all methods.  Note that for the peak algorithms there is no correlation
between number of stars associated with an element of substructure
and the total number of spectra obtained along that line of sight.
This fact combined with the result of our completeness calculation implies
that our detections do not cluster just at our detection thresholds.
Therefore, it is not likely that the inner halo is comprised of an array
of substructures below our detection thresholds.\label{fig15}}
\end{figure}

\begin{figure}
\plotone{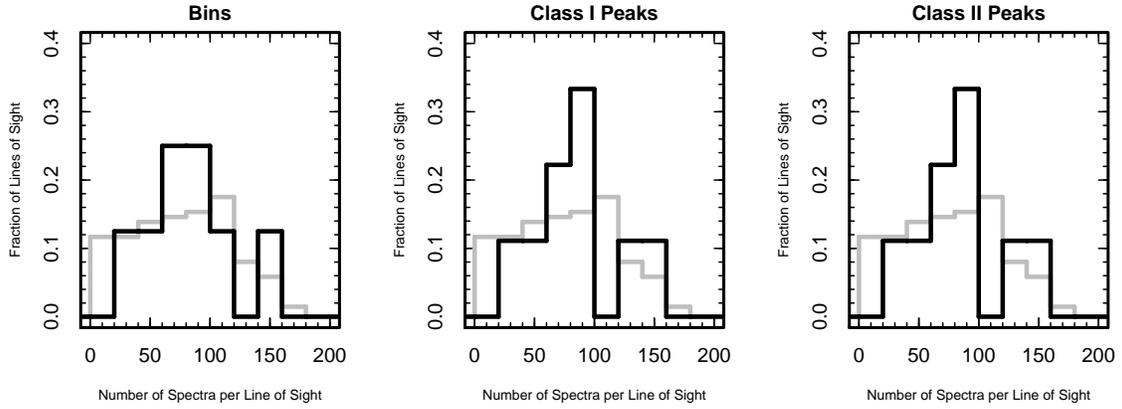} 
\caption{Histograms of the distribution of the number of spectra for each
line of sight.  We plot the distribution of our entire set of lines of
sight in gray and the subset of lines of sight with detections in black.
Note that the distribution of the total number of SEGUE spectra obtained
along lines of sight with substructure detections is similar to the
same distribution for all lines of sight in our sample.  That is,
our detections are not just found along the lines of sight where SEGUE
most densely sampled the inner halo MPMSTO population.  This suggests
that the inner halo is not made up of a population of diffuse
substructure below our sensitivity thresholds.\label{fig16}}
\end{figure}

Figure~\ref{fig14} confirms the results of our completeness calculation:
the peak algorithms can discover lower fractional overdensity
ECHOS along lines of sight with many stars than the bin algorithm.
Figures~\ref{fig15} and \ref{fig16} demonstrate that our detections
are not all clustered at the edge of detectability.  Specifically,
Figure~\ref{fig15} demonstrates that for the peak algorithms there is
no correlation between the number of stars associated with ECHOS and
the total number of spectra obtained along the line of sight where
the ECHOS was discovered.  Recall the results of our completeness
calculation (presented in Figures~\ref{fig12} ans \ref{fig13}) that
the detectable fractional overdensity is correlated with the number
of stars per line of sight.  In other words, we are most sensitive to
low fractional overdensity substructure along the most well-sampled
lines of sight.  If there were a substantial population just below
our detection thresholds, then there would be many detections at small
absolute number of stars associated with elements of substructure at large
total numbers of spectra in Figure~\ref{fig14}.  This is not observed,
so these observations imply that the ECHOS we find are not all close to
the lowest fractional overdensity we can detect along any line of sight.
On the other hand, Figure~\ref{fig15} indicates that our detections are
not just found along the lines of sight where SEGUE most densely sampled
the inner halo MPMSTO population.  Together, these facts suggest that
at our velocity resolution the inner halo is not composed of a multitude
of diffuse substructures beyond our sensitivity threshold.

\section{Extension to the Full Inner Halo}

We would like to generalize our search for ECHOS along 115 SEGUE
lines of sight to the full inner halo.  If each line of sight were
targeted randomly, then the average of our upper limits on the fraction
of substructure along each line of sight would be an unbiased
estimator for the fraction of substructure in the entire inner halo.
To ensure that our final estimate of the fraction of the inner halo in
ECHOS is not biased by the lines of sight that were targeted at previously
known elements of substructure, we exclude those lines of sight from
the analyses in sections \S4.1 and \S4.2.  When the exceptions listed
in Table~\ref{tbl-4} are excluded, the RA and Dec of each SEGUE line of
sight center was selected without regard to the presence of substructure
in the Milky Way's inner halo.  If we make the following well-justified
assumptions (see \S2):

\begin{enumerate}
\item
SDSS photometrically detected all MPMSTO stars in our sensitivity range
\item
The color-color cut used to select MPMSTO candidates accurately classified
them as MPMSTO stars, such that the radial velocities belong to stars
within our definition of the inner halo given in \S2
\end{enumerate}

\noindent
and we recall that fibers were allocated randomly to the photometric
MPMSTO candidates without prior knowledge of their radial velocities,
then the SEGUE MPMSTO sample is an unbiased tracer of the inner
halo.  Therefore we can assert that, on average, the fraction of
spectroscopically observed MPMSTO stars that reside in ECHOS along a
given SEGUE line of sight is the same as the fraction of all MPMSTO
stars in ECHOS in the volume searched along that line of sight.  Again,
by ECHOS we mean single elements of substructure that are unresolved
(or barely resolved) in radial velocity in the SEGUE data.

\subsection{A Limiting Case}

Imagine the worst-case scenario: our algorithms miss all ECHOS
in the inner halo just below the 95\% detection limits given in
Table~\ref{tbl-4}.  That is, they fail to identify literally all
ECHOS that they would normally detect 94.9\% (or less) of the time.
The probability of this occurrence is vanishingly small -- $P(X)
\sim 0.05^{115}$ -- and the probability of missing two or more such
ECHOS in each line of sight is smaller still; nevertheless, it is
a useful limiting case.  In that situation, any line of sight would
harbor an ECHOS just below the line of sight specific 95\% thresholds
given in Table~\ref{tbl-4}.  If we imagine that every line of sight
in fact intersects an ECHOS just below the threshold, then the true
fraction of the MPMSTO population in ECHOS along a given line of sight
would be identical to the threshold given for that line of sight in
Table~\ref{tbl-4}.  We have 105 lines of sight that were not pointed
at known elements of substructure and that possess the more than the 30
spectra necessary for a potential detection.  Those 105 lines of sight
provide us with 105 independent estimates of the fraction of the MPMSTO
population in ECHOS.  We combine these independent estimates in an average
value with an appropriate weighting scheme.  We can compute the volume
of the inner halo (as defined in \S2) scanned along every line of sight
using Monte Carlo integration; we report that volume in Table~\ref{tbl-4}.
Therefore, we weight the contribution to the upper limit of each line of
sight by its volume to give lines of sight that scanned a larger volume
of the inner halo greater leverage in determining our average limit.
Finally, for lines of sight along which we have a high-confidence
detection we substitute the observed fraction in substructure from
Tables~\ref{tbl-1} or \ref{tbl-2} for the estimated upper limit given
in Table~\ref{tbl-4}.  We use bootstrap resampling to estimate the errors
on our upper limits.  Ultimately, we find that the bin algorithm produces
a 95\% upper limit of $0.52^{+0.04}_{-0.03}$ for the fraction of MPMSTO
stars in the inner that belong to ECHOS.  This slightly higher than
the upper limit of $0.42^{+0.01}_{-0.02}$ produced by the peak algorithm.
We regard our limit from the peak algorithm as more accurate because the
bin algorithm is the least sensitive of our methods.  If we restrict
our calculation to only those 41 lines of sight with more than 100
spectra, the limits quoted above become $0.38^{+0.02}_{-0.02}$ for the
bins and $0.34^{+0.02}_{-0.02}$ for the peaks.  The discrepancy between
these two is due to the fact that our volume weighting scheme does not
penalize lines of sight with poor sensitivity enough in the average.
That is, we do not get improved upper limits from adding more lines
of sight because our sensitivity to ECHOS is a function of the number
of MPMSTO spectra obtained by SEGUE along a given line of sight.  As a
result, the upper limit is weaker if lines of sight with poor sampling
are included in the calculation.  In any case, we present a much more
precise calculation of the fraction of the halo in substructure in \S4.2.
Again note that our algorithms are not necessarily sensitive to the
possibility that the inner halo is made up of innumerable diffuse,
fully phase-mixed, or violently-relaxed elements of substructure below
our sensitivity thresholds.

Equivalently, we can estimate the total number of ECHOS in the inner
halo by computing the ratio of the volume scanned by SEGUE to the total
volume of the inner halo as defined in \S2.  Using the same assumptions
given above, that ratio should be the same as the ratio between the total
number of ECHOS we identify (less the expected number of false positives)
and the total number of ECHOS in the entire inner halo.  We find that
we searched about 0.54\% of the volume of the inner halo as defined is
\S2 and found seven new ECHOS.  This suggests that there should be of
order 10$^3$ ECHOS like those we have identified in the entire volume
of the inner halo.

\subsection{A More General Calculation}

\begin{figure}
\plottwo{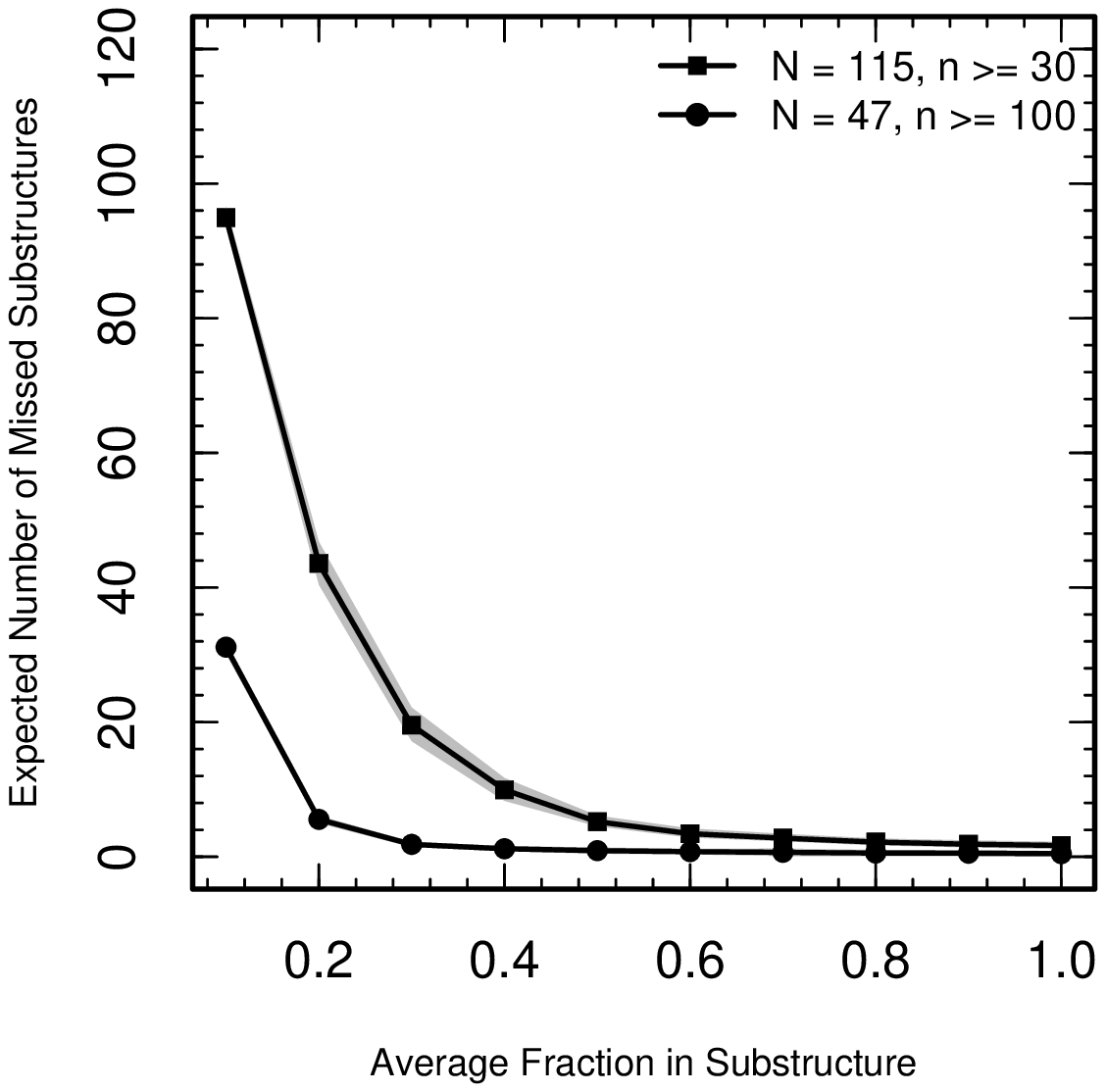}{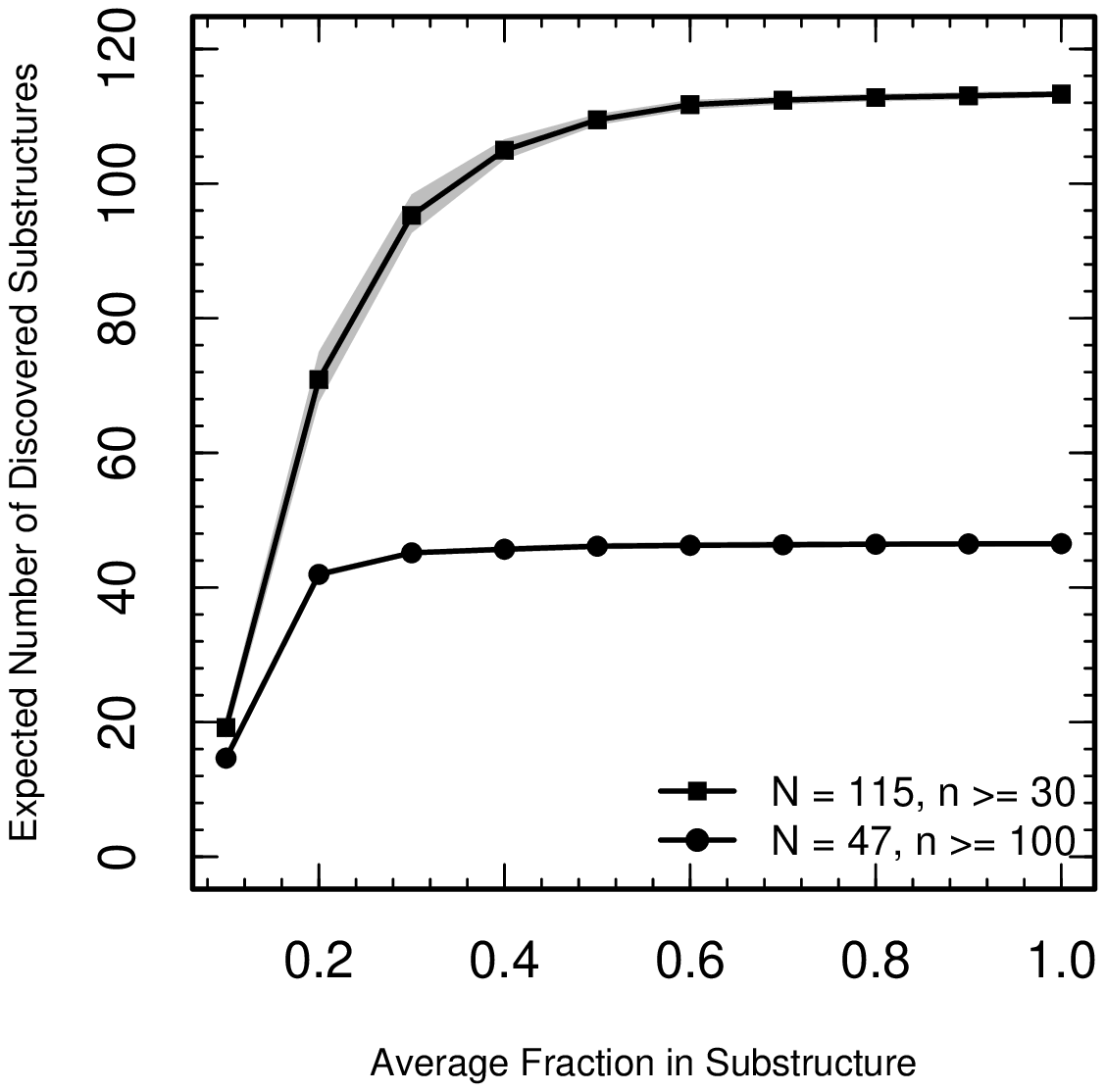}
\caption{\emph{Left}: The expected number of undetected substructures
as a function of the average fraction in substructure for the class II
peak algorithm; the gray area is the 1-$\sigma$ region.  \emph{Right}:
The expected number of detected substructures as a function of the
average fraction in substructure for the class II peak algorithm; the
gray area is the 1-$\sigma$ region.\label{fig17}}
\end{figure}

We can also compute the total number of ECHOS we expect to miss or
discover over our entire search as a function of the average fraction of
the halo in ECHOS.  In our previous calculation, we assumed that there
was an ECHOS just below our 95\% detection threshold along every line of
sight and we showed that this event was extremely unlikely.  That is not
to say that we don't miss anything; in fact, even for ECHOS beyond our
95\% thresholds it is almost certain that we miss at least one ECHOS
over our entire search, as the probability of that event is $P(X) \sim 1
- 0.95^{115}$.  To address this point, we used a Monte Carlo simulation
to predict the expected number of ECHOS we both miss and discover as
a function of the average fraction of the inner halo in substructure.
For every line of sight and for each of ten steps in the fraction of the
field that is a part of an ECHOS we have already computed the average
detection probability for an ECHOS at that fractional overdensity.
Therefore, we can model the detection process for every line of sight
and for every one of those ten fractional overdensities by drawing
a random number with uniform probability from the interval $[0,1]$.
If that number is greater than the average detection probability, it
counts as a missed substructure; if it is less, it counts as a discovered
substructure.  We can repeat this process for every line of sight in our
search 100 times such that we compute both the expected number of missed
and discovered ECHOS over our entire search, as well as the distribution
of both quantities.  We plot the result of these calculations for the
class II peak algorithms in Figure~\ref{fig17}; the results for the bin
algorithm and the class I peak algorithm are similar.

\begin{figure}
\plotone{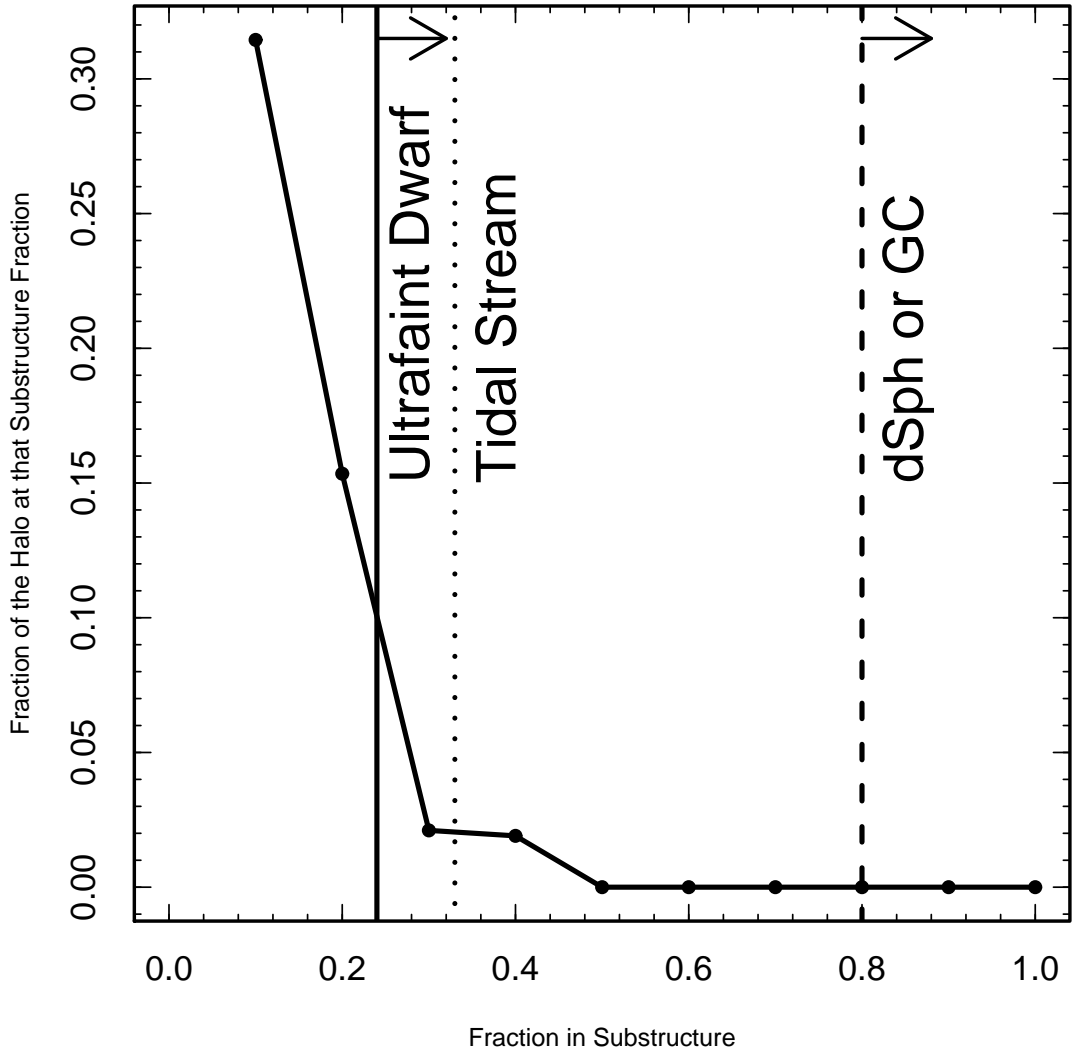}
\caption{The volume fraction of the halo at a given fraction in
substructure.  Note that about 1/3 of the halo (by volume) has 10\% of
its MPMSTO population in ECHOS and about 1/6 of the halo (by volume)
has 20\% of its MPMSTO population in ECHOS; the fraction of the halo
(by volume) with more than 20\% of its MPMSTO population in ECHOS is
just a few percent.  We also plot the expected fraction of the halo in
ECHOS with properties similar to ultrafaint dwarf galaxies, known tidal
streams like Monoceros and \citet{gri06b}, and classical dwarf spheroidal
galaxies and globular clusters.  There are unlikely to be ECHOS like
undiscovered classical dwarf spheroidal galaxies or globular clusters in
the inner halo, and only a few percent of the halo hosts ECHOS like the
Monoceros or \citet{gri06b} tidal streams.  Our search does not rule out
the possibility that there could be ECHOS like ultrafaint dwarf galaxies
in the inner halo.\label{fig18}}
\end{figure}

We can use our list of class II detections combined with the results of
this more general completeness calculation to compute a prediction for
the fraction of the halo in ECHOS of a given overdensity (we use our list
of class II detections because that list is likely more representative
of the whole inner halo ECHOS population than either of the other two
lists).  That is, Table~\ref{tbl-3} tells us how many ECHOS we found
at each fractional overdensity and we know from Figure~\ref{fig17}
how many ECHOS our search would have yielded if the entire halo was
at a particular fraction in substructure.  The ratio between the
observed number of ECHOS at a given fraction in substructure and the
number predicted by the completeness calculation at the fraction is an
estimate of the fraction of the halo that has substructure at that level.
We find that we expect about 1/3 of the halo (by volume) to have 10\%
of its MPMSTO population in ECHOS and about 1/6 of the halo (by volume)
to have 20\% of its MPMSTO population in ECHOS; the fraction of the halo
(by volume) with more than 20\% of its MPMSTO population in ECHOS is very
small.  We plot the results of this calculation in Figure~\ref{fig18}.
We include in Figure~\ref{fig18} the location of ECHOS with properties
like ultrafaint dwarf galaxies, known tidal streams like Monoceros and the
\citet{gri06b} stream, and classical dwarf spheroidal galaxies and
globular clusters.  There are unlikely to be ECHOS like the classical
dwarf spheroidal galaxies or globular clusters in the inner halo, and only
a few percent of the halo volume hosts ECHOS like the Monoceros stream
or the \citet{gri06b} stream.  On the other hand, our search does not
rule out the possibility of ECHOS comparable to ultrafaint dwarf galaxies.

\section{Discussion}

\begin{figure}
\plotone{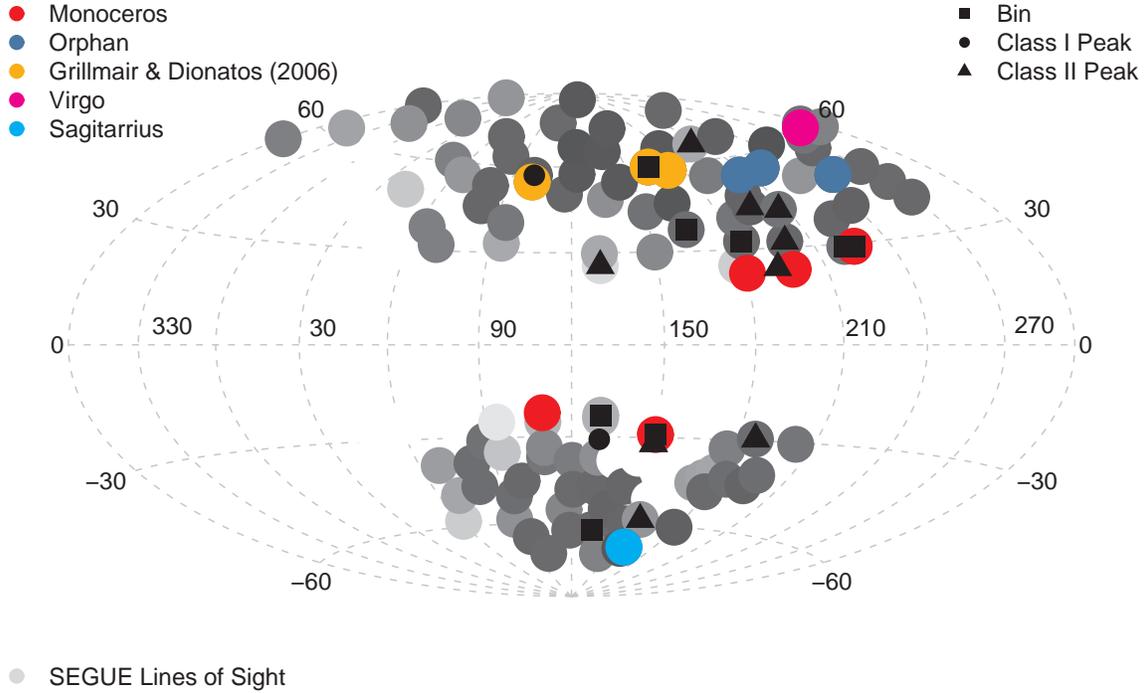}
\caption{A Hammer projection of our detections in galactic coordinates.
We plot all 137 of the SEGUE lines of sight we analyze in a gray-scale.
The darker circles are those lines of where we are the most complete
according to Table~\ref{tbl-4} and lighter circles where we are least
complete.  We indicate lines of sight that were pointed at pieces of the
sky expected to potentially intersect a known element of substructure
by coloring that line of sight according to the legend.  We plot our
detections from the bin algorithm as black squares, our class I detections
from the peak algorithm as black circles, and our class II detections from
the peak algorithm as black triangles.  The distribution of our ECHOS in
galactic coordinates is consistent with an isotropic distribution given
our completeness.  If a piece of substructure was discovered by more
than one algorithm we only plot the symbol corresponding to the most
robust algorithm.  In order of decreasing robustness: bins $\approx$
class I peaks $>$ class II peaks.\label{fig19}}
\end{figure}

We plot our ECHOS on the sky in galactic coordinates in Figure~\ref{fig19}
and we indicate lines of sight targeted at known elements of substructure.
The distribution of ECHOS on the sky is consistent with an isotropic
distribution given our completeness, simply because more stars fall within
our definition of the inner halo toward the Galactic anticenter; for that
reason, we are more sensitive to lower fractional density substructures
in that direction.  More quantitatively, imagine that the fraction of the
halo in cold substructure is uniform in galactic longitude and latitude.
Under that assumption, we can use the line of sight specific sensitivity
thresholds given in Table~\ref{tbl-4} to compute the expected $l$ and $b$
distribution under the assumption of isotropy.  The distribution that
results from that analysis is statistically indistinguishable from the
distribution of our ECHOS.

\begin{figure}
\plotone{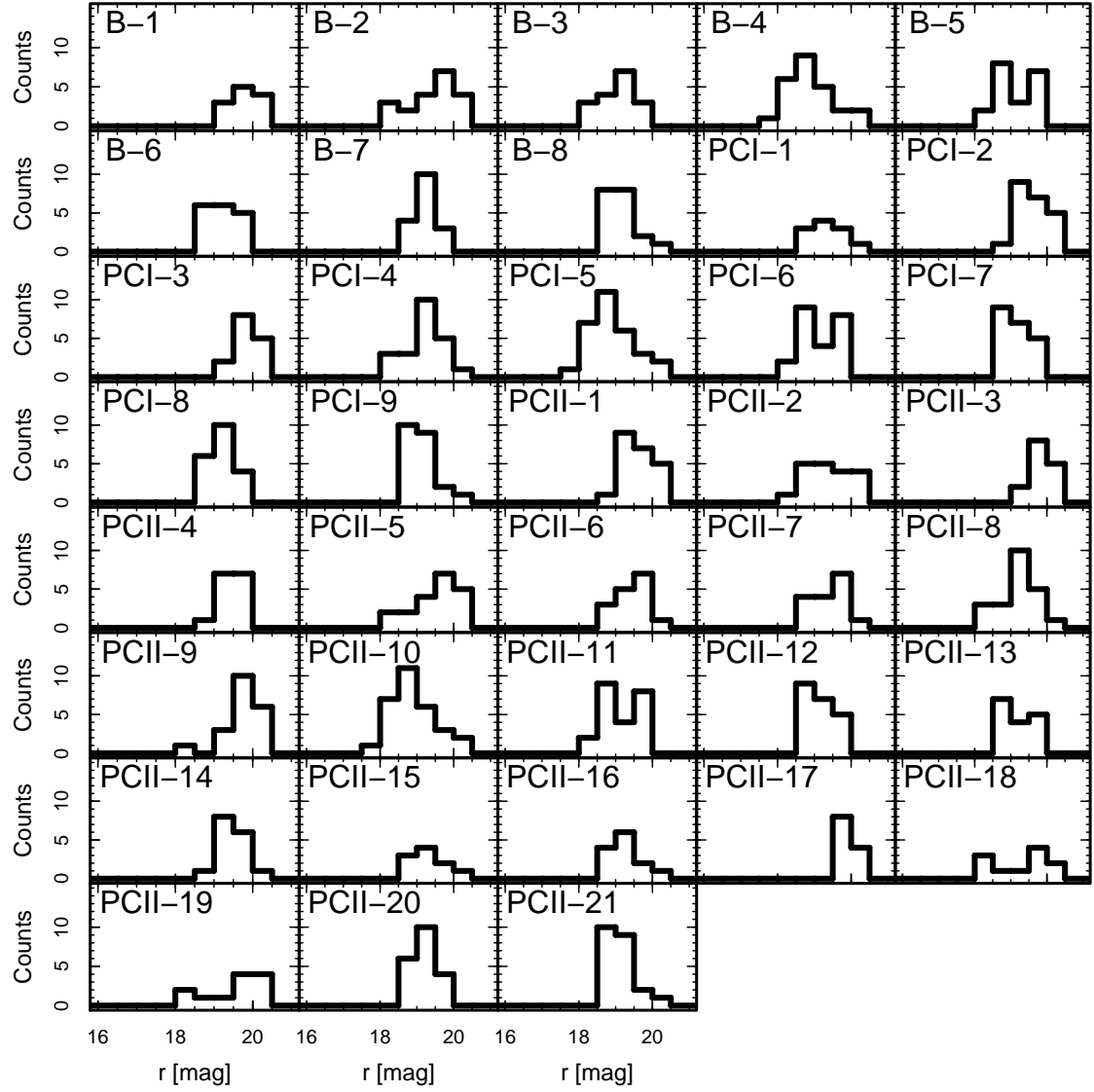}
\caption{The $r$-magnitude histograms for stars in our detections listed
in Tables~\ref{tbl-1}, \ref{tbl-2}, and \ref{tbl-3} -- the identifier
in each panel corresponds to its ID number in Tables~\ref{tbl-1},
\ref{tbl-2}, and \ref{tbl-3}.  For the bin detections the histograms
include all stars with radial velocities that place them in the
significant bin while for the peak detections the histograms include all
stars with radial velocities that place them within one median velocity
error of the significant radial velocity peak.\label{fig20}}
\end{figure}

\begin{figure}
\plotone{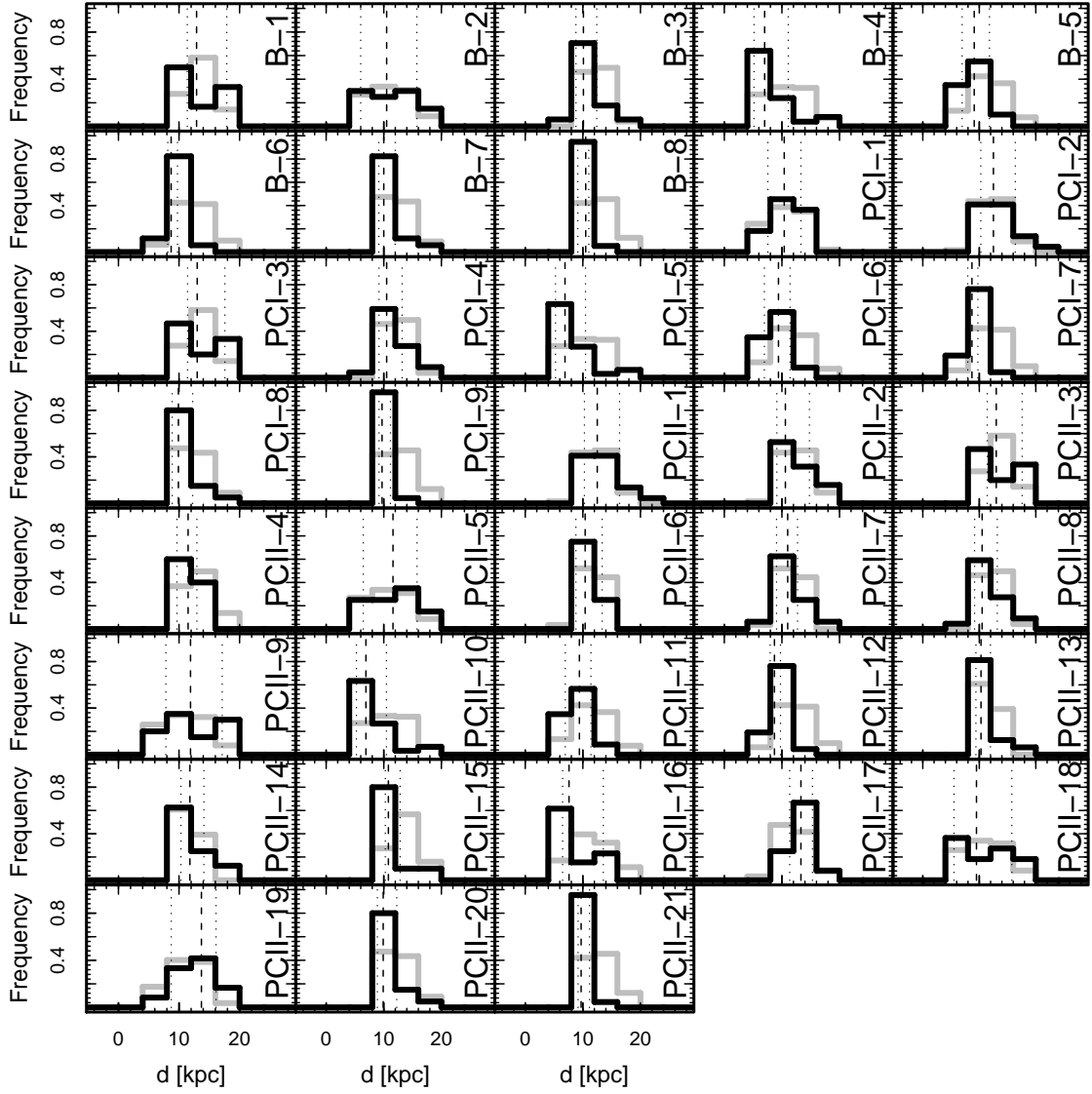}
\caption{Approximate heliocentric distance histograms for stars in our
detections listed in Tables~\ref{tbl-1}, \ref{tbl-2}, and \ref{tbl-3}
-- the identifier in each panel corresponds to its ID number in
Tables~\ref{tbl-1}, \ref{tbl-2}, and \ref{tbl-3}.  In black we plot
approximate heliocentric distance distributions for all stars with radial
velocities that place them in the significant bin (for bin detections)
or within one median velocity error of the significant radial velocity
peak (for the peak detections).  The dashed vertical line denotes the
median heliocentric distance and the two vertical dotted lines delimit
the interval that contains 95\% of the distribution.  In the same panels
we plot in gray the average heliocentric distance distribution of all
stars in our mock catalog along that line of sight.\label{fig21}}
\end{figure}

\begin{figure}
\plotone{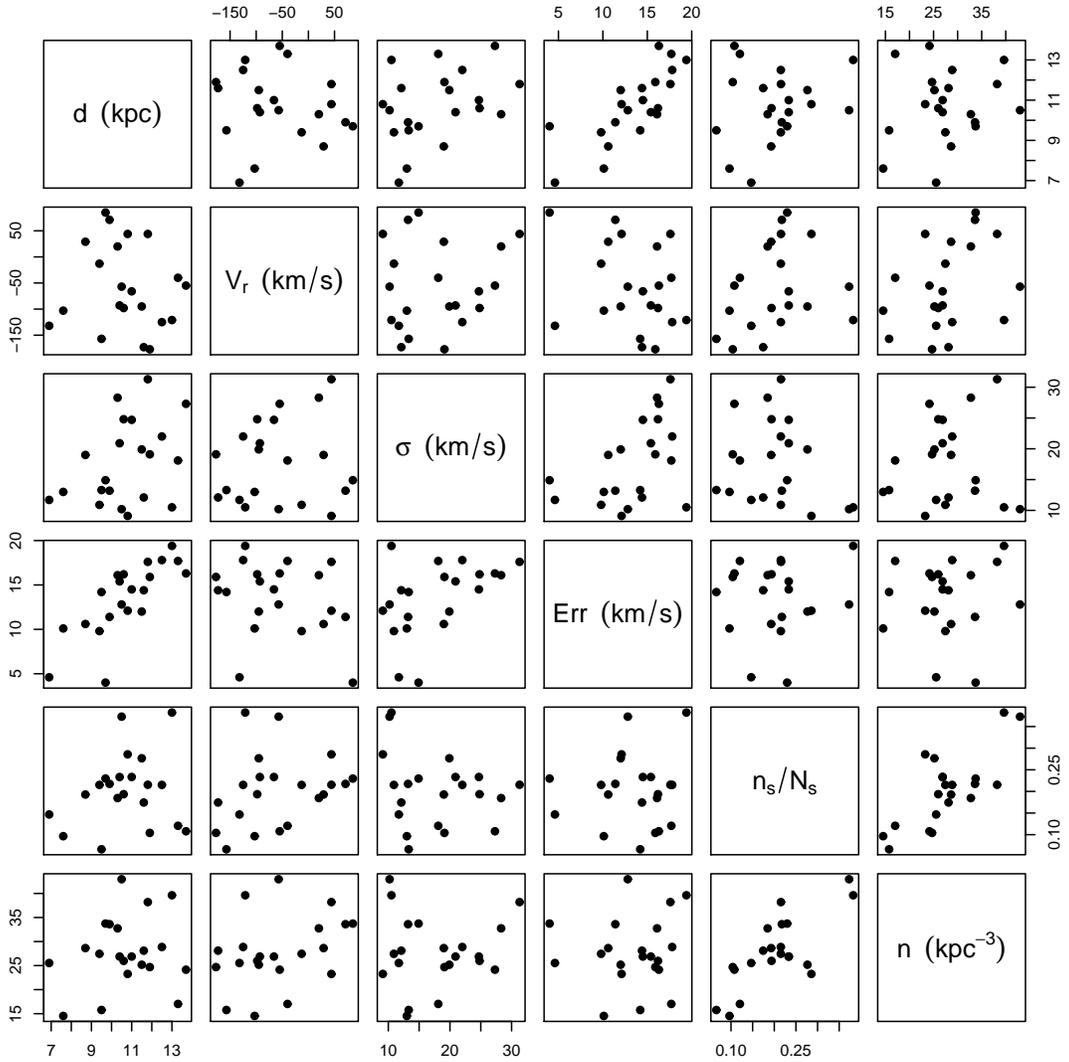}
\caption{Multiplot for the properties of our class II detections:
median distance $d$ in kiloparsecs, radial velocity $v_r$ in km s$^{-1}$,
velocity dispersion $\sigma$ in km s$^{-1}$, median radial velocity error
in km s$^{-1}$, the fraction in substructure $n_s/N_s$, and number density
$n$ in kpc$^{-3}$.  There are no obvious trends save for those expected
from instrumental limitations and basic physics.  Substructures at greater
distance appear to have larger velocity dispersions (and larger median
errors) because of decreasing radial velocity precision for faint stars.
Substructures with larger fractional overdensities also tend to have
larger physical number densities.\label{fig22}}
\end{figure}

We give $r$-magnitude histograms and approximate heliocentric distance
distributions for all of our detections in Figures~\ref{fig20} and
\ref{fig21}.  We plot in Figure~\ref{fig22} a multiplot for the relevant
physical properties of our class II detections given in Table~\ref{tbl-3}
and Table~\ref{tbl-4}.  We plot our class II detections because they
provide the largest self-consistent sample and we believe the population
of class II detections is more representative of the inner halo ECHOS
population collectively.  We find no obvious non-trivial correlations.
Finally, we note that the stars belonging to all of our ECHOS are
spread uniformly over the solid angle sampled along the line of sight
where they were discovered.  In other words, our ECHOS appear to have
sheet-like (as opposed to stream-like) morphologies.  This observation
is consistent with the prediction in \citet{joh08} that substructures
within 20 kpc of the Galactic center are more likely to have sheet-like
``cloudy" morphologies than stream-like ``great circle" morphologies.

Note in Figure~\ref{fig22} that we expect the 10\% fractional overdensity
ECHOS to both have MPMSTO star number densities $n \approx 15$ kpc$^{-3}$
and velocity dispersions no larger then the floor set by our radial
velocity errors.  We previously showed in \S4.2 that these 10\%
overdensity ECHOS are likely to be found in 1/3 of the halo volume.
That is, we expect a significant but not unlimited population of low
density ECHOS in the inner halo of the Milky Way.  For comparison,
\cite{sea08} used RAVE data to show that there are no vertical streams
in the solar neighborhood with total (not just MPMSTO) stellar number
densities $n \gtrsim 10^3$ kpc$^{-3}$.  By itself, the fraction of the
Milky Way's halo in substructure provides an independent, qualitative
measure of the intensity of the Milky Way's stellar accretion history
over the past few Gyr: more substructure indicates a more intense stellar
accretion accretion history while less substructure indicates a less
intense stellar accretion history.  We make a more quantitative estimate
in the next subsection.

\subsection{Comparison To Previous Studies}

We identify seven new substructures as well as rediscover all known
substructures in our search volume, so in that way our search is
more sensitive than past studies.  Moreover, our search for cold
radial velocity substructure in the inner halo bridges a gap in
galactocentric distance between solar neighborhood searches using 6D
phase-space information and more distant in situ halo searches using
surface brightness.  At the same galactocentric distance, substructures
discovered in surface brightness are likely younger than substructures
discovered in radial velocity.  Those radial velocity substructures
are themselves likely younger than substructures discovered using 6D
phase-space information.  For that reason, we argue that we are sensitive
to older substructure than those identified by \citet{bell08}.  Note that
\citet{bell08} statistically quantified the degree of surface brightness
substructure using the same tracer population in an $r$-magnitude range
(and therefore distance range) that partially overlaps with our search.
As a result, our searches are complimentary in relative look-back time --
\citet{bell08} quantified the level of debris left behind by more recent
accretion events and we have quantified the level of debris left behind
by less recent accretion events.  The fact that both studies find that
about 30\% of the MPMSTO population in the inner halo is in substructure
suggests that the level of stellar accretion into the Milky Way has
been relatively constant over the past few Gyr.  More quantitatively,
the typical velocity of stars in the inner halo of the Milky Way at $z =
0$ is $v \sim 200$ km s$^{-1}$ and the characteristic distance from the
center of the Galaxy is 10 kpc.  Therefore, a crossing time is $t_{c} \sim
50$ Myr and radial velocity substructure will likely dissolve below our
limit of detectability in $t_{d} \sim 100~t_{c} \sim 5$ Gyr.  The nascent
Milky Way was less massive than today, so the crossing time was likely
longer in the past and this estimate is probably a lower limit on the
look-back time to which our search is sensitive to.  This observation
also implies that there have been no significant mass-ratio mergers
in the past $\sim5$ Gyr.  This is consistent with observations of
the scale height of the thin disk, and suggests that the Milky Way has
a less active merger history than might be expected for a halo of its
mass \citep[e.g.][]{ste08}.  In summary, the union of the \citet{bell08}
result with out work suggests the stellar accretion history of the Milky
Way has been more or less constant with no significant mass-ratio mergers
over the past $\sim5$ Gyr.

Direct comparison of our results with more local searches using
6D phase-space \citep[e.g.][]{hel99a,gou03,kle08,sea08,kle09,mor09}
information is not as simple.  While it's true that in a given volume 6D
phase-space searches are sensitive to the oldest extant substructures
and capable of probing the accretion history farthest into the past,
the fact that there is currently no proper motion database that is
both sufficiently large and sufficiently precise in the volume we
search precludes a 6D search.  Published searches for substructure in
6D phase-space are limited to the local volume -- a region that does
not overlap with the volume we searched in the inner halo.  The rate
at which substructure phase mixes is a function of galactocentric
distance, so constructing a consistent accretion history including
both the radial velocity substructures we detect in the inner halo and
phase-space substructures discovered in the local volume is problematic.
While it appears that there is more substructure in 6D solar neighborhood
searches, perhaps to the point that there is no smooth component
as in \citet{mor09}, the dynamical effects of the disk on the local
volume population makes comparison between our result and local volume
results difficult \citep[e.g.][]{deh98,bov09}.  As a result, placing the
debris of the accretion activity identified in local volume samples in
the same accretion timeline as our results will require more detailed
modeling of the Milky Way's accretion and dynamical evolution than is
currently available.  In the future, Gaia measurements of the full 6D
phase-space distribution of SEGUE MPMSTO will allow us to construct a
self-consistent inner halo stellar accretion timeline all the way back
to the last major instance of violent relaxation.

It is also difficult to assess the relative performance of our algorithm
optimized to work on densely sampled in situ data with precise radial
velocities but imprecise distance estimates with algorithms designed
to work on very sparsely sampled in situ data with precise distance
estimates.  Searches like the latter are appropriate for distant,
luminous, and rare tracers of the outer halo \citep[e.g.][]{sta09}.

\subsection{Previously Known Substructure}

Thirteen lines of sight from the 137 total lines of sight in our sample
were targeted at known substructures: the \citet{gri06b} stream, the
Monoceros stream, the Orphan Stream, the Sagittarius stream, and the
Virgo stream.  Table~\ref{tbl-5} lists the radial velocities and mean
heliocentric distances associated with each detection for the lines
of sight targeted at known substructure for which we found an ECHOS.
For the lines of sight targeted at known substructure along which we have
non-detections, we list in Table~\ref{tbl-6} upper limits on the fraction
of the MPMSTO star population in ECHOS along those lines of sight.
We discuss each case in detail below.

\subsubsection{\citet{gri06b} Stream}

We detect the \citet{gri06b} stream along one line of sight listed
in Table~\ref{tbl-5}.  Our estimate of its radial velocity is based
on a line of sight with equatorial coordinates substantially different
from the fiducial radial velocity point given for the best fit model in
\citet{gri06b}.  Those authors used a line of sight centered at (RA,Dec)
= (202.0,58.4) and found $v_r = -208 \pm 30$ km s$^{-1}$; more precise
modeling is necessary to determine if the two observations are consistent.
Our estimate of $6.9^{+3.6}_{-1.6}$ kpc to the \citet{gri06b} stream is
in agreement with the heliocentric distance obtained by those authors
of 7.7 kpc.  \citet{wil09} used SEGUE spectroscopic data to derive the
properties of the \citet{gri06b} stream along the line of sight listed
in Table~\ref{tbl-5}; they found $v_r = -124$ km s$^{-1}$ and
a heliocentric distance of 8.8 kpc, both consistent with our measurements.
We also note that we resolve the \citet{gri06b} stream in
radial velocity and find that its radial velocity dispersion ($\sigma =
11.7$ km s$^{-1}$) is much hotter than the estimated radial velocity
errors (Err = 4.6 km s$^{-1}$) associated with the stars in our detection.

\begin{figure}
\epsscale{.40}
\plotone{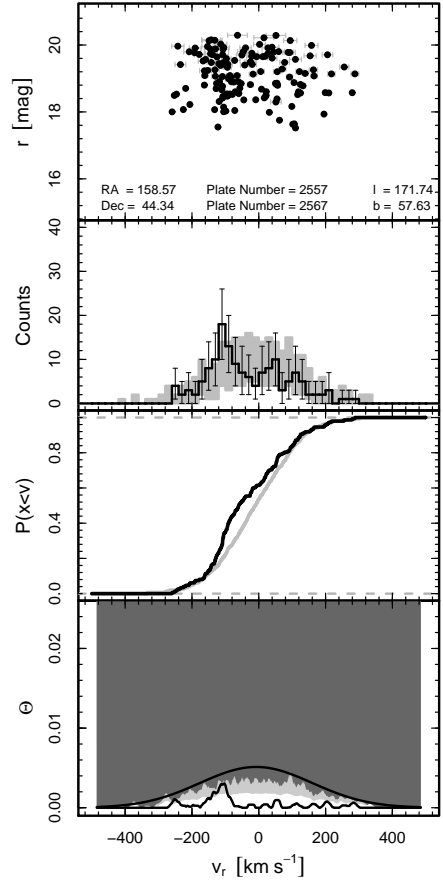}
\caption{Data and analyses for the line of sight target at (RA,Dec) =
(158.6,44.3) expected to intersect the \citet{gri06b} stream.  Note
the feature obvious to the naked eye but invisible to our detection
algorithms because of the strong dependence of the mean radial velocity
of the feature on $r$-magnitude and therefore distance.  The feature
has a velocity dispersion of at least 40 km s$^{-1}$ so it cannot be
described as cold.  That large velocity dispersion is not the reason
for its non-detection (as we showed in \S3.5.1); its non-detection is
due to the fact that the mean radial velocity of the apparent feature at
the bright end is offset by 40 km s$^{-1}$ from its mean radial velocity
at the faint end.  \label{fig23}}
\end{figure}

We fail to detect the \citet{gri06b} stream along two lines of
sight listed in Table~\ref{tbl-6}.  Our non-detection along the
line of sight targeted at (RA,Dec) = (217.7,58.2) is consistent with
Figure 1 of \citet{gri06b}, as the stream is nearly invisible in their
matched-filter analysis at those coordinates.  We plot the data for the
line of sight targeted at (RA,Dec) = (158.6,44.3) in Figure~\ref{fig23}
-- the substructure present in the data is manifestly not cold as it has
a velocity dispersion of at least 40 km s$^{-1}$.  That large velocity
dispersion is not the reason for its non-detection (as we showed in
\S3.5.1); its non-detection is due to the fact that the mean radial
velocity of the apparent feature at the bright end is offset by 40 km
s$^{-1}$ from its mean radial velocity at the faint end.

\subsubsection{Monoceros Stream}

We detect the Monoceros stream along two lines of sight listed in
Table~\ref{tbl-5}.  Our estimates for its heliocentric distance and radial
velocity are consistent with the comprehensive model for the Monoceros
stream given in \citet{pen05}, as well as with previous observational
results referenced therein.  We resolve the Monoceros stream in radial
velocity and find that its radial velocity dispersion ($\sigma = 10.2$ km
s$^{-1}$) is much hotter than the estimated radial velocity errors (Err =
4.0 km s$^{-1}$) associated with the stars in our detection.  We also note
that many of our low galactic latitude detections in Tables~\ref{tbl-1},
\ref{tbl-2}, and \ref{tbl-3} are plausibly related to Monceros as well.

We fail to detect the Monoceros stream along three lines of sight listed
in Table~\ref{tbl-6}.  Our non-detections along those lines of sight are
almost certainly because of our lack of sensitivity to substructure along
those lines of sight, primarily because most of the spectra obtained by
SEGUE along those lines of sight belonged to MPMSTO stars that did not
fall within the inner halo as defined in \S2.  As a result, we had too
few radial velocities to find anything significant.

\subsubsection{Orphan Stream}

We fail to detect the Orphan stream along three lines of sight listed
in Table~\ref{tbl-6}.  However, our non-detections along those lines of
sight are consistent with the \citet{bel07} analysis because they find
that the stream should be beyond our heliocentric distance 
threshold of 17.5 kpc at heliocentric distance between 20 and 30 kpc
at those coordinates.

\subsubsection{Sagittarius Stream}

We fail to detect the Sagittarius stream along one line of sight listed
in Table~\ref{tbl-6}.  However, our non-detection along that line of
sight is consistent with the comprehensive model for the Sagittarius
stream given in \citet{law05} as well as with previous observational
results referenced therein.  That is, \citet{law05} predict that the
Sagittarius tidal stream should be beyond our heliocentric distance
threshold of 17.5 kpc at a heliocentric distance $\gtrsim 20$ kpc.

\subsubsection{Virgo Stream}

We fail to detect the Virgo stream along one line of sight listed
in Table~\ref{tbl-6}.  However, our non-detection along that line of
sight is consistent with the analysis in \citet{duf06} because their RR
Lyrae were all beyond our heliocentric distance threshold of 17.5 kpc
at heliocentric distances greater than 18 kpc.

\subsection{Implications for the Formation of the Milky Way}

Our seven new high-confidence ECHOS greatly expand the known number
of inner halo substructures, and our detections of previously known
elements of substructure can be used to further constrain models of
the substructures themselves and their progenitors.  The existence of a
substantial population of low density ECHOS in the inner halo provides
a strong constraint that theoretical models must meet.  At the same
time, we note that our observed radial velocity distributions taken as
a whole give us little reason to reject a smooth model for the radial
velocity distribution of the inner halo.  These observations are therefore
consistent with inner halo formation scenarios in which relatively massive
protogalaxies accrete into the nascent Milky Way early on.  As a result
of the massive mergers, the potential of the nascent Milky Way changes
on short timescales, so violent relaxation smooths-out the stellar
phase-space distribution.  The substructure that remains is mostly
erased as the number of crossing times since accretion grows large.
On the other hand, \citet{mor09} examined a solar neighborhood sample
with 6D phase-space information and concluded that violent relaxation
was not efficient.  We expect that the efficiency of violent relaxation
in the Milky Way's past will remain an active area of research.

We cannot assess the relative probability that a single accretion event
produced the seven ECHOS we observe as compared to seven unique accretion
events, nor can we unambiguously determine a class of progenitors.
Our ECHOS could result from the disruption of globular clusters, small
dwarf spheroidal galaxies, large LMC-like progenitors of the bulk of the
inner halo, or from dynamical interactions of any of those three classes
of objects with the stellar disk of the nascent Milky Way. Nevertheless,
there is an enormous amount of information left to be extracted from
our detections.  Many authors \citep[e.g.][]{whe89,nis94,carr00} have
observed that inner halo stars in the solar neighborhood are enriched
in $\alpha$-elements relative to stars in surviving classical dwarf
spheroidal galaxies at constant [Fe/H].  \citet{rob05} and \citet{fon06}
explained this observation in the context of the hierarchical paradigm
by noting that most of the stars in the inner halo were formed in a few
relatively massive ($\sim 5 \times 10^{10} M_{\odot}$) protogalaxies that
merged with the nascent Milky Way $\sim$10 Gyr in the past.  The star
formation histories of those protogalaxies would therefore have been
sharply truncated, resulting in enrichment mostly by Type II supernovae.
On the other hand, the surviving classical dwarf spheroidals are
lower mass ($\sim 10^9 M_{\odot}$) with more sustained star formation
histories that allow for chemical enrichment by Type Ia supernovae.
Even moderate-resolution spectroscopic follow-up of our ECHOS using
the techniques presented in \citet{kir08} should reveal the degree
of $\alpha$-enhancement in the stars in each ECHOS.  This may even be
feasible using the subset of existing SEGUE spectra with sufficiently high
S/N \citep{lee09}.  In any case, individual stellar [Fe/H] and
[$\alpha$/Fe] measurements within an ECHOS will reveal the distribution
in composition within single, massive, long-ago disrupted inner halo
progenitors.  That information has the potential to uniquely inform not
only models of Milky Way formation, but also the physics of the high
redshift universe and the star formation environments in the ancient
massive stellar systems that merged with the nascent Milky Way to form
the inner halo.

\section{Conclusion}

We used the observed spatial and radial velocity distribution of
metal-poor main sequence turnoff (MPMSTO) stars in 137 Sloan Extension
for Galactic Understanding and Exploration (SEGUE) lines of sight to
identify ten -- seven for the first time -- high-confidence elements
of cold halo substructure (ECHOS) in the inner halo of the Milky Way,
none of which we expect to be false positives.  We also found 21 lower
confidence ECHOS of which we expect three to be false positives. ECHOS
are the debris of ancient merger events, and we used our detections and
completeness estimates to infer that at most $0.34^{+0.02}_{-0.02}$ of the
MPMSTO stars in the inner halo belong to ECHOS.  Our result also implies
that there exists a significant population of low fractional overdensity
ECHOS in the inner halo; we predict that 1/3 of the inner halo (by volume)
hosts low density ECHOS with number densities $n \approx 15$ kpc$^{-3}$
and that there are of order 10$^3$ ECHOS in the entire inner halo.
When combined with the work of \citet{bell08}, our result suggests that
there has been a constant rate of merger activity over the past $\sim5$
Gyr with no accretion of single stellar systems with mass more than a
few percent of a Milky Way mass in that time.

\acknowledgments We thank James Bullock, J\"{u}rg Diemand, Evan Kirby,
David Lai, Doug Lin, and Piero Madau for useful comments and conversation.
We are especially grateful to Carlos Allende Prieto, Heather Morrison, and
Matthias Steinmetz for their insightful comments on an earlier draft of
this paper.  This research has made use of NASA's Astrophysics Data System
Bibliographic Services.  This material is based upon work supported under
a National Science Foundation Graduate Research Fellowship.  T.C.B. and
Y.S.L. acknowledge partial support for this work from PHY 02-16783 and
PHY 08-22648: Physics Frontiers Center / Joint Institute for Nuclear
Astrophysics (JINA), awarded by the U.S. National Science Foundation.
Funding for the SDSS and SDSS-II has been provided by the Alfred
P. Sloan Foundation, the Participating Institutions, the National Science
Foundation, the U.S. Department of Energy, the National Aeronautics and
Space Administration, the Japanese Monbukagakusho, the Max Planck Society,
and the Higher Education Funding Council for England. The SDSS Web Site
is http://www.sdss.org/.

The SDSS is managed by the Astrophysical Research Consortium for the
Participating Institutions. The Participating Institutions are the
American Museum of Natural History, Astrophysical Institute Potsdam,
University of Basel, University of Cambridge, Case Western Reserve
University, University of Chicago, Drexel University, Fermilab, the
Institute for Advanced Study, the Japan Participation Group, Johns
Hopkins University, the Joint Institute for Nuclear Astrophysics, the
Kavli Institute for Particle Astrophysics and Cosmology, the Korean
Scientist Group, the Chinese Academy of Sciences (LAMOST), Los Alamos
National Laboratory, the Max-Planck-Institute for Astronomy (MPIA),
the Max-Planck-Institute for Astrophysics (MPA), New Mexico State
University, Ohio State University, University of Pittsburgh, University
of Portsmouth, Princeton University, the United States Naval Observatory,
and the University of Washington.

{\it Facilities:} \facility{Sloan}

\appendix

\section{A. Phase-Space Structure of the Smooth Component of the Inner Halo as
         Viewed from the Sun}

\begin{enumerate}
   \item
   We model the galactocentric position-space distribution of stars in
   the inner halo by a spherically symmetric power-law in radius with
   index $\alpha = -3.5$  \citep[e.g.][]{mor00,yan00,bell08}

   \begin{eqnarray} \label{Aeq1}
   \rho & \propto & r^{\alpha}
   \end{eqnarray}

   We draw $n_s$ radial coordinates $r'$ from the distribution described
   by Eq.~(\ref{Aeq1}).  We then draw random $\theta'$ and $\phi'$
   coordinates such that the points are spread uniformly over $4 \pi$
   steradians.  Together these three coordinates define the standard
   spherical coordinate vector $\mathbf{r}'$.

   \item
   We model the galactocentric spherical velocity-space
   distribution of stars in inner halo as a multivariate
   normal with mean $\mathbf{\mu}_{r,\theta,\phi}$ and
   variance-covariance matrix $\mathbf{\Sigma}_{r,\theta,\phi}$
   \citep[e.g.][]{som97,sir04a,sir04b,xue08}

   \begin{mathletters} \label{Aeq2}
   \begin{eqnarray}
   \mathbf{v}_{r,\theta,\phi}' & \sim &
   \mathcal{N}(\mu_{r,\theta,\phi},
   \mathbf{\Sigma}_{r,\theta,\phi})
   \end{eqnarray}

   \begin{eqnarray}
   \mathbf{\mu}_{r,\theta,\phi} & = & \left( \begin{array}{c}
   0 \\
   0 \\
   0 \end{array} \right)
   \end{eqnarray}

   \begin{eqnarray}
   \mathbf{\Sigma}_{r,\theta,\phi} & = & \left( \begin{array}{ccc}
   120^2 & 0 & 0 \\
   0 & 100^2 & 0 \\
   0 & 0 & 100^2 \end{array} \right)
   \end{eqnarray}
   \end{mathletters}

   We draw $n_s$ galactocentric spherical velocities
   $\mathbf{v}_{r,\theta,\phi}'$ from the distribution described
   by above and associate them with the position-space
   distribution derived in step \#1.

   \item
   We transform the galactocentric spherical velocities into Cartesian
   velocities using the transformation $\mathbf{v}_{x,y,z}' = \mathbf{A}
   \mathbf{v}_{r,\theta,\phi}'$ defined by the matrix $\mathbf{A}$

   \begin{eqnarray}
   \mathbf{A} & = & \left( \begin{array}{ccc}
   \cos \theta' \sin \phi' & -\sin \theta' & \cos \theta' \cos \phi'\\
   \sin \theta' \sin \phi' &  \cos \theta' & \sin \theta' \cos \phi'\\
                \cos \phi' &             0 &             -\sin \phi' \end{array}
   \right)
   \end{eqnarray}

   Then transform the velocity distribution into the Sun's standard of
   rest \citep[e.g.][]{deh98}

   \begin{mathletters}
   \begin{eqnarray}
   v_{x} & = & v_{x}' - v_{\odot,x}\\
   v_{y} & = & v_{y}' - v_{\odot,y}\\
   v_{z} & = & v_{z}' - v_{\odot,z}
   \end{eqnarray}
   \end{mathletters}

   \item
   We transform the galactocentric spherical coordinates $\mathbf{r}'$
   into galactocentric Cartesian coordinates $\mathbf{x}'$ with the
   usual transformation

   \begin{mathletters}
   \begin{eqnarray}
   x' & = & r' \cos \theta' \sin \phi'\\
   y' & = & r' \sin \theta' \sin \phi'\\
   z' & = & r' \cos \phi'
   \end{eqnarray}
   \end{mathletters}

   We translate the distribution such that the zero point of the
   $x$-coordinate corresponds with the position of the Sun

   \begin{mathletters}
   \begin{eqnarray}
   x & = & x' + 8\\
   y & = & y'\\
   z & = & z'
   \end{eqnarray}
   \end{mathletters}

   We compute Sun-centered spherical coordinates $\mathbf{r}$ from the
   Sun-centered Cartesian coordinates $x,y,z$.

   \item
   We project the Sun-centered Cartesian velocities onto the line of
   sight between the synthetic star and the Sun using the transformation
   $\mathbf{v}_{r,\theta,\phi} = \mathbf{B} \mathbf{v}_{x,y,z}$ defined
   by the matrix $\mathbf{B}$

   \begin{eqnarray}
   \mathbf{B} & = & \left( \begin{array}{ccc}
    \cos \theta \sin \phi             & \sin \theta \sin \phi             &              cos \phi\\
   \frac{-\sin \theta}{K r \sin \phi} & \frac{\cos \theta}{K r \sin \phi} &                     0\\
   \frac{ \cos \theta \cos \phi}{K r} & \frac{\sin \theta \cos \phi}{K r} & \frac{-\sin \phi}{K r}\end{array}
   \right)
   \end{eqnarray}

   where $K$ is a constant of proportionality between kilometers and
   kiloparsecs.  Finally, we eliminate all synthetic stars that would
   fall outside of our definition of the inner halo.

\end{enumerate}

\section{B. Detailed Description of the Bin Algorithm}

\begin{enumerate}
   \item
   Consider each SEGUE line of sight in sequence and let $n$ be the total
   number of MPMSTO star spectra obtained along that line of sight.
   We compute the histogram describing the MPMSTO star radial velocity
   distribution along that line of sight.  We use bootstrap resampling to
   estimate the uncertainty in the number of counts in each bin.
   \item
   Under the null hypothesis, the radial velocity distribution of
   the Milky Way's inner halo can be calculated as discussed in \S3.1.
   For each SEGUE line of sight, we determine which synthetic stars from our
   mock catalog fall within the volume scanned by SEGUE along that line
   of sight.  There are typically more than an order-of-magnitude more
   synthetic stars $m$ in a given patch of sky than the number of MPMSTO
   star spectra $n$ observed along the corresponding SEGUE line of sight.
   We select a random subsample $S$ of $n$ synthetic stars from the $m$
   available and compute the histogram of that subsample.
   \item
   We repeat step \#2 a large number of times.  In this way, we calculate
   the median histogram that results from observing the mock catalog
   a large number of times as well as distributions for the number of
   counts in each bin.  In this analysis, we have always repeated step
   \#2 $10^4$ times.
   \item
   We identify bins for which the distribution of counts estimated in step
   \#1 is inconsistent with the distribution calculated in step \#3 and flag
   the stars in that radial velocity bin as a potential element of cold
   substructure.  An inconsistent bin is one for which the 95\% confidence
   interval on the number of counts in the bin from the observed MPMSTO
   population from bootstrap resampling does not overlap with the 95\%
   confidence interval for the expected number of counts in the bin from
   the mock catalog under the null hypothesis.
\end{enumerate}

\section{C. Detailed Description of the Peak Algorithm}

\begin{enumerate}
   \item
   Consider each SEGUE line of sight in sequence and let $n$ be the total
   number of MPMSTO star spectra obtained along that line of sight.
   Let $F(v_r)$ and $F'(v_r)$ denote the CDF of the radial velocities
   observed along that line of sight and its slope respectively.
   \item
   Under the null hypothesis, the radial velocity distribution of
   the Milky Way's inner halo can be calculated as discussed in \S3.1.
   For each SEGUE line of sight, we determine which synthetic stars from our
   mock catalog fall within the volume scanned by SEGUE along that line
   of sight.  There are typically more than an order-of-magnitude more
   synthetic stars $m$ in a given patch of sky than the number of MPMSTO
   star spectra $n$ observed along the corresponding SEGUE line of sight.
   We select a random subsample $S$ of $n$ synthetic stars from the $m$
   available and compute the CDF $F_S(v_r)$ and its slope $F_S'(v_r)$
   of that subsample.
   \item
   We repeat step \#2 a large number of times.  In this way, we calculate
   the distribution of the CDF and its slope at each point in radial
   velocity-space.  Specifically, its average value $\overline{F_S}(v_r)$
   and an estimate of its average slope $\overline{F_S'}(v_r)$.  In this
   analysis, we have always repeated step \#2 $10^4$ times.
   \item
   Again we select a random subsample of $n$ stars from the $m$ available,
   and compute the CDF $F_S(v_r)$ and slope $F_S'(v_r)$ of this subsample.
   We then calculate the difference $\Theta_S(v_r) = F_S'(v_r) -
   \overline{F_S'}(v_r)$ and smooth it using a moving average kernel
   with its width set to 10 km s$^{-1}$, very close to the median velocity
   error of the MPMSTO sample.
   \item
   We repeat step \#4 a large number of times.  As a result, we calculate
   the distribution of $\Theta_S(v_r)$, or in other words, the differences
   between the average value of the slope $\overline{F_S'}(v_r)$ and
   a single random realization $F_S'(v_r)$ under the null hypothesis.
   In particular, we compute formal significance contours that correspond
   to 1 in 10$^2$, 1 in 10$^3$, and 1 in 10$^4$ events.  In this analysis,
   we have always repeated step \#2 $10^4$ times.  We emphasis that
   the distribution is not Gaussian; nevertheless, these significance
   thresholds would naively correspond to 2.33-$\sigma$, 3.09-$\sigma$,
   and 3.72-$\sigma$.
   \item
   We compute $\Theta(v_r) = F'(v_r) - \overline{F_S'}(v_r)$, the difference
   between the observed slope along a single SEGUE line of sight and
   the average slope under the null hypothesis, and smooth as before.
   Note that since we normalize the number of synthetic stars in $S$
   to the number of stars $n$ observed along the SEGUE line of sight,
   every interval in which $\Theta(v_r) < 0$ must necessarily correspond
   to an interval in which $\Theta(v_r) > 0$; only the intervals with
   $\Theta(v_r) > 0$ correspond to an overdensity.
   \item
   We flag any radial velocity $v_r$ at which $\Theta(v_r)$, the difference
   between the slope of the CDF of the MPMSTO radial velocity distribution
   and the average CDF of the smooth model, is significant at more than
   the 1 in 10$^4$ level as a potential element of cold substructure.
   We are formally limited to 1 in 10$^4$ events because of computational
   limits on the number of Monte Carlo iterations we can execute.
\end{enumerate}

\noindent
We can self-consistently estimate the radial velocity dispersion of the
candidate ECHOS identified by the peak algorithm by fitting a Gaussian
to the overdensity in $\Theta(v_r)$ in a window centered on the peak of
the overdensity with width six times our median velocity resolution, such
that the window contains 99\% of the signal from the detection.

\clearpage

\clearpage

\clearpage

\clearpage

\clearpage

\clearpage

\clearpage

\clearpage

\clearpage

\clearpage

\clearpage

\clearpage

\clearpage

\clearpage

\clearpage

\clearpage

\clearpage

\clearpage

\clearpage

\clearpage

\clearpage

\clearpage 

\clearpage

\clearpage
\begin{deluxetable}{lrrrrrrrrrc}
\tablecaption{Summary of Bin Detections\label{tbl-1}}
\tablewidth{0pt}
\tablehead{\colhead{ID} & \colhead{RA} & \colhead{Dec} & \colhead{l} &
\colhead{b} & \colhead{$d$\tablenotemark{a}} &
\colhead{$v_r$\tablenotemark{b}} & \colhead{$n_s$\tablenotemark{c}} &
\colhead{$N_s$\tablenotemark{d}} & \colhead{Figure} & \colhead{Comment}}
\startdata
B-1 & 21.3 & 39.6 & 130 & -22.8 & $12.9^{+5.0}_{-1.5}$ & -130 & 12 & 34 & 2 & - \\
B-2 & 17 & 0 & 132 & -62.6 & $10.5^{+5.2}_{-4.5}$ & -170 & 20 & 109 & 3a & - \\
B-3 & 39.7 & 28.2 & 150 & -29 & $10.1^{+2.3}_{-1.3}$ & -50 & 17 & 59 & 3b & Mon Strm \\
B-4 & 163.8 & 48 & 162.4 & 59.2 & $7^{+4.1}_{-1.8}$ & -130 & 25 & 150 & 4a & GD Strm \\
B-5 & 129.6 & 53.9 & 164.3 & 37.2 & $9.1^{+2.7}_{-2.1}$ & -10 & 20 & 93 & 4b & - \\
B-6 & 124.5 & 38 & 183.4 & 32.6 & $8.7^{+1.0}_{-0.6}$ & 30 & 17 & 83 & 5a & - \\
B-7 & 132.6 & 6.1 & 221.5 & 29.2 & $10^{+2.0}_{-0.9}$ & 70 & 17 & 69 & 5b & - \\
B-8 & 134 & 3.2 & 225.2 & 29 & $10.5^{+0.5}_{-1.4}$ & 90 & 19 & 74 & 6a & Mon Strm \\
\enddata
\tablenotetext{a}{Median heliocentric distance in kpc of MPMSTO stars in the
significant bin}
\tablenotetext{b}{Central radial velocity of the significant bin in km s$^{-1}$}
\tablenotetext{c}{Number of MPMSTO star radial velocities within the
significant bin}
\tablenotetext{d}{Number of MPMSTO star spectra obtained along that
line of sight that lie in the inner halo as defined in \S2}
\end{deluxetable}

\clearpage
\begin{deluxetable}{ccccccccccccc}
\tablecaption{Summary of Class I Peak Detections\label{tbl-2}}
\tablewidth{0pt}
\tablehead{\colhead{ID} & \colhead{RA} & \colhead{Dec} & \colhead{l} &
\colhead{b} & \colhead{$d$\tablenotemark{a}} &
\colhead{$v_r$\tablenotemark{b}} & \colhead{$\sigma$\tablenotemark{c}} &
\colhead{Err\tablenotemark{d}} & \colhead{$n_s$\tablenotemark{e}} &
\colhead{$N_s$\tablenotemark{f}} & \colhead{Figure} & \colhead{Comment}}
\startdata
PCI-1 & 214.8 & 56.4 & 100.7 & 56.8 & $10.4^{+2.9}_{-2.8}$ & -328 & 15.1 & 11.5 & 8 & 122 & 6b & - \\
PCI-2 & 20 & 31.7 & 130 & -30.8 & $12.5^{+3.9}_{-2.2}$ & -125 & 22 & 17.8 & 20 & 93 & 7a & - \\
PCI-3 & 21.3 & 39.6 & 130 & -22.8 & $13^{+4.6}_{-1.6}$ & -121 & 10.5 & 19.4 & 13 & 34 & 2 & - \\ 
PCI-4 & 39.7 & 28.2 & 150 & -29 & $10.5^{+2.7}_{-1.3}$ & -57 & 10.2 & 12.8 & 22 & 59 & 3b & Mon Strm \\ 
PCI-5 & 163.8 & 48 & 162.4 & 59.2 & $6.9^{+3.6}_{-1.6}$ & -132 & 11.7 & 4.6 & 22 & 150 & 4a & GD Strm \\
PCI-6 & 129.6 & 53.9 & 164.3 & 37.2 & $9.4^{+2.0}_{-2.4}$ & -13 & 10.9 & 9.8 & 20 & 93 & 4b & - \\
PCI-7 & 124.5 & 38 & 183.4 & 32.6 & $8.7^{+1.1}_{-0.7}$ & 29 & 19 & 10.6 & 16 & 83 & 5a & - \\
PCI-8 & 132.6 & 6.1 & 221.5 & 29.2 & $9.9^{+2.4}_{-1.0}$ & 71 & 13.2 & 11.4 & 15 & 69 & 5b & - \\
PCI-9 & 134 & 3.2 & 225.2 & 29 & $9.7^{+1.5}_{-0.5}$ & 85 & 14.9 & 4 & 17 & 74 & 6a & Mon Strm \\
\enddata
\tablenotetext{a}{Median heliocentric distance in kpc of MPMSTO stars with
radial velocities within one velocity resolution of the identified peak}
\tablenotetext{b}{Radial velocity in km s$^{-1}$ at which $\Theta(v_r)$ peaks}
\tablenotetext{c}{Velocity dispersion in km s$^{-1}$ of the significant peak}
\tablenotetext{d}{Median radial velocity error in km s$^{-1}$ for all MPMSTO
                  radial velocities within 12 km s$^{-1}$ of the peak in
                  $\Theta(v_r)$}
\tablenotetext{e}{Number of MPMSTO star radial velocities within one velocity
resolution of the identified peak}
\tablenotetext{f}{Number of MPMSTO star spectra obtained along that
line of sight that lie in the inner halo as defined in \S2}
\end{deluxetable}

\clearpage
\begin{deluxetable}{ccccccccccccc}
\tablecaption{Summary of Class II Peak Detections\label{tbl-3}}
\tablewidth{0pt}
\tablehead{\colhead{ID} & \colhead{RA} & \colhead{Dec} & \colhead{l} &
\colhead{b} & \colhead{$d$\tablenotemark{a}} &
\colhead{$v_r$\tablenotemark{b}} & \colhead{$\sigma$\tablenotemark{c}} &
\colhead{Err\tablenotemark{d}} & \colhead{$n_s$\tablenotemark{e}} &
\colhead{$N_s$\tablenotemark{f}} & \colhead{Figure} & \colhead{Comment}}
\startdata
PCII-1 & 20 & 31.7 & 130 & -30.8 & $12.5^{+3.9}_{-2.2}$ & -125 & 22 & 17.8 & 20 & 93 & 7a & - \\ 
PCII-2 & 20 & 31.7 & 130 & -30.8 & $10.6^{+4.2}_{-1.5}$ & -98 & 24.8 & 16.2 & 18 & 93 & 7a & - \\
PCII-3 & 21.3 & 39.6 & 130 & -22.8 & $13^{+4.6}_{-1.6}$ & -121 & 10.5 & 19.4 & 13 & 34 & 2 & - \\
PCII-4 & 91.8 & 83.5 & 130 & 25.7 & $11.5^{+1.5}_{-1.8}$ & -95 & 19.9 & 12 & 13 & 47 & 7b & - \\ 
PCII-5 & 17 & 0 & 132 & -62.6 & $11.6^{+4.2}_{-5.2}$ & -173 & 12.1 & 14.4 & 19 & 109 & 3a & - \\ 
PCII-6 & 38.2 & 25.5 & 150 & -32 & $10.4^{+3.0}_{-1.6}$ & -93 & 20.9 & 15.4 & 14 & 60 & 8a & - \\
PCII-7 & 38.2 & 25.5 & 150 & -32 & $11^{+1.7}_{-1.9}$ & -66 & 24.7 & 14.5 & 14 & 60 & 8a & - \\
PCII-8 & 39.7 & 28.2 & 150 & -29 & $10.5^{+2.7}_{-1.3}$ & -57 & 10.2 & 12.8 & 22 & 59 & 3b & Mon Strm \\
PCII-9 & 30 & 0 & 157 & -58.3 & $11.9^{+5.2}_{-4.1}$ & -177 & 19.1 & 15.9 & 18 & 173 & 8b & - \\ 
PCII-10 & 163.8 & 48 & 162.4 & 59.2 & $6.9^{+3.6}_{-1.6}$ & -132 & 11.7 & 4.6 & 22 & 150 & 4a & GD Strm \\
PCII-11 & 129.6 & 53.9 & 164.3 & 37.2 & $9.4^{+2.0}_{-2.4}$ & -13 & 10.9 & 9.8 & 20 & 93 & 4b & - \\
PCII-12 & 124.5 & 38 & 183.4 & 32.6 & $8.7^{+1.1}_{-0.7}$ & 29 & 19 & 10.6 & 16 & 83 & 5a & - \\ 
PCII-13 & 64.8 & 6.6 & 187 & -29.5 & $10.3^{+1.7}_{-0.9}$ & 20 & 28.3 & 16.1 & 12 & 65 & 9a & - \\
PCII-14 & 64.8 & 6.6 & 187 & -29.5 & $11.8^{+2.3}_{-1.5}$ & 44 & 31.3 & 17.6 & 14 & 65 & 9a & - \\
PCII-15 & 116.9 & 28 & 192.4 & 23.9 & $10.8^{+2.0}_{-0.5}$ & 44 & 9.1 & 12.1 & 10 & 35 & 9b & - \\
PCII-16 & 139.4 & 30.4 & 195.6 & 43.5 & $7.6^{+6.0}_{-1.1}$ & -103 & 13 & 10.1 & 11 & 114 & 10a & - \\
PCII-17 & 127.7 & 24.4 & 199.8 & 32 & $13.3^{+2.0}_{-1.9}$ & -40 & 18.1 & 17.7 & 10 & 83 & 10b & - \\
PCII-18 & 165.6 & 28.6 & 203.1 & 65.9 & $9.5^{+6.3}_{-4.0}$ & -157 & 13.3 & 14.2 & 10 & 151 & 11a & - \\
PCII-19 & 139.9 & 22.2 & 206.6 & 41.9 & $13.7^{+2.4}_{-5.0}$ & -55 & 27.3 & 16.3 & 11 & 102 & 11b & - \\
PCII-20 & 132.6 & 6.1 & 221.5 & 29.2 & $9.9^{+2.4}_{-1.0}$ & 71 & 13.2 & 11.4 & 15 & 69 & 5b & - \\
PCII-21 & 134 & 3.2 & 225.2 & 29 & $9.7^{+1.5}_{-0.5}$ & 85 & 14.9 & 4 & 17 & 74 & 6a & Mon Strm \\
\enddata
\tablenotetext{a}{Median heliocentric distance in kpc of MPMSTO stars with
radial velocities within one velocity resolution of the identified peak}
\tablenotetext{b}{Radial velocity in km s$^{-1}$ at which $\Theta(v_r)$ peaks}
\tablenotetext{c}{Velocity dispersion in km s$^{-1}$ of the significant peak}
\tablenotetext{d}{Median radial velocity error in km s$^{-1}$ for all MPMSTO
                  radial velocities within 12 km s$^{-1}$ of the peak in
                  $\Theta(v_r)$}
\tablenotetext{e}{Number of MPMSTO star radial velocities within one velocity
resolution of the identified peak}
\tablenotetext{f}{Number of MPMSTO star spectra obtained along that
line of sight that lie in the inner halo as defined in \S2}
\end{deluxetable}

\clearpage
\LongTables
\begin{deluxetable}{ccccccccccc}
\tablecaption{Summary of Completeness Calculation\label{tbl-4}}
\tablewidth{0pt}
\tablehead{\colhead{RA} & \colhead{Dec} & \colhead{l} & \colhead{b} &
\colhead{$N_s$\tablenotemark{a}} & \colhead{$N_p$\tablenotemark{b}} &
\colhead{Volume\tablenotemark{c}} & 
\colhead{$\Omega_b$\tablenotemark{d}} &
\colhead{$\Omega_{I}$\tablenotemark{e}} &
\colhead{$\Omega_{II}$\tablenotemark{f}} & \colhead{Comment}}
\startdata
207.2 & 18.6 & 3.2 & 74.3 & 61 & 359 & 2.7 & 0.55 & 0.48 & 0.43 & - \\
229.4 & 7.2 & 9.8 & 50 & 16 & 82 & 1.08 & $>$1 & 0.97 & $>$1 & - \\
243.8 & 16.7 & 31.4 & 41.9 & 20 & 89 & 1.11 & $>$1 & 0.82 & $>$1 & - \\
238.5 & 26.5 & 42.9 & 49.5 & 29 & 108 & 2.22 & 0.8 & 0.69 & 0.75 & - \\
253.1 & 24 & 44 & 36.1 & 16 & 135 & 1.28 & $>$1 & $>$1 & $>$1 & - \\
320.6 & -7.2 & 44.8 & -36.7 & 19 & 101 & 2.16 & $>$1 & $>$1 & $>$1 & - \\
311 & 0 & 46.6 & -24.8 & 10 & 437 & 1.06 & $>$1 & $>$1 & $>$1 & - \\
271.6 & 23.7 & 50 & 20 & 11 & 125 & 1.41 & $>$1 & $>$1 & $>$1 & - \\
266.5 & 25.4 & 50 & 25 & 12 & 188 & 1.07 & $>$1 & $>$1 & $>$1 & - \\
261.2 & 27 & 50 & 30 & 16 & 160 & 2.37 & $>$1 & $>$1 & $>$1 & - \\
317 & 0 & 50.1 & -30 & 12 & 434 & 1.64 & $>$1 & $>$1 & $>$1 & - \\
226.4 & 32.2 & 51 & 60.6 & 53 & 292 & 2.43 & 0.57 & 0.49 & 0.39 & - \\
263.1 & 33.2 & 57.4 & 30.1 & 22 & 321 & 1.73 & $>$1 & 0.84 & $>$1 & - \\
319 & 10.5 & 61.2 & -25.6 & 21 & 833 & 2.33 & $>$1 & 0.82 & $>$1 & - \\
344.7 & -9.4 & 61.3 & -58.1 & 49 & 566 & 3.22 & 0.7 & 0.77 & 0.77 & - \\
254.9 & 39.6 & 63.6 & 37.7 & 53 & 370 & 2.54 & 0.54 & 0.47 & 0.39 & - \\
231.4 & 39.4 & 64 & 55.8 & 41 & 439 & 2.49 & 0.66 & 0.54 & 0.51 & - \\
212.8 & 36.6 & 67.1 & 70.7 & 94 & 428 & 3.78 & 0.39 & 0.35 & 0.28 & - \\
332.8 & 6.4 & 67.8 & -38.8 & 40 & 322 & 2.66 & 0.64 & 0.56 & 0.54 & - \\
341 & 0 & 69.2 & -49.1 & 40 & 522 & 2.64 & 0.71 & 0.61 & 0.59 & - \\
263.4 & 44.2 & 70 & 32 & 53 & 455 & 2.83 & 0.57 & 0.47 & 0.39 & - \\
332.5 & 21.5 & 80.1 & -27.7 & 54 & 461 & 2.68 & 0.85 & $>$1 & $>$1 & - \\
344.8 & 7 & 80.4 & -46.4 & 80 & 482 & 2.94 & 0.45 & 0.38 & 0.3 & - \\
340.3 & 13.7 & 81 & -38.4 & 73 & 880 & 3.37 & 0.44 & 0.38 & 0.3 & - \\
231.8 & 49.9 & 81.1 & 52.7 & 110 & 561 & 3.19 & 0.37 & 0.31 & 0.27 & - \\
242.5 & 52.4 & 81.4 & 45.5 & 99 & 536 & 3.67 & 0.38 & 0.33 & 0.27 & - \\
217.2 & 45.3 & 82.5 & 63.5 & 92 & 579 & 3.42 & 0.39 & 0.36 & 0.28 & - \\
341 & 23.1 & 88.3 & -31.1 & 73 & 431 & 3.21 & 0.47 & 0.39 & 0.32 & - \\
356 & 0 & 89.3 & -58.4 & 68 & 419 & 3.17 & 0.57 & 0.49 & 0.48 & - \\
347.5 & 22.1 & 94 & -35 & 41 & 321 & 2.5 & 0.76 & 0.73 & 0.73 & - \\
262.6 & 64.4 & 94 & 33 & 71 & 551 & 2.76 & 0.59 & 0.6 & 0.6 & - \\
1 & -4.8 & 94 & -65 & 109 & 533 & 2.87 & 0.37 & 0.34 & 0.28 & - \\
247.2 & 62.8 & 94 & 40 & 85 & 601 & 3.01 & 0.45 & 0.39 & 0.35 & - \\
342.1 & 30.9 & 94 & -25 & 31 & 311 & 2.08 & 0.87 & 0.7 & 0.89 & - \\
354.5 & 8.7 & 94 & -50 & 73 & 486 & 2.37 & 0.44 & 0.39 & 0.3 & - \\
355.7 & 14.8 & 99.2 & -44.9 & 101 & 861 & 3.22 & 0.39 & 0.32 & 0.28 & - \\
217.7 & 58.2 & 100.6 & 54.4 & 93 & 664 & 3.02 & 0.39 & 0.35 & 0.28 & GD Strm \\
214.8 & 56.4 & 100.7 & 56.8 & 122 & 673 & 2.79 & 0.34 & 0.29 & 0.25 & - \\
6 & -10 & 101 & -71.7 & 90 & 905 & 3.47 & 0.39 & 0.37 & 0.29 & - \\
198 & 39.3 & 104.9 & 77.1 & 98 & 637 & 3.53 & 0.39 & 0.36 & 0.29 & - \\
1.2 & 25 & 109.8 & -36.7 & 61 & 309 & 2.5 & 0.48 & 0.4 & 0.34 & - \\
357.3 & 39.3 & 110 & -22 & 14 & 152 & 1.37 & $>$1 & $>$1 & $>$1 & Mon Strm \\
358.3 & 36.4 & 110 & -25 & 35 & 376 & 2.58 & 0.73 & 0.65 & 0.66 & - \\
311.2 & 76.2 & 110 & 20 & 10 & 30 & 0.596 & $>$1 & $>$1 & $>$1 & - \\
0.6 & 28.1 & 110 & -33.5 & 49 & 256 & 2.63 & 0.57 & 0.48 & 0.43 & - \\
9 & 7.5 & 116.3 & -55.2 & 99 & 406 & 3.35 & 0.45 & 0.42 & 0.42 & - \\
202.8 & 66.5 & 116.8 & 50.2 & 113 & 616 & 3.15 & 0.38 & 0.3 & 0.27 & - \\
11 & 0 & 118.9 & -62.8 & 104 & 488 & 3.21 & 0.39 & 0.34 & 0.29 & - \\
10.5 & 24.9 & 120.2 & -37.9 & 77 & 335 & 2.5 & 0.49 & 0.39 & 0.37 & - \\
11.2 & 14.9 & 120.6 & -47.9 & 89 & 739 & 2.82 & 0.4 & 0.35 & 0.29 & - \\
193 & 59.8 & 122.8 & 57.4 & 142 & 658 & 3.17 & 0.3 & 0.28 & 0.21 & - \\
192.8 & 49.7 & 123.1 & 67.4 & 142 & 696 & 3.18 & 0.29 & 0.27 & 0.2 & - \\
91.8 & 83.5 & 130 & 25.7 & 47 & 223 & 2.45 & 0.67 & 0.83 & 0.83 & - \\
20 & 31.7 & 130 & -30.8 & 93 & 349 & 2.6 & 0.53 & $>$1 & $>$1 & - \\
21.1 & 38.6 & 130 & -23.8 & 33 & 241 & 2.35 & 0.76 & 0.61 & 0.66 & - \\
17.9 & 15.6 & 130 & -47 & 81 & 352 & 3.04 & 0.41 & 0.36 & 0.28 & - \\
21.3 & 39.6 & 130 & -22.8 & 34 & 228 & 2.2 & 0.73 & 0.59 & 0.64 & - \\
19.1 & 25.7 & 130 & -36.8 & 53 & 262 & 2.51 & 0.59 & 0.49 & 0.47 & - \\
127.1 & 83.3 & 130 & 29.7 & 58 & 317 & 2.57 & 0.66 & 0.6 & 0.6 & - \\
17 & 0 & 132 & -62.6 & 109 & 561 & 3.48 & 0.38 & 0.34 & 0.29 & - \\
172.1 & 67 & 134.9 & 48.2 & 113 & 513 & 2.95 & 0.39 & 0.47 & 0.47 & - \\
24.7 & 23.7 & 136.7 & -37.9 & 52 & 241 & 2.87 & 0.93 & $>$1 & $>$1 & - \\
21.1 & 7.2 & 137.2 & -54.7 & 108 & 401 & 3.49 & 0.37 & 0.31 & 0.26 & - \\
181.9 & 50 & 140.2 & 65.7 & 132 & 639 & 3.33 & 0.31 & 0.28 & 0.21 & - \\
18.7 & -9.7 & 141.6 & -71.7 & 138 & 659 & 3.68 & 0.39 & 0.4 & 0.4 & - \\
26.7 & 14 & 142.7 & -46.8 & 105 & 679 & 3.17 & 0.38 & 0.31 & 0.27 & - \\
169.3 & 59 & 143.5 & 54.2 & 132 & 584 & 3.38 & 0.3 & 0.28 & 0.21 & - \\
32.2 & 22.5 & 145.5 & -36.9 & 68 & 294 & 2.74 & 0.65 & $>$1 & $>$1 & - \\
191.5 & 29.8 & 147 & 87 & 132 & 656 & 4.12 & 0.32 & 0.29 & 0.21 & - \\
39.7 & 28.2 & 150 & -29 & 59 & 265 & 2.3 & 0.52 & 0.44 & 0.38 & Mon Strm \\
116.2 & 66.1 & 150 & 30 & 59 & 286 & 2.3 & 0.56 & 0.46 & 0.44 & - \\
38.2 & 25.5 & 150 & -32 & 60 & 273 & 2.37 & 0.76 & $>$1 & $>$1 & - \\
43.6 & 34.3 & 150 & -22 & 22 & 263 & 2.23 & $>$1 & $>$1 & $>$1 & - \\
26 & 0 & 150 & -60.1 & 112 & 491 & 3.78 & 0.36 & 0.31 & 0.27 & - \\
146.4 & 62.1 & 150.9 & 43.6 & 110 & 395 & 3.29 & 0.39 & 0.36 & 0.32 & - \\
182.4 & 40 & 154.3 & 74.5 & 139 & 729 & 3.94 & 0.31 & 0.29 & 0.22 & - \\
33.2 & 6.6 & 156.2 & -50.9 & 105 & 409 & 3.21 & 0.66 & $>$1 & $>$1 & - \\
24.3 & -9.5 & 156.4 & -69.3 & 142 & 570 & 4.04 & 0.3 & 0.28 & 0.2 & - \\
30 & 0 & 157 & -58.3 & 173 & 987 & 4.16 & 0.38 & 0.5 & 0.5 & - \\
25.3 & -9.4 & 158.8 & -68.7 & 112 & 630 & 3.88 & 0.38 & 0.31 & 0.27 & Sgr Strm \\
163.8 & 48 & 162.4 & 59.2 & 150 & 672 & 3.86 & 0.29 & 0.28 & 0.2 & GD Strm \\
144.7 & 52.9 & 163.5 & 46.2 & 114 & 414 & 3.58 & 0.34 & 0.29 & 0.2 & - \\
129.6 & 53.9 & 164.3 & 37.2 & 93 & 425 & 3.33 & 0.4 & 0.34 & 0.3 & - \\
45.2 & 5.7 & 171.4 & -44.6 & 61 & 412 & 3.39 & 0.57 & 0.56 & 0.55 & - \\
158.6 & 44.3 & 171.7 & 57.6 & 148 & 636 & 3.45 & 0.29 & 0.27 & 0.21 & GD Strm \\
48.2 & 5.5 & 174.6 & -42.7 & 40 & 335 & 2.9 & 0.7 & 0.58 & 0.57 & - \\
51.2 & 5.2 & 177.7 & -40.8 & 53 & 353 & 2.64 & 0.6 & 0.59 & 0.59 & - \\
57.2 & 10.3 & 178 & -33 & 55 & 297 & 2.66 & 0.49 & 0.43 & 0.39 & - \\
113.5 & 40.9 & 178 & 25 & 35 & 164 & 2.35 & 0.78 & 0.61 & 0.78 & - \\
167 & 38.6 & 178.4 & 65.5 & 135 & 634 & 3.87 & 0.32 & 0.29 & 0.21 & - \\
37.4 & -8.5 & 178.7 & -60.2 & 156 & 513 & 3.9 & 0.3 & 0.29 & 0.24 & - \\
47 & 0 & 179 & -47.4 & 102 & 460 & 3 & 0.39 & 0.32 & 0.29 & - \\
111.3 & 37.6 & 180.9 & 22.4 & 18 & 109 & 1.62 & $>$1 & $>$1 & $>$1 & Mon Strm \\
112.5 & 36 & 182.9 & 22.9 & 17 & 146 & 1.6 & $>$1 & $>$1 & $>$1 & - \\
124.5 & 38 & 183.4 & 32.6 & 83 & 514 & 3.46 & 0.39 & 0.35 & 0.29 & - \\
53 & 0 & 184.5 & -42.9 & 73 & 414 & 3.57 & 0.43 & 0.38 & 0.3 & - \\
134.4 & 37.1 & 185.9 & 40.3 & 103 & 472 & 3.01 & 0.38 & 0.35 & 0.34 & - \\
64.8 & 6.6 & 187 & -29.5 & 65 & 353 & 1.99 & 0.48 & 0.39 & 0.3 & - \\
110.7 & 31.4 & 187 & 20 & 10 & 56 & 1.58 & $>$1 & $>$1 & $>$1 & - \\
152.5 & 35.3 & 189.4 & 54.8 & 126 & 570 & 3.7 & 0.37 & 0.38 & 0.37 & - \\
116.9 & 28 & 192.4 & 23.9 & 35 & 254 & 3.12 & 0.89 & $>$1 & $>$1 & - \\
55.4 & -6.4 & 193.7 & -44.6 & 100 & 402 & 4 & 0.37 & 0.3 & 0.26 & - \\
139.4 & 30.4 & 195.6 & 43.5 & 114 & 527 & 3.5 & 0.35 & 0.29 & 0.26 & - \\
59.4 & -5.9 & 195.9 & -40.9 & 78 & 400 & 3.15 & 0.46 & 0.39 & 0.3 & - \\
144 & 30.1 & 197 & 47.3 & 117 & 618 & 3.53 & 0.36 & 0.29 & 0.26 & - \\
118 & 23.2 & 197.7 & 23.2 & 27 & 128 & 2.59 & 0.89 & 0.78 & 0.87 & Mon Strm \\
127.7 & 24.4 & 199.8 & 32 & 83 & 431 & 3.05 & 0.39 & 0.33 & 0.28 & - \\
116 & 18.2 & 202 & 19.5 & 11 & 61 & 1.3 & $>$1 & $>$1 & $>$1 & - \\
71.4 & -5.7 & 203 & -30.5 & 66 & 343 & 2.41 & 0.45 & 0.39 & 0.34 & - \\
165.6 & 28.6 & 203.1 & 65.9 & 151 & 764 & 3.21 & 0.39 & 0.6 & 0.6 & - \\
152.4 & 25.9 & 205.4 & 53.9 & 126 & 644 & 3.61 & 0.35 & 0.29 & 0.23 & Orph Strm \\
139.9 & 22.2 & 206.6 & 41.9 & 102 & 609 & 2.72 & 0.4 & 0.36 & 0.33 & - \\
156.5 & 17.7 & 220.9 & 55.3 & 124 & 736 & 3.29 & 0.35 & 0.29 & 0.26 & Orph Strm \\
132.6 & 6.1 & 221.5 & 29.2 & 69 & 470 & 3.04 & 0.47 & 0.39 & 0.33 & - \\
134 & 3.2 & 225.2 & 29 & 74 & 514 & 3.5 & 0.45 & 0.38 & 0.3 & Mon Strm \\
141.6 & 7.3 & 225.3 & 37.6 & 114 & 625 & 3.86 & 0.36 & 0.29 & 0.27 & - \\
169.1 & 19.3 & 227.6 & 66.8 & 125 & 864 & 3.53 & 0.33 & 0.29 & 0.24 & - \\
128 & -4.3 & 229 & 20 & 11 & 101 & 1.34 & $>$1 & $>$1 & $>$1 & - \\
156.6 & 8.8 & 234.2 & 51.2 & 117 & 752 & 3.51 & 0.43 & 0.5 & 0.5 & - \\
150 & 0 & 239.1 & 40.7 & 101 & 723 & 3.19 & 0.36 & 0.29 & 0.25 & - \\
181.8 & 20 & 245.9 & 77.6 & 91 & 797 & 3.63 & 0.39 & 0.35 & 0.28 & - \\
168.8 & 9.6 & 246 & 61.3 & 159 & 858 & 3.75 & 0.29 & 0.27 & 0.19 & - \\
162 & 0 & 250.3 & 49.8 & 165 & 2083 & 3.69 & 0.28 & 0.26 & 0.19 & Orph Strm \\
174 & 0 & 266.1 & 57.4 & 102 & 913 & 3.15 & 0.37 & 0.32 & 0.27 & - \\
169.7 & -11.9 & 270 & 45 & 114 & 1039 & 3.14 & 0.38 & 0.3 & 0.29 & - \\
167.2 & -16.2 & 270 & 40 & 91 & 997 & 3.13 & 0.39 & 0.36 & 0.28 & - \\
172.2 & -7.5 & 270 & 50 & 95 & 1018 & 3.01 & 0.37 & 0.34 & 0.28 & - \\
181 & 0 & 278.2 & 60.6 & 68 & 778 & 2.6 & 0.48 & 0.39 & 0.33 & - \\
186 & 0 & 288.2 & 62.1 & 82 & 774 & 2.89 & 0.47 & 0.4 & 0.36 & Vir Strm \\
189 & 0 & 294.5 & 62.6 & 73 & 749 & 2.99 & 0.45 & 0.39 & 0.32 & - \\
191 & -2.5 & 299.2 & 60.3 & 53 & 631 & 2.29 & 0.56 & 0.47 & 0.4 & - \\
191.2 & -7.8 & 300 & 55 & 71 & 577 & 2.42 & 0.49 & 0.43 & 0.39 & - \\
193.1 & 9.9 & 303.8 & 72.8 & 100 & 685 & 2.79 & 0.38 & 0.34 & 0.27 & - \\
198 & 0 & 314.1 & 62.4 & 38 & 476 & 2.43 & 0.68 & 0.57 & 0.57 & - \\
194.6 & 19.7 & 315.3 & 82.5 & 94 & 573 & 3.05 & 0.48 & 0.5 & 0.5 & - \\
205.3 & 9.4 & 338.8 & 68.7 & 44 & 297 & 2.64 & 0.6 & 0.54 & 0.48 & - \\
217.4 & 8.5 & 358.7 & 60.2 & 22 & 135 & 2 & $>$1 & 0.91 & $>$1 & - \\
\enddata
\tablenotetext{a}{Number of spectra obtained along that line of sight
that lie in the inner halo as defined in \S2}
\tablenotetext{b}{Number of photometrically classified MPMSTO stars in the
volume scanned by SEGUE along that line of sight that lie in the inner halo as
defined in \S2}
\tablenotetext{c}{In kpc$^{3}$}
\tablenotetext{d}{The fraction of the total MPMSTO spectra sample that
must belong to a single cold element of substructure for it to be
classified as a bin detection 95\% of the time.}
\tablenotetext{e}{The fraction of the total MPMSTO spectra sample that
must belong to a single cold element of substructure for the it to be
classified as a class I detection 95\% of the time}
\tablenotetext{f}{The fraction of the total MPMSTO spectra sample that
must belong to a single cold element of substructure for the it to be
classified as a class II detection 95\% of the time}
\end{deluxetable}

\clearpage
\begin{deluxetable}{lcccccccc}
\tablecaption{Known Substructure Our Algorithms Recover\label{tbl-5}}
\tablewidth{0pt}
\tablehead{\colhead{Stream} & \colhead{RA} & \colhead{Dec} & \colhead{l} &
\colhead{b} & \colhead{$d$\tablenotemark{a}} & \colhead{$v_r$\tablenotemark{b}} &
\colhead{$\sigma$\tablenotemark{c}} & \colhead{Err\tablenotemark{d}}}
\startdata
Grillmair \& Dionatos & 163.8 & 48 & 162.4 & 59.2 & $6.9^{+3.6}_{-1.6}$ & -132 & 11.7 & 4.6 \\
Monoceros & 39.7 & 28.2 & 150 & -29 & $10.5^{+2.7}_{-1.3}$ & -57 & 10.2 & 12.8\\
Monoceros & 134 & 3.2 & 225.2 & 29 & $9.7^{+1.5}_{-0.5}$ & 85 & 14.9 & 4 \\
\enddata
\tablenotetext{a}{Median heliocentric distance in kpc}
\tablenotetext{b}{Radial velocity in km s$^{-1}$ at which $\Theta(v_r)$ peaks}
\tablenotetext{c}{Velocity dispersion in km s$^{-1}$ of the significant peak}
\tablenotetext{d}{Median radial velocity error in km s$^{-1}$ for all MPMSTO
                  radial velocities within 12 km s$^{-1}$ of the peak in
                  $\Theta(v_r)$}
\end{deluxetable}

\clearpage
\begin{deluxetable}{lccccccc}
\tablecaption{Known Substructure Our Algorithms Do Not Recover\label{tbl-6}}
\tablewidth{0pt}
\tablehead{\colhead{Stream} & \colhead{RA} & \colhead{Dec} & \colhead{l} &
\colhead{b} & \colhead{$\Omega_b$\tablenotemark{a}} &
\colhead{$\Omega_{I}$\tablenotemark{b}} &
\colhead{$\Omega_{II}$\tablenotemark{c}}}
\startdata
Grillmair \& Dionatos & 217.7 & 58.2 & 100.6 & 54.4 & 0.39 & 0.35 & 0.28 \\
Grillmair \& Dionatos & 158.6 & 44.3 & 171.7 & 57.6 & 0.29 & 0.27 & 0.21 \\
Monoceros & 357.3 & 39.3 & 110 & -22 & $>$1 & $>$1 & $>$1 \\
Monoceros & 111.3 & 37.6 & 180.9 & 22.4 & $>$1 & $>$1 & $>$1 \\
Monoceros & 118 & 23.2 & 197.7 & 23.2 & 0.89 & 0.78 & 0.87 \\
Orphan & 152.4 & 25.9 & 205.4 & 53.9 & 0.35 & 0.29 & 0.23 \\
Orphan & 156.5 & 17.7 & 220.9 & 55.3 & 0.35 & 0.29 & 0.26 \\
Orphan & 162 & 0 & 250.3 & 49.8 & 0.28 & 0.26 & 0.19 \\
Sagittarius & 25.3 & -9.4 & 158.8 & -68.7 & 0.38 & 0.31 & 0.27 \\
Virgo & 186 & 0 & 288.2 & 62.1 & 0.47 & 0.4 & 0.36 \\
\enddata
\tablenotetext{a}{The fraction of the total MPMSTO spectra sample that
must belong to a single cold element of substructure for the bin algorithm
to have a 95\% chance of detecting it.}
\tablenotetext{b}{The fraction of the total MPMSTO spectra sample that
must belong to a single cold element of substructure for the it to be
classified as a class I detection 95\% of the time.}
\tablenotetext{c}{The fraction of the total MPMSTO spectra sample that
must belong to a single cold element of substructure for the it to be
classified as a class II detection 95\% of the time.}
\end{deluxetable}

\end{document}